\UseRawInputEncoding
\documentclass[
 aps,prx,twocolumn,superscriptaddress,
 amsmath,amssymb,amsfonts,
 floatfix
]{revtex4-2}
\usepackage{graphicx}
\usepackage{bm}
\usepackage{appendix}
\usepackage{soul}
\usepackage[dvipsnames]{xcolor}
\usepackage{hyperref}
\usepackage{listings}
\usepackage{xcolor}

\definecolor{codegreen}{rgb}{0,0.6,0}
\definecolor{codegray}{rgb}{0.5,0.5,0.5}
\definecolor{codepurple}{rgb}{0.58,0,0.82}
\definecolor{backcolour}{rgb}{0.95,0.95,0.92}

\lstdefinestyle{mystyle}{
    backgroundcolor=\color{backcolour},   
    commentstyle=\color{codegreen},
    keywordstyle=\color{magenta},
    numberstyle=\tiny\color{codegray},
    stringstyle=\color{codepurple},
    basicstyle=\ttfamily\footnotesize,
    breakatwhitespace=false,         
    breaklines=true,                 
    captionpos=b,                    
    keepspaces=true,                 
    numbers=left,                    
    numbersep=5pt,                  
    showspaces=false,                
    showstringspaces=false,
    showtabs=false,                  
    tabsize=2
}

\hypersetup{
    colorlinks=true,
    citecolor=Green,
    linkcolor=Red,
    urlcolor=NavyBlue
}
\usepackage[capitalise]{cleveref}
\usepackage[italicdiff]{physics}
\usepackage{relsize}

\usepackage{ulem}

\begin{abstract}
It is well know that many full atomic multiplet codes are available for experimentalists to check x-ray absorption or emission spectra against known valence materials to identify effect valence configuration of transition metal ions as well as their ligands. This has grown recently with the continued development of resonant inelastic x-ray scattering as a general tool that can characterize orbital, spin, charge, and lattice excitations in quantum materials. In this note, I show that all multiplet eigenstates for most $f^n$ configurations can be obtained analytically. I correct prior misprints in the literature and present new results for $f-$ electrons. These results can serve as checks against new and developed numerical codes, and further provide experimentalists with deeper insights into the many configurations probed by advanced x-ray spectroscopies.
\end{abstract}

\begin{document}

\title{Analytic Expressions for Most \texorpdfstring{$f^n$}{TEXT}  Valence Multiplet Eigenvalues}

\author{Thomas P. Devereaux}
\email{tpd@stanford.edu}
\affiliation{Stanford Institute for Materials and Energy Sciences,
SLAC National Accelerator Laboratory, 2575 Sand Hill Road, Menlo Park, CA 94025, USA}
\affiliation{
Department of Materials Science and Engineering, Stanford University, Stanford, CA 94305, USA}
\affiliation{
Geballe Laboratory for Advanced Materials, Stanford University, Stanford, CA 94305, USA}

\maketitle

\section{Introduction}
X-ray spectroscopy is a vital tool to determine local atomic coordination number via Extended X-ray Absorption Fine Structure (EXAFS). X-rays are used to excite an inner-shell electron in an atom, creating a photoelectron. This photoelectron then scatters off neighboring atoms, creating an interference pattern that modulates the absorption coefficient, and the oscillations of the signal beyond the absorption edge can reveal information about the neighboring atoms' identity, distance, and coordination number. Numerical toolboxes such as FEFF have enjoyed widespread use across many scientific disciplines for many years.

Likewise, X-ray absorption (XAS) at the element-specific edges can reveal information about oxidation states, local atomic environments, and chemical bonding. Such broad applicability of detecting local atomic and electronic structure 
XAS finds homes in various scientific fields, including materials science, chemistry, and biology. It offers unique insights into the local structure of materials, making it a valuable tool for studying a wide range of materials, including amorphous solids, crystalline materials, solutions, and biological samples. There are many available application softwares, such as CTM4XAS and Quanty, that offer versatile platforms for calculation XAS for comparison with experiments. 

More recently, the emergence of resonant inelastic x-ray scattering
(RIXS) has impacted materials science investigations of collective excitations in quantum materials, sensing and characterizing momentum-dependent excited charge, lattice, orbital, and spin degrees of freedom with atomic specificity. Codes, such as EDRIXS and WebXRS have become available for RIXS, albeit for small cluster applications. More advanced codes, such as OCEAN, have moving into the direction of calculating RIXS for wider applicability to different materials.

Thus, the field of x-ray spectroscopy is well advanced. These advances have been aided by deep insights into the understanding of multiplets - the states obtained for valence electrons with full Coulomb matrix elements - that form the core of all of the above techniques. Since the early works of many in the 1930's, 40's and 50's, such as Slater and Racah, these multiplet states have been well-known for $p, d$ and some low $f$ configurations, and often feature in advanced classes on spectroscopy \cite{Slater,Slater1929,Racah1942, Racah1949,Ballhausen,Sugano,Cowan,degroot}. Multiplet-averaged Coulomb interactions were derived by Slater \cite{Slater} and the Hund's rule ground states and effective $U$ parameter were derived by van der Marel and Sawatzky in Ref. \cite{vanderMarel1988}. However, to the author's knowledge, no listing or table exists for all $f$ valence configurations. 

X-ray tools are valuable to probe $f$-electron occupations, as well as the spin and orbital moments, and the degree of covalency in actinide or so-called "critical" materials, essential for understanding their magnetic and charge properties. X-rays are particularly well suited for determining configurations representing so called mixed-or intermediate valent states of $f-$electrons, as well as their hybridization and covalency with $p, d$ electrons.  The lack of analytic results to be used to check against numerical methods has contributed to why this part of the periodic table remains least understood.

Therefore, the goal of this paper is to provide such a table for all multiplet states analytically. It is almost entirely successful, and importantly, we correct misprints in the literature for $d-$ and existing $f-$ electron multiplets.

\section{Method}
The general form for the Coulomb operator describing how two particles scatter from $a,\sigma,b;\sigma'$ to $c,\sigma,d;\sigma'$ can be written as
\begin{align*}
\hat H_{int}=\frac{1}{2} \sum_{c,d,b,a,\sigma,\sigma'} U_{\sigma,\sigma'}(c,d,b,a) ~\hat c^{\dagger}_{c,\sigma} \hat c^{\dagger}_{d,\sigma'} \hat c_{b,\sigma'} \hat c_{a,\sigma},
\label{Eq:Hint}
\end{align*}
with $\hat c, \hat c^\dagger$ are the annihilation, creation operators, and the order of the operators  defines the vacuum. 
The Coulomb matrix elements, which feature prominently here, are written in terms of single-electron basis functions $\psi$ as
\begin{widetext}
\begin{equation}
U_{\sigma,\sigma'}(c,d,b,a) = \int d{\bf r} \int d{\bf r'} \psi^*_{c,\sigma}({\bf r}) \psi^*_{d,\sigma'}({\bf r'}) \psi_{b,\sigma'}({\bf r'}) \psi_{a,\sigma}({\bf r}) ~\frac{e^2}{\mid {\bf r-r'}\mid} .
\end{equation}
As is typically done, I will take the orbital part of the basis functions to be described in terms of spherical harmonics, leaving the radial part independent of spin and characterized by Slater integrals.

The Coulomb matrix element $U$ for particles having quantum numbers $n,l,m$ can be expressed by Gaunt coefficients \cite{Slater,Sugano}
\begin{equation}
c^k(l,m;l',m')=\sqrt{\frac{4\pi}{2k+1}} \int_0^{2\pi} d\phi \int_{-1}^1 d\cos(\theta) Y_{l,m}^* (\theta,\phi)
Y_{k,m-m'}(\theta,\phi)Y_{l',m'}(\theta,\phi).
\end{equation}
as
\begin{eqnarray}
&&U(n_1,l_1,m_1;n_2,l_2,m_2;n_3,l_3,m_3;n_4,l_4,m_4)=\delta_{m_1+m_2,m_3+m_4} \nonumber\\
&\times&\sum_{k=0}^{\infty} c^k(l_1,m_1;l_4,m_4) c^k(l_3,m_3;l_2,m_2) R^k(n_1,l_1;n_2,l_2;n_3,l_3;n_4,l_4),
\label{Eq:UME}
\end{eqnarray} 
where $R^k$ are the radial integrals of the orbitals depending on quantum numbers $\{n,l\}$. 
\begin{align}
R^k(n_1,l_1;n_2,l_2;n_3,l_3;n_4,l_4)=\int_{0}^{\infty} dr~r^2 \int_{0}^{\infty} dr' r'^2~R_{n_1,l_1}(r)R_{n_2,l_2}(r') \frac{r^k_<}{r^{k+1}_>}R_{n_4,l_4}(r)R_{n_3,l_3}(r')
\label{Eq:SlaterCondon}    
\end{align}
The delta function in Eq. \ref{Eq:UME} conserves total azimuthal angular momentum of the pairs of particles.
\end{widetext}

For like orbitals, the radial integrals are expressed in terms of the Slater integrals \cite{Slater1929}:
\begin{align*}
F^k(n,l;n,l)&=
\int_{0}^{\infty} r^2dr \int_{0}^{\infty} r'^2dr' \frac{r^k_<}{r^{k+1}_>}R^2_{n,l}(r)R^2_{n,l}(r')
\end{align*}
I note that the Coulomb matrix elements can be in terms of different choices of parameters that differ in terms of their normalization coming from the Gaunt coefficients. Slater integrals appearing with k as a super- (Slater-Condon) or sub-script (Slater) are related as shown in Table \ref{table0}.

Tables for Coulomb matrix elements are expressed in terms of Slater integrals $F_k$ for l = 1 (Table \ref{table1}), l = 2 (Table \ref{table2}), and l = 3 (Table \ref{table3}) for different
combinations of azimuthal angular momentum quantum numbers $m_l$, derived from Eq. \ref{Eq:UME} and the Gaunt coefficients. The results for the multiplets will be given in terms of these Slater parameters.

\begin{table}[]
\centering
\begin{tabular}{|c|c c c c|}
\hline
 & $s$  &$p$  & $d$ & $f$ \\\hline
$F_0$ & $F^0$ & $F^0$ &$F^0$ &$F^0$ \\ 
$F^2$ & $-$ & $25F_2$ & $49F_2$ & $225F_2$\\ 
$F^4$ & $-$ &$-$ &   $441F_4$  &  $1089F_4$   \\
$F^6$ & $-$ & $-$ & $-$ & $\frac{184041}{25}F_6$  \\ 
\hline
\end{tabular}
\vskip 0.5cm
\caption{Relation between Slater-Condon $F^k$ and Slater $F_k$ parameters.}
\label{table0}
\end{table}

\begin{table}[]
\centering
\begin{tabular}{|c|c|c|c||c|}
\hline
$i$ & $j$ & $k$ & $l$ & $U(i,j,k,l)$\\\hline
-1 & -1 & -1 & -1 & $F_0+F_2$\\\hline
-1 & 0 & -1 & 0 & $3F_2$\\\hline
-1 & 0 & 0 & -1 & $F_0-2F_2$\\\hline
-1 & 1 & -1 & 1 & $6F_2$\\\hline
-1 & 1 & 0 & 0 & $3F_2$\\\hline
-1 & 1 & 1 & -1 & $F_0+F_2$\\\hline
0 & -1 & 0 & -1 & $3F_2$\\\hline
0 & -1 & -1 & 0 & $F_0-2F_2$\\\hline
0 & 0 & 0 & 0 & $F_0+4F_2$\\\hline
\end{tabular}
\vskip 0.5cm
\caption{Coulomb matrix elements for $l=1$. $U(i,j,k,l)=U(-i,-j,-k,-l)$.}
\label{table1}
\end{table}

\begin{table}[]
\centering
\resizebox{\columnwidth}{!}{%
\begin{tabular}{|c|c|c|c||c|}
\hline
$i$ & $j$ & $k$ & $l$ & $U(i,j,k,l)$\\\hline
-2 & -2 & -2 & -2 & $F_0+4F_2+F_4$\\\hline
-2 & -1 & -2 & -1 & $6F_2+5F_4$\\\hline
-2 & -1 & -1 & -2 & $F_0-2F_2-4F_4$\\\hline
-2 & 0 & -2 & 0 & $4F_2+15F_4$\\\hline
-2 & 0 & -1 & -1 & $\sqrt{6}F_2-5\sqrt{6}F_4$\\\hline
-2 & 0 & 0 & 2 & $6\sqrt{70}F_4$\\\hline
-2 & 1 & -2 & 1 & $35F_4$\\\hline
-2 & 1 &  -1 & 0 & $2\sqrt{6}F_2-10\sqrt{6}F_4$\\\hline
-2 & 1 & 0 & -1 & $-\sqrt{6}F_2+5\sqrt{6}F_4$\\\hline
-2 & 1 & 1 & -2 & $F_0-2F_2-4F_4$\\\hline
-2 & 2 & -2 & 2 & $70F_4$\\\hline
-2 & 2 &  -1 & 1 & $-35F_4$\\\hline
-2 & 2 & 0 & 0 & $4F_2+15F_4$\\\hline
-2 & 2 & 1 & -1 & $-6F_2-5F_4$\\\hline
-2 & 2 & 2 & -2 & $F_0+4F_2+F_6$\\\hline
\end{tabular}
\begin{tabular}{|c|c|c|c||c|}
\hline
$i$ & $j$ & $k$ & $l$ & $U(i,j,k,l)$\\\hline
-1 & -2 & -2 & -1 & $F_0-2F_2-4F_4$\\\hline
-1 & -2 & -1 & -2 & $6F_2+5F_4$\\\hline
-1 & -1 & -2 & 0 & $\sqrt{6}F_2-5\sqrt{6}F_4$\\\hline
-1 & -1 & -1 & -1 & $F_0+F_2+16F_4$\\\hline
-1 & -1 & 0 & -2 & $\sqrt{6}F_2-5\sqrt{6}F_4$\\\hline
-1 & 0 & -2 & 1 & $2\sqrt{3}F_2-10\sqrt{6}F_4$\\\hline
-1 & 0 & -1 & 0 & $F_2+30F_4$\\\hline
-1 & 0 & 0 & -1 & $F_0+2F_2-24F_4$\\\hline
-1 & 0 & 1 & -2 & $-\sqrt{6}F_2+5\sqrt{6}F_4$\\\hline
-1 & 1 & -2 & 2 & $-35F_4$\\\hline
-1 & 1 & -1 & 1 & $6F_2+40F_4$\\\hline
-1 & 1 & 0 & 0 & $-F_2-30F_4$\\\hline
-1 & 1 & 1 & -1 & $F_0+F_2+16F_4$\\\hline
-1 & 1 & 2 & -2 & $-70F_4$\\\hline
-1 & 2 & -1 & 2 & $35F_4$\\\hline
-1 & 2 & 0 & 1 & $2\sqrt{6}F_21-10\sqrt{6}F_4$\\\hline
-1 & 2 & 1 & 0 & $-\sqrt{6}F_2+5\sqrt{6}F_4$\\\hline
-1 & 2 & 2 & -1 & $F_0-2F_2-4F_4$\\\hline
\end{tabular}
\begin{tabular}{|c|c|c|c||c|}
\hline
$i$ & $j$ & $k$ & $l$ & $U(i,j,k,l)$\\
\hline
0 & -2 & -2 & 0 & $F_0-4F_2+6F_4$\\
\hline
0 & -2 & -1 & -1 & $\sqrt{6}F_2-5\sqrt{6}F_4$\\\hline
0 & -2 & 0 & -2 & $4F_2+15F_4$\\\hline
0 & -1 & -2 & 1 & $-\sqrt{6}F_2+5\sqrt{6}F_4$\\\hline
0 & -1 & -1 & 0 & $F_0+2F_2-24F_4$\\\hline
0 & -1 & 0 & -1 & $F_2+30F_4$\\\hline
0 & -1 & 1 & -2 & $2\sqrt{6}F_2-10\sqrt{6}F_4$\\\hline
0 & 0 & -2 & 2 & $4F_2+15F_4$\\\hline
0 & 0 & -1 & 1 & $F_2+30F_4$\\\hline
0 & 0 & 0 & 0 & $F_0+4F_2+36F_4$\\\hline
\end{tabular}}
\vskip 0.5cm
\caption{Coulomb matrix elements for $l=2$. $U(i,j,k,l)=U(-i,-j,-k,-l)$.}
\label{table2}
\end{table}

\begin{widetext}
\begin{table}[]
\centering
\resizebox{\columnwidth}{!}{%
\begin{tabular}{|c|c|c|c||c|}
\hline
$i$ & $j$ & $k$ & $l$ & $U(i,j,k,l)$\\\hline
-3 & -3 & -3 & -3 & $F_0+25F_2+9F_4+F_6$\\\hline
-3 & -2 & -3 & -2 & $25F_2+30F_4+7F_6$\\\hline
-3 & -2 & -2 & -3 & $F_0-21F_4-6F_6$\\\hline
-3 & -1 & -3 & -1 & $10F_2+54F_4+28F_6$\\\hline
-3 & -1 & -2 & -2 & $5\sqrt{15}F_2-8\sqrt{15}F_4-7\sqrt{15}F_6$\\\hline
-3 & -1 & -1 & -3 & $F_0-15F_2+3F_4+15F_6$\\\hline
-3 & 0 & -3 & 0 & $63F_4+84F_6$\\\hline
-3 & 0 & -2 & -1 & $10\sqrt{2}F_2-9\sqrt{2}F_4-56\sqrt{2}F_6$\\\hline
-3 & 0 & -1 & -2 & $5\sqrt{2}F_2-15\sqrt{2}F_4+35\sqrt{2}F_6$\\\hline
-3 & 0 & 0 & -3 & $F_0-20F_2+18F_4-20F_6$\\\hline
-3 & 1 & -3 & 1 & $42F_4+210F_6$\\\hline
-3 & 1 & -2 & 0 & $21\sqrt{2}F_4-126\sqrt{2}F_6$\\\hline
-3 & 1 & -1 & -1 & $4\sqrt{15}F_2-12\sqrt{15}F_4+28\sqrt{15}F_6$\\\hline
-3 & 1 & 0 & -2 & $-5\sqrt{2}F_2+15\sqrt{2}F_4-35\sqrt{2}F_6$\\\hline
-3 & 1 & 1 & -3 & $F_0-15F_2+3F_4+15F_6$\\\hline
-3 & 2 & -3 & 2 & $462F_6$\\\hline
-3 & 2 & -2 & 1 & $14\sqrt{15}F_4-84\sqrt{15}F_6$\\\hline
-3 & 2 & -1 & 0 & $-21\sqrt{2}F_4+126\sqrt{2}F_6$\\\hline
-3 & 2 & 0 & -1 & $10\sqrt{2}F_2-9\sqrt{2}F_4-56\sqrt{2}F_6$\\\hline
-3 & 2 & 1 & -2 & $-5\sqrt{15}F_2+8\sqrt{15}F_4+7\sqrt{15}F_6$\\\hline
-3 & 2 & 2 & -3 & $F_0-21F_4-6F_6$\\\hline
-3 & 3 & -3 & 3 & $924F_6$\\\hline
-3 & 3 & -2 & 2 & $-462F_6$\\\hline
-3 & 3 & -1 & 1 & $42F_4+210F_6$\\\hline
-3 & 3 & 0 & 0 & $-63F_4-84F_6$\\\hline
-3 & 3 & 1 & -1 & $10F_2+54F_4+28F_6$\\\hline
-3 & 3 & 2 & -2 & $25F_2-30F_4-7F_6$\\\hline
-3 & 3 & 3 & -3 & $F_0+25F_2+9F_4+F_6$\\\hline
\end{tabular}
\begin{tabular}{|c|c|c|c||c|}
\hline
$i$ & $j$ & $k$ & $l$ & $U(i,j,k,l)$\\\hline
-2 & -2 & -3 & -1 & $5\sqrt{15}F_2-8\sqrt{15}F_4-7\sqrt{15}F_6$\\\hline
-2 & -2 & -2 & -2 & $F_0+49F_4+36F_6$\\\hline
-2 & -2 & -1 & -3 & $5\sqrt{15}F_2-8\sqrt{15}F_4-7\sqrt{15}F_6$\\\hline
-2 & -1 & -3 & 0 & $10\sqrt{2}F_2-9\sqrt{2}F_4-56\sqrt{2}F_6$\\\hline
-2 & -1 & -2 & -1 & $15F_2+32F_4+105F_6$\\\hline
-2 & -1 & -1 & -2 & $F_0-7F_4-90F_6$\\\hline
-2 & -1 & 0 & -3 & $5\sqrt{2}F_2-15\sqrt{2}F_4+35\sqrt{2}F_6$\\\hline
-2 & 0 & -3 & 1 & $21\sqrt{2}F_4-126\sqrt{2}F_6$\\\hline
-2 & 0 & -2 & 0 & $20F_2+3F_4+224F_6$\\\hline
-2 & 0 & -1 & -1 & $\sqrt{30}F_2+4\sqrt{30}F_4-35\sqrt{30}F_6$\\\hline
-2 & 0 & 0 & -2 & $F_0-42F_4+120F_6$\\\hline
-2 & 0 & 1 & -3 & $-5\sqrt{2}F_2+15\sqrt{2}F_4-35\sqrt{2}F_6$\\\hline
-2 & 1 & -3 & 2 & $14\sqrt{15}F_4-84\sqrt{15}F_6$\\\hline
-2 & 1 & -2 & 1 & $14F_4+378F_6$\\\hline
-2 & 1 & -1 & 0 & $4\sqrt{30}F_2+2\sqrt{30}F_4-56\sqrt{30}F_6$\\\hline
-2 & 1 & 0 & -1 & $-\sqrt{30}F_2-4\sqrt{30}F_4+35\sqrt{30}F_6$\\\hline
-2 & 1 & 1 & -2 & $F_0-7F_4-90F_6$\\\hline
-2 & 1 & 2 & -3 & $-5\sqrt{15}F_2+8\sqrt{15}F_4+7\sqrt{15}F_6$\\\hline
-2 & 2 & -3 & 3 & $-462F_6$\\\hline
-2 & 2 & -2 & 2 & $70F_4+504F_6$\\\hline
-2 & 2 & -1 & 1 & $-14F_4-378F_6$\\\hline
-2 & 2 & 0 & 0 & $20F_2+3F_4+224F_6$\\\hline
-2 & 2 & 1 & -1 & $-15F_2-32F_4-105F_6$\\\hline
-2 & 2 & 2 & -2 & $F_0+49F_4+36F_6$\\\hline
-2 & 2 & 3 & -3 & $-25F_2-30F_4-7F_6$\\\hline
-2 & 3 & -2 & 3 & $462F_6$\\\hline
-2 & 3 & -1 & 2 & $14\sqrt{15}F_4-84\sqrt{15}F_6$\\\hline
-2 & 3 & 0 & 1 & $-21\sqrt{2}F_4+126\sqrt{2}F_6$\\\hline
-2 & 3 & 1 & 0 & $10\sqrt{2}F_2-9\sqrt{2}F_4-56\sqrt{2}F_6$\\\hline
-2 & 3 & 2 & -1 & $-5\sqrt{15}F_2+8\sqrt{15}F_4+7\sqrt{15}F_6$\\\hline
-2 & 3 & 3 & -2 & $F_0-21F_4-6F_6$\\\hline
\end{tabular}
\begin{tabular}{|c|c|c|c||c|}
\hline
$i$ & $j$ & $k$ & $l$ & $U(i,j,k,l)$\\\hline
-1 & -3 & -3 & -1 & $F_0-15F_2+3F_4+15F_6$\\\hline
-1 & -3 & -2 & -2 & $5\sqrt{15}F_2-8\sqrt{15}F_4-7\sqrt{15}F_6$\\\hline
-1 & -3 & -1 & -3 & $10F_2+54F_4+28F_6$\\\hline
-1 & -2 & -3 & 0 & $5\sqrt{2}F_2-15\sqrt{2}F_4+35\sqrt{2}F_6$\\\hline
-1 & -2 & -2 & -1 & $F_0-7F_4-90F_6$\\\hline
-1 & -2 & -1 & -2 & $15F_2+32F_4+105F_6$\\\hline
-1 & -2 & 0 & -3 & $10\sqrt{2}F_2-9\sqrt{2}F_4-56\sqrt{2}F_6$\\\hline
-1 & -1 & -3 & 1 & $4\sqrt{15}F_2-12\sqrt{15}F_4+28\sqrt{15}F_6$\\\hline
-1 & -1 & -2 & 0 & $\sqrt{30}F_2+4\sqrt{30}F_4-35\sqrt{30}F_6$\\\hline
-1 & -1 & -1 & -1 & $F_0+9F_2+F_4+225F_6$\\\hline
-1 & -1 & 0 & -2 & $\sqrt{30}F_2+4\sqrt{30}F_4-35\sqrt{30}F_6$\\\hline
-1 & -1 & 1 & -3 & $4\sqrt{15}F_2-12\sqrt{15}F_4+28\sqrt{15}F_6$\\\hline
-1 & 0 & -3 & 2 & $-21\sqrt{2}F_4+126\sqrt{2}F_6$\\\hline
-1 & 0 & -2 & 1 & $4\sqrt{30}F_2+2\sqrt{30}F_4-56\sqrt{30}F_6$\\\hline
-1 & 0 & -1 & 0 & $2F_2+15F_4+350F_6$\\\hline
-1 & 0 & 0 & -1 & $F_0+12F_2+6F_4-300F_6$\\\hline
-1 & 0 & 1 & -2 & $-\sqrt{30}F_2-4\sqrt{30}F_4+35\sqrt{30}F_6$\\\hline
-1 & 0 & 2 & -3 & $10\sqrt{2}F_2-9\sqrt{2}F_4-56\sqrt{2}F_6$\\\hline
-1 & 1 & -3 & 3 & $42F_4+210F_6$\\\hline
-1 & 1 & -2 & 2 & $-14F_4-378F_6$\\\hline
-1 & 1 & -1 & 1 & $24F_2+40F_4+420F_6$\\\hline
-1 & 1 & 0 & 0 & $-2F_2-15F_4-350F_6$\\\hline
-1 & 1 & 1 & -1 & $F_0+9F_2+F_4+225F_6$\\\hline
-1 & 1 & 2 & -2 & $-15F_2-32F_4-105F_6$\\\hline
-1 & 1 & 3 & -3 & $10F_2+54F_4+28F_6$\\\hline
-1 & 2 & -2 & 3 & $14\sqrt{15}F_4-84\sqrt{15}F_6$\\\hline
-1 & 2 & -1 & 2 & $14F_4+378F_6$\\\hline
-1 & 2 & 0 & 1 & $4\sqrt{30}F_2+2\sqrt{30}F_4-56\sqrt{30}F_6$\\\hline
-1 & 2 & 1 & 0 & $-\sqrt{30}F_2-4\sqrt{30}F_4+35\sqrt{30}F_6$\\\hline
-1 & 2 & 2 & -1 & $F_0-7F_4-90F_6$\\\hline
-1 & 2 & 3 & -2 & $-5\sqrt{15}F_2+8\sqrt{15}F_4+7\sqrt{15}F_6$\\\hline
-1 & 3 & -1 & 3 & $42F_4+210F_6$\\\hline
-1 & 3 & 0 & 2 & $21\sqrt{2}F_4-126\sqrt{2}F_6$\\\hline
-1 & 3 & 1 & 1 & $4\sqrt{15}F_2-12\sqrt{15}F_4+28\sqrt{15}F_6$\\\hline
-1 & 3 & 2 & 0 & $-5\sqrt{2}F_2+15\sqrt{2}F_4-35\sqrt{2}F_6$\\\hline
-1 & 3 & 3 & -1 & $F_0-15F_2+3F_4+15F_6$\\\hline
\end{tabular}
\begin{tabular}{|c|c|c|c||c|}
\hline
$i$ & $j$ & $k$ & $l$ & $U(i,j,k,l)$\\\hline
0 & -3 & -3 & 0 & $F_0-20F_2+18F_4-20F_6$\\\hline
0 & -3 & -2 & -1 & $5\sqrt{2}F_2-15\sqrt{2}F_4+35\sqrt{2}F_6$\\\hline
0 & -3 & -1 & -2 & $10\sqrt{2}F_2-9\sqrt{2}F_4-56\sqrt{2}F_6$\\\hline
0 & -3 & 0 & -3 & $63F_4+84F_6$\\\hline
0 & -2 & -3 & 1 & $-5\sqrt{2}F_2+15\sqrt{2}F_4-35\sqrt{2}F_6$\\\hline
0 & -2 & -2 & 0 & $F_0-42F_4+120F_6$\\\hline
0 & -2 & -1 & -1 & $\sqrt{30}F_2+4\sqrt{30}F_4-35\sqrt{30}F_6$\\\hline
0 & -2 & 0 & -2 & $20F_2+3F_4+224F_6$\\\hline
0 & -2 & 1 & -3 & $21\sqrt{2}F_4-126\sqrt{2}F_6$\\\hline
0 & -1 & -3 & 2 & $10\sqrt{2}F_2-9\sqrt{2}F_4-56\sqrt{2}F_6$\\\hline
0 & -1 & -2 & 1 & $-\sqrt{30}F_2-4\sqrt{30}F_4+35\sqrt{30}F_6$\\\hline
0 & -1 & -1 & 0 & $F_0+12F_2+6F_4-300F_6$\\\hline
0 & -1 & 0 & -1 & $2F_2+15F_4+350F_6$\\\hline
0 & -1 & 1 & -2 & $4\sqrt{30}F_2+2\sqrt{30}F_4-56\sqrt{30}F_6$\\\hline
0 & -1 & 2 & -3 & $-21\sqrt{2}F_4+126\sqrt{2}F_6$\\\hline
0 & 0 & -3 & 3 & $-63F_4-84F_6$\\\hline
0 & 0 & -2 & 2 & $20F_2+3F_4+224F_6$\\\hline
0 & 0 & -1 & 1 & $-2F_2-15F_4-350F_6$\\\hline
0 & 0 & 0 & 0 & $F_0+16F_2+36F_4+400F_6$\\\hline
0 & 0 & 1 & -1 & $-2F_2-15F_4-350F_6$\\\hline
0 & 0 & 2 & -2 & $20F_2+3F_4+224F_6$\\\hline
0 & 0 & 3 & -3 & $-63F_4-84F_6$\\\hline
\end{tabular}}
\vskip 0.5cm
\caption{Coulomb matrix elements for $l=3$. $U(i,j,k,l)=U(-i,-j,-k,-l)$.}
\label{table3}
\end{table}
\end{widetext}

There are many ways to obtain the eigenstates for a given configuration of orbitals, using matrices \cite{Slater,Ballhausen}, or coefficients of fractional parentage expansions \cite{Racah1942,Cowan,degroot}, and different coupling schemes for addition of angular momenta. Moreover, many useful formula can simplify evaluation of angular integrals, such as Gaunt coefficients, Clebsh-Gordan, Wigner 3J, 6J, etc coefficients and tensor algebra. Given that most functions are now tabulated in symbolic languages like sympy in python or matlab, in this paper I will use a simple Fock basis of formed from individual electron states and enumerate them. I will use matrix mechanics for Hilbert subspaces carrying quantum numbers of total angular momentum $L_z$ and spin $S_z$, and perform diagonalizations within these subspaces. 

A python code to generate the matrix elements for a given orbital level $l=1,2,3$, particle number $n$, and total azimuthal angular momentum $L_z$ and spin $S_z$ is given in the Appendix.

There are two key tools to use generally for matrix diagonalization. One is that the trace of a matrix is a sum of its eigenvalues:
\begin{align*}
    Tr {\bf A} = \sum_i \lambda_i
\end{align*}
and the other is that the determinant of a matrix is a product of those eigenvalues:
\begin{align*}
    \mid {\bf A}\mid =\Pi_i \lambda_i.
\end{align*}
This is sufficient to obtain all multiplets for $p^n$ and $d^n$ configurations.

For $f^n$ configurations the matrices become quite big, so big that even Matlab and python have trouble determining even the determinant, yet alone the eigenvalues directly. It is much more efficient to calculate powers of the matrix to generate values that can used as solutions to a characteristic polynomial to obtain eigenstates.

Often we have a subset of the full set of eigenvalues. Let's say that we have a $N\times N$ matrix with $N$ values that we wish to obtain. Yet suppose that we only have $N-\nu$ known eigenvalues $\lambda$ determined via previous considerations for example. To obtain the missing $\nu$ eigenvalues we utilize the trace of powers of subblock Hamiltonians $\hat H$ and subtract off the $N-\nu$ known eigenvalues
\begin{align*}
T_k=Tr(\hat H^k)-\sum_{i=1}^{N-\nu}\lambda_i^k     
\end{align*}
Then Newton's identities can be used with the coefficients
\begin{align*}
    e_1&=T_1\\
    e_2&=\frac{1}{2}(e_1T_1-T_2)\\
    e_3&=\frac{1}{3}(e_2T_1-e_1T_2+T_3)\\
    e_4&=\frac{1}{4}(e_3T_1-e_2T_2+e_1T_3-T_4)\\
    \dots&
\end{align*}
to obtain the roots of the characteristic equation from the monic polynomials
\begin{align}
    P_3(x)&=x^3-e_1x^2+e_2x-e_3=0\\
    P_4(x)&=x^4-e_1x^3+e_2x^2-e_3x+e_4=0
    \label{Eq:poly}
\end{align}
for example to obtain the three or four missing eigenvalues, respectively. Closed form expressions for the real roots of cubic and quartic polynomials have been obtained via applications of Vi\`ete's formula \cite{Nickalls2006}. For example, the cubic polynomial $P_3(x)$ generally has three real roots that can be expressed in trigonomic form as
\begin{align*}
    \lambda_{k=0,1,2}=\frac{e_1}{3}+2\sqrt{\frac{p}{3}}\cos\left[\frac{1}{3}\arccos\left(\frac{3q}{2p}\sqrt{\frac{3}{p}}\right)-\frac{2\pi k}{3}\right],
\end{align*}
with the symbols
\begin{align*}
    p=\frac{e_1^2-3e_2}{3}; ~~ q=\frac{2e_1^3+9e_1e_2e_3+27e_3}{27}.
\end{align*}
The quartic polynomial $P_4(x)$ will have four real roots expressed as 
\begin{align*}
    \lambda_{1,2}&=\frac{e_1}{4}-S \pm\frac{1}{2}\sqrt{-4S^2-2p+\frac{q}{S}}\\
    \lambda_{3,4}&=\frac{e_1}{4}+S \pm\frac{1}{2}\sqrt{-4S^2-2p-\frac{q}{S}},
\end{align*}
with the symbols
\begin{align*}
    p&=\frac{8e_2-3e_1^2}{8}, ~ q=\frac{-e_1^3+4e_1e_2-8e_3}{8}\\
    \Delta_0&=e_2^2-3e_1e_3+12e_4\\
    \Delta_1&=2e_2^3-9e_1e_2e_3+27e_1^2e_4+27e_3^2-72e_2e_4\\
       Q&=\sqrt[3]{\frac{\Delta_1+\sqrt{\Delta_1^2-4\Delta_0^3}}{2}},~S=\frac{1}{2}\sqrt{-\frac{2}{3}p+\frac{1}{3}\left(Q+\frac{\Delta_0}{Q}\right)}
\end{align*}
Quintic equations $x^5-e_1x^4+e_2x^3-e_3x^2+e_4x-e_5=0$ and higher order polynomials can be obtained via coefficients determined from traces of powers of matrices. Yet, even if so determined, known analytic solutions to quintic polynomials having real roots are very hard to obtain.  The Abel-Ruffini (Impossibility) theorem states that there is no general algebraic solution in radicals to polynomial equations of degree five or higher. Thus, these eigenvalues cannot be determined analytically but can easily obtained numerically with basic root finding routines. 

A python code given in the Appendix evaluates the traces up to four powers of matrices, but can easily be augmented to higher dimensions at higher computational cost.

\section{Results}

\subsection{\texorpdfstring{$p^n$}{TEXT} Configurations}
The $p^n$ configurations are tabulated in many places, so the results will only be given for completeness in Table (\ref{Table:ATMp}). 

\begin{table}[]
\centering
\begin{tabular}{|c|c|c|c|}
\hline
configuration & term symbol & eigenvalue & degeneracy\\
\hline
 &$^1S$ &$F_0+10F_2$ &$1$\\
 $p^2/p^4$ &$^1D$ &$F_0+F_2$ &$5$\\
 &$^3P$ &$F_0-5F_2$&$9$\\
\hline
 &$^2P$ &$3F_0$&$6$\\
$p^3$ & $^2D$ & $3F_0-6F_2$&$10$\\
& $^4S$ & $3F_0-15F_2$&$4$\\
\hline
\end{tabular}
\vskip 0.4cm
\caption{Table of multiplet energies for $p^n$ configurations derived in the text. The term symbols are to be read as $^{2S+1}L$, and the multiplicity $(2L+1)\times(2S+1)$. Note for $p$ electrons $F_0=F^0$ and $F_2=F^2/25$.}
\label{Table:ATMp}
\end{table}

\subsection{\texorpdfstring{$d^n$}{TEXT} Configurations}
\begin{widetext}
    
\begin{table}[]
\centering
\begin{tabular}{|c|c|c|c|}
\hline
configuration & term symbol & eigenvalue & degeneracy\\
\hline
 &$^1S$ &$F_0+14F_2+126F_4$&$1$\\
 &$^1G$ &$F_0+4F_2+F_4$&$9$\\
$d^2/d^8$ &$^3P$ &$F_0+7F_2-84F_4$&$9$\\
 &$^1D$ &$F_0-3F_2+36F_4$&$5$\\
 &$^3F$ &$F_0-8F_2-9F_4$&$21$\\
\hline
 &$^2P$ &$3F_0-6F_2-12F_4$&$6$\\
& $^2D$ & $3F_0+5F_2+3F_4\pm\sqrt{193F_2^2-1650F_2F_4+8325F_4^2}$&$10$\\
& $^2F$ & $3F_0+9F_2-87F_4$&14\\
$d^3/d^7$& $^2G$ & $3F_0-11F_2+13F_4$&18\\
& $^2H$ & $3F_0-6F_2-12F_4$&22\\
& $^4P$ & $3F_0-147F_4$&12\\
& $^4F$ & $3F_0-15F_2-72F_4$&28\\
\hline
 & $^1S$ & $6F_0+10F_2+6F_4\pm\frac{1}{2}\sqrt{3088F_2^2-26400F_2F_4+133200F_4^2}$&1\\
                 & $^1D$ & $6F_0+9F_2-76.5F_4\pm\frac{1}{2}\sqrt{1296F_2^2-10440F_2F_4+30825F_4^2}$&5\\
                 & $^1F$ & $6F_0-84F_4$&7\\
                 & $^1G$ & $6F_0-5F_2-6.5F_4\pm\frac{1}{2}\sqrt{708F_2^2-7500F_2F_4+30825F_4^2}$&9\\
                 & $^1I$ & $6F_0-15F_2-9F_4$&13\\
                $d^4/d^6$ & $^3P$ & $6F_0-5F_2-76.5F_4\pm\frac{1}{2}\sqrt{912F_2^2-9960F_2F_4+38025F_4^2}$&9\\
                 & $^3D$ & $6F_0-5F_2-129F_4$&15\\
                 & $^3F$ & $6F_0-5F_2-76.5F_4\pm\frac{1}{2}\sqrt{612F_2^2-4860F_2F_4+20025F_4^2}$&21\\
                 & $^3G$ & $6F_0-12F_2-94F_4$&27\\
                 & $^3H$ & $6F_0-17F_2-69F_4$&33\\
                 & $^5D$ & $6F_0-21F_2-189F_4$&25\\
\hline
 &$^2S$ &$10F_0-3F_2-195F_4$&2\\
           &$^2P$ & $10F_0+20F_2-240F_4$&6\\
           &$^2D$ & $10F_0-3F_2-90F_4\pm\sqrt{513F_2^2-4500F_2F_4+20700F_4^2}$&10\\
           &$^2D$ & $10F_0-4F_2-120F_4$&10\\
           &$^2F$ & $10F_0-9F_2-165F_4$&14\\
           &$^2F$ & $10F_0-25F_2-15F_4$&14\\
           &$^2G$ & $10F_0+3F_2-155F_4$&18\\
           $d^5$ &$^2G$ & $10F_0-13F_2-145F_4$&18\\
           &$^2H$ & $10F_0-22F_2-30F_4$&22\\
           &$^2I$ & $10F_0-24F_2-90F_4$&26\\
           &$^4P$ & $10F_0-28F_2-105F_4$&12\\
           &$^4D$ & $10F_0-18F_2-225F_4$&20\\
           &$^4F$ & $10F_0-13F_2-180F_4$&28\\
           &$^4G$ & $10F_0-25F_2-190F_4$&36\\
           &$^6S$ & $10F_0-35F_2-315F_4$&6\\
\hline
\end{tabular}
\vskip 0.4cm
\caption{Table of multiplet energies for $d^n$ configurations. The term symbols are to be read as $^{2S+1}L$, and the multiplicity $(2L+1)\times(2S+1)$. Note for $d$-electrons, $F^2=49F_2$ and $F^4=441F_4$. }
\label{Table:ATMd}
\end{table}

Likewise,the $d^n$ multiplets have been tabulated in a few places, although some errors have occurred in various texts. Our results are shown in Table \ref{Table:ATMd} for all $d^n$ multiplets. I note that there are two misprints in Ballhausen's book \cite{Ballhausen}, reprinted from the same errors in O. Laporte and J. R. Platt \cite{Laporte}. One is for the $^1F$ configuration for $d^4/d^6$, and the other is for one of the $^2G$ configurations for $d^5$. This has already been noted by Racah \cite{Racah1942}.

\subsection{\texorpdfstring{$f^n$}{TEXT} Configurations}
Most of the newer results are presented for the case of $f^n$ multiplets with $n\ge 4$. These are enumerated as follows.

\begin{table}[]
\centering
\resizebox{\columnwidth}{!}{
\begin{tabular}{|c|c|c|c|}
\hline
configuration & term symbol & eigenvalue & degeneracy\\
\hline
 &$^1S$ &$F_0+60F_2+198F_4+1716F_6$&$1$\\
 &$^3P$ &$F_0+45F_2+33F_4-1287F_6$&$9$\\
 &$^1D$ &$F_0+19F_2-99F_4+715F_6$&$5$\\
 $f^2/f^{12}$ &$^3F$ &$F_0-10F_2-33F_4-286F_6$&$21$\\
 &$^1G$ &$F_0-30F_2+97F_4+78F_6$&$9$\\
 &$^3H$ &$F_0-25F_2-51F_4-13F_6$&$33$\\
 &$^1I$ &$F_0+25F_2+9F_4+F_6$&$13$\\
\hline
& $^4S$ &$3F_0-30F_2-99F_4-858F_6$ &$4$\\
& $^2P$ &$3F_0-25F_2-44F_4 +143F_6$ &$6$\\
& $^2D$ &$3F_0 - 7F_2 - 31.5F_4 - 130F_6$ &$20$\\
&  &$\pm \sqrt{2176F_2^2-20112F_2F_4+72800F_2F_6+118809F_4^2-1204476F_4F_6+3213028F_6^2}$ &\\
& $^4D$ &$3F_0+25F_2-33F_4-1859F_6$ &$20$\\
& $^2F$ &$3F_0 + 55F_2 + 55.5F_4 + 52F_6$ &$28$\\
&  &$\pm \frac{1}{2}\sqrt{9700F_2^2+11940F_2F_4+36081F_4^2- 451360F_2F_6+283920F_4F_6+7684768F_6^2}$ &\\
$f^3/f^{11}$ & $^4F$ &$3F_0-30F_2-99F_4-858F_6$ &$28$\\
& $^2G$ &$3F_0+7F_2+24.5F_4-620F_6$ &$36$\\
&  &$\pm \sqrt{12676F_2^2 - 13924F_2F_4 - 195328F_2F_6 + 35929F_4^2 - 277984F_4F_6 + 1908256F_6^2}$ &\\
& $^4G$ &$3F_0-10F_2-75F_4-1222F_6$ &$36$\\
& $^2H$ &$3F_0 - 23F_2 - 46.5F_4 + 136F_6
$ &$44$\\
& & $\pm \frac{1}{2}\sqrt{
5056F_2^2 - 14208F_2F_4 - 25984F_2F_6 + 32841F_4^2 - 237804F_4F_6 + 856324F_6^2}$ & \\
& $^4I$ &$3F_0-65F_2-141F_4-221F_6$ &$52$\\
& $^2I$ &$3F_0-5F_2-6F_4-305F_6$ &$26$\\
& $^2J$ &$3F_0-40F_2+F_4+38F_6$ &$30$\\
& $^2K$ &$3F_0-63F_4-18F_6$ &$34$\\
\hline
\end{tabular}}
\vskip 0.4cm
\caption{Table of multiplet energies for $f^{2,3}/f^{12,11}$ configurations. The term symbols are to be read as $^{2S+1}L$, and the multiplicity $(2L+1)\times(2S+1)$. Note for $f$-electrons, $F^2=15^2F_2, F^4=33^2F_4$, and $F^6=\frac{429^2}{5^2}F_6$.} 
\label{Table:ATMfa}
\end{table}
\end{widetext}

\subsubsection{\texorpdfstring{$f^{2,12}$}{TEXT} Configurations}
For the $f^2/f^{12}$ configurations there are ${14 \choose{2}}= 91$ states. The allowed term symbols are $^1I, ^3H,^1G, ^3F, ^1D, ^3P,^1S$. The $^1I$ state having $L_z=6$ is $E(^1I)=U(3,3,3,3)=F_0+25F_2+9F_4+F_6$. 

The $L_z=5$ sector has 4 states, with one them the preceding singlet, and the rest triplets. Taking the trace and subtracting the singlet we obtain $E(^3H)=F_0-25F_2-51F_4-13F_6$. 

For $L_z=4$ there are 5 states. Three are triplets having $E(^3H)$, plus two singlets, one of which is $E(^1I)$. Taking the trace and subtracting the others yields $E(^1G)=F_0-30F_2+97F_4+78F_6$. 

The $L_z=3$ sector has 8 states, one having energy $E(^1I)$, three having $E(^3H)$, one having energy $E(^1G)$, leaving three as part of a triplet. Subtracting off the others and dividing by 3 gives $E(^3F)=F_0-10F_2-33F_4-286F_6$.

The $L_z=2$ sector has one additional singlet, giving $E(^1D)=F_0+19F_2-99F_4+715F_6$.

The $L_z=1$ sector has an additional triplet, giving $E(^3P)=F_0+45F_2+33F_4-1287F_6$.

Lastly, the $L_z=0$ sector has an additional singlet, giving $E(^1S)=F_0+60F_2+198F_4+1716F_6$.

These results have been obtained by Racah \cite{Racah1942,Racah1949}.
The ground state is $^3H$, following Hund's rules.

\subsubsection{\texorpdfstring{$f^{3,11}$}{TEXT} Configurations}
For the $f^3/f^{11}$ configurations there are ${14 \choose{3}}= 364$ states. The allowed term symbols are $^2K, ^2J, ^2I, ^4I, ^2H, ^4G, ^2G, ^4F, ^2F, ^2D, ^2P,$ and $^4S$. 

The $^2K$ state having $L_z=8, S_z=1/2$ is $E(^2K)=U(3,3,3,3)+2U(3,2,2,3)-U(3,2,3,2)=3F_0-63F_4-18F_6$. 

There are two states having $L_z=7, S_z=1/2$. Taking the trace and subtracting off $E(^2K)$ gives $E(^2J)=U(2,2,2,2)+2U(3,1,1,3)-U(3,1,3,1)=3F_0-40F_2+F_4+38F_6$.

The one state having $L_z=6, S_z=3/2$ gives $E(^4I)=U(2,1,1,2)-U(2,1,2,1)+U(3,1,1,3)-U(3,1,3,1)+U(3,2,2,3)-U(3,2,3,2)=3F_0-65F_2-141F_4-221F_6$. 

There are four states having $L_z=6, S_z=1/2$. Taking the trace and subtracting off $E(^4I), E(^2K),$ and $E(^2J)$ gives $E(^2I)=2U(3,0,0,3)-U(3,0,3,0)+2U(2,1,1,2)+U(3,2,3,2)+U(3,1,3,1)-U(2,2,2,2)=3F_0-5F_2-6F_4-305F_6$.

The one state having $L_z=5, S_z=3/2$ belongs to $^4I$.

\begin{widetext}
There are six states having $L_z=5, S_z=1/2$, one each belonging to $^2K, ^2J, ^2I, ^4I$, and therefore there must be two contributing to $^2H$. We need to utilize the trace and the determinant. This gives 
\small
\begin{align*}
E(^2H)=3F_0 - 23F_2 - 46.5F_4 + 136F_6
\pm \frac{1}{2}\sqrt{5056F_2^2 - 14208F_2F_4 - 25984F_2F_6 + 32841F_4^2 - 237804F_4F_6 + 856324F_6^2}.    
\end{align*}
\normalsize

For $L_z=4, S_z=3/2$ there are only two states, one of which is $^4I$ and the other is $^4G$. Using the trace and $E(^4I)$ gives $E(^4G)=3F_0-10F_2-75F_4-1222F_6$. 

There are nine states in $L_z=4, S_z=1/2$. Therefore, there are two $^2G$ states. This gives
\small
\begin{align*}
E(^2G)=3F_0+7F_2+24.5F_4-620F_6 
\pm \sqrt{12676F_2^2 - 13924F_2F_4 - 195328F_2F_6 + 35929F_4^2 - 277984F_4F_6 + 1908256F_6^2}.
\end{align*}
\normalsize

For $L_z=3, S_z=3/2$ there are three states, one of which is $^4I$, one $^4G$, and one $^4F$. Utilizing the trace gives $E(^4F)=3F_0-30F_2-99F_4-858F_6$.

For $L_z=3, S_z=1/2$ there are 12 states, meaning there are two $^2F$ states. 
This gives 
\small
\begin{align*}
E(^2F)=3F_0 + 55F_2 + 55.5F_4 + 52F_6\pm \sqrt{9700F_2^2+11940F_2F_4+36081F_4^2- 451360F_2F_6+283920F_4F_6+7684768F_6^2}.
\end{align*}
\normalsize

For $L_z=2, S_z=3/2$ there are four states, one of each in $^4I, ^4G, ^4F$, and $^4D$. Utilizing the trace gives $E(^4D)=3F_0+25F_2-33F_4-1859F_6$.

For $L_z=2, S_z=1/2$ there are 15 states, meaning there are two $^2D$ states. 
This gives
\small
\begin{align*}
E(^2D)=3F_0-7F_2-31.5F_4-130F_6
\pm \sqrt{2176F_2^2-20112F_2F_4+72800F_2F_6+118809F_4^2-1204476F_4F_6+3213028F_6^2}
\end{align*}
\normalsize

For $L_z=1, S_z=3/2$ there are four states, already accounted for. For $L_z=1, S_z=1/2$ there are 16 states, giving one $^2P$ state. Taking the trace and subtracting off other eigenvalues gives $E(^2P)=3F_0-25F_2-44F_4+143F_6$.

For $L_z=0, S_z=3/2$ there are five states, yield the last as $^4S$. Utilizing the trace gives $E(^4S)=3F_0-30F_2-99F_4-858F_6$. This is exactly degenerate with $E(^4F)$.

For $L_z=0, S_z=1/2$ there are 16 states, all accounted for. These reproduce the results found by Racah \cite{Racah1942}.

The Hund's rule ground state is $^4I$.

\begin{table}[]
\centering
\resizebox{\columnwidth}{!}{
\begin{tabular}{|c|c|c|c|}
\hline
configuration & term symbol & eigenvalue & degeneracy\\
\hline
& $^5S$ & $6F_0-60F_2-198F_4-1716F_6$ & $5$\\
& $^1S$ & $6F_0+110F_2+111F_4+104F_6$& $2$\\
&  & $\pm\frac{1}{2}\sqrt{38800F_2^2+47760F_2F_4-1805440F_2F_6+144324F_4^2+1135680F_4F_6+30739072F_6^2}$& \\
& $^3P_{k=0,1,2}$ & $6F_0 + \frac{19}{3}F_2 - \frac{29}{3}F_4 - \frac{2873}{3}F_6+\lambda_P\cos(\theta_P/3+2\pi k/3)$ & $27$\\
& $^5D$ & $6F_0-5F_2-132F_4-2717F_6$ & $25$\\
& $^3D$ & $6F_0-3F_2-85.5F_4-1170F_6$ & $30$\\
&  & $\pm\frac{1}{2}\sqrt{17296F_2^2-25320F_2F_4-228592F_2F_6+11289F_4^2+143052F_4F_6+828100F_6^2}$ & \\
& $^1D$ & $6F_0 + 17F_2 + 22.5F_4 - 536.5F_6- S_D\pm\frac{1}{2}\sqrt{-4S_D^2-2p_D+\frac{q_D}{S_D}}$ & $20$\\
&  & $6F_0 + 17F_2 + 22.5F_4 - 536.5F_6+ S_D\pm\frac{1}{2}\sqrt{-4S_D^2-2p_D-\frac{q_D}{S_D}}$ & \\
& $^5F$ & $6F_0-60F_2-198F_4-1716F_6$ & $35$\\
& $^3F$ &  $6F_0 - 14F_2 - 73.5F_4 - 543.5F_6- S_F\pm\frac{1}{2}\sqrt{-4S_F^2-2p_F+\frac{q_F}{S_F}}$& $84$\\
&  &$6F_0 - 14F_2 - 73.5F_4 - 543.5F_6+ S_F\pm\frac{1}{2}\sqrt{-4S_F^2-2p_F-\frac{q_F}{S_F}}$  & \\
& $^1F$ & $6F_0 + 20F_2 + 87F_4 - 1534F_6$ & $7$\\
& $^5G$ & $6F_0-40F_2-174F_4-2080F_6$ & $45$\\
& $^3G_{k=0,1,2}$ & $6F_0-18F_2-65F_4-1128F_6+\lambda_G\cos(\theta_G/3+2\pi k/3)$& $81$\\
$f^4/f^{10}$ & $^1G$ & $6F_0 + 6.5F_2 - 19.5F_4 - 169F_6 - S_G\pm\frac{1}{2}\sqrt{-4S_G^2-2p_G+\frac{q_G}{S_G}}$& $36$\\
&  &$6F_0 + 6.5F_2 - 19.5F_4 - 169F_6 + S_G\pm\frac{1}{2}\sqrt{-4S_G^2-2p_G-\frac{q_G}{S_G}}$ & \\
 & $^3H$ & $6F_0 - 11F_2 - 51F_4 - 711.5F_6 - S_H\pm\frac{1}{2}\sqrt{-4S_H^2-2p_H+\frac{q_H}{S_H}}$& 132\\
&  & $6F_0 - 11F_2 - 51F_4 - 711.5F_6 + S_H\pm\frac{1}{2}\sqrt{-4S_H^2-2p_H-\frac{q_H}{S_H}}$& \\
& $^1H$ & $6F_0+13F_2-91.5F_4-386F_6$ & 22\\
&  &$\pm \frac{1}{2}\sqrt{13456F_2^2-7320F_2F_4-252112F_2F_6+60057F_4^2-640164F_4F_6+3307108F_6^2}$ & \\
& $^5I$ & $6F_0-95F_2-240F_4-1079F_6$ & $65$\\
& $^3I$ & $6F_0 -15F_2 -81F_4-1065F_6$ & $78$\\
&  & $\pm\frac{1}{2}\sqrt{6400F_2^2-14880F_2F_4-51520F_2F_6+12324F_4^2+15792F_4F_6+235984F_6^2}$ & \\
& $^1I_{k=0,1,2}$ & $6F_0-\frac{1}{3}F_2+29F_4-\frac{577}{3}F_6+\lambda_I\cos(\theta_I/3+2\pi k/3)$ & $39$\\
& $^3J$ & $6F_0-43F_2-119.5F_4-526F_6$ & $90$\\
& & $\pm\frac{1}{2}\sqrt{7716F_2^2+548F_2F_4+26569F_4^2-173040F_2F_6-324856F_4F_6+1926288F_6^2}$ & \\
& $^1J$ & $6F_0 - 46F_2 - 23F_4 - 148F_6$ & $15$\\
& $^3K$ & $6F_0-70F_2-105F_4-316F_6$ & $51$\\
& $^1K$ & $6F_0-43F_2-70.5F_4+104F_6$ & $34$\\
&  & $\pm \frac{1}{2}\sqrt{5476F_2^2-12948F_2F_4+25281F_4^2-42784F_2F_6-160944F_4F_6+718144F_6^2}$ & \\
& $^3L$ & $6F_0-55F_2-150F_4-211F_6$ & $57$\\
& $^1M$ & $6F_0-25F_2-86F_4-F_6$ & $21$\\
\hline
\end{tabular}}
\vskip 0.4cm
\caption{Table of multiplet energies for $f^{4}/f^{10}$ configurations. The term symbols are to be read as $^{2S+1}L$, and the multiplicity $(2L+1)\times(2S+1)$. Note for $f$-electrons, $F^2=15^2F_2, F^4=33^2F_4$, and $F^6=\frac{429^2}{5^2}F_6$.} 
\label{Table:ATMfa}
\end{table}

\subsubsection{\texorpdfstring{$f^4/f^{10}$}{TEXT} Configurations}
For the $f^4/f^{10}$ configurations there are ${14 \choose{4}}= 1001$ states. The allowed term symbols are $^1M$, $^3L$, $^1K$, $^3K$, $^1J$, $^3J$, $^1I$, $^3I$, $^5I$, $^1H$, $^3H$, $^3G$, $^5G$, $^3F$, $^5F$, $^1D$, $^3D$, $^5D$, $^3P$, $^1S$, $^5S$. 

The matrices are now becoming quite big, so big that even Matlab and python have trouble determining even the determinant, yet alone the eigenvalues directly. It is much more efficient to calculate powers of the matrix to generate values that can used as solutions to a characteristic polynomial to obtain eigenstates.

Starting with the $^1M$ state that has the maximum $L_z=10$, its energy is $E(^1M)=6F_0-25F_2-86F_4-F_6.$

The $^3L$ state triplet has $L_z=9$ and energy $E(^3L)=6F_0-55F_2-150F_4-211F_6.$

The $L_z=8$ sector has 9 states, 4 from above plus another triplet and two singlets. The triplet energy is $E(^3K)=6F_0-70F_2-105F_4-316F_6.$ The two singlets are
\small
\begin{align*}
    E(^1K_\pm)=6F_0-43F_2-70.5F_4 + 104F_6
\pm\frac{1}{2}\sqrt{5476F_2^2-12948F_2F_4 + 25281F^2_4-42784F_2F_6-160944F_4F_6 + 718144F^2_6}
\end{align*}
\normalsize

The $L_z=7$ sector has 16 states, 9 from above plus two triplets and one singlet. The singlet has eigenvalue $E(^1J)=6F_0-46F_2-23F_4-148F_6$. The two triplets are
\small
\begin{align*}
  E(^3J_\pm)=6F_0-43F_2-119.5F_4-526F_6\pm\frac{1}{2}\sqrt{7716F_2^2 + 548F_2F_4 + 26569F_4^2
-173040F_2F_6-324856F_4F_6 + 1926288F_6^2}  
\end{align*}
\normalsize

The $L_z=6$ sector has 30 states, including an additional quintet, two triplets, and three singlets. The quintet eigenvalue is $E(^5I)=6F_0-95F_2-240F_4-1079F_6$, and the two triplets are
\small
\begin{align*}
  E(^3I_\pm)=6F_0-15F_2-81F_4-1065F_6\pm\frac{1}{2}\sqrt{6400F^2_2-14880F_2F_4-51520F_2F_6 + 12324F^2_4 + 15792F_4F_6 + 235984F^2_6}  
\end{align*}
\normalsize
For the three singlets, I have used Eq. \ref{Eq:poly} where 
\small
\begin{align*}
e_1&=18F_0-F_2+87F_4-577F_6\\
e_2&=108F_0^2-12F_0F_2+1044F_0F_4-6924F_0F_6-3205F_2^2+2262F_2F_4+16942F_2F_6-7488F_4^2+32502F_4F_6-1650541F_6^2\\
e_3&=216F_0^3-36F_0^2F_2 + 3132F_0^2F_4-20772F_0^2F_6-19230F_0F_2^2 + 13572F_0F_2F_4 + 101652F_0F_2F_6- 44928F_0F_4^2 + 195012F_0F_4F_6\\
&-9903246F_0F_6^2-6275F_2^3 + 100575F_2^2F_4-1560945F_2^2F_6 + 185580F_2F_4^2 + 6109830F_2F_4F_6 + 12306951F_2F_6^2-273780F_4^3\\
&-8530020F_4^2F_6 - 799929F_4F_6^2 + 835297525F_6^3\\
\end{align*}
\normalsize
This gives the three singlet eigenvalues
$E(^1I_{k=0,1,2})=6F_0-13F_2 + 29F_4-5773 F_6 + \lambda_I \cos(\theta_I/3+2\pi k/3)$.

The moments, which will be useful later, are
\small
\begin{align*}
    \sum_{i=1}^3E_i(^1I)^2&=108F_0^2 - 12F_0F_2 + 1044F_0F_4 - 6924F_0F_6 + 6411F_2^2 - 4698F_2F_4 - 32730F_2F_6 + 22545F_4^2 - 165402F_4F_6\\& + 3634011F_6^2\\
    \sum_{i=1}^3E_i(^1I)^3&=648F_0^3 - 108F_0^2F_2 + 9396F_0^2F_4 - 62316F_0^2F_6 + 115398F_0F_2^2 - 84564F_0F_2F_4 - 589140F_0F_2F_6 + 405810F_0F_4^2 \\&- 2977236F_0F_4F_6 + 65412198F_0F_6^2 - 28441F_2^3 + 1145277F_2^2F_4 - 10181595F_2^2F_6 - 78813F_2F_4^2 + 18221850F_2F_4F_6\\& + 60297045F_2F_6^2 + 1791531F_4^3 - 60136749F_4^2F_6 + 571546845F_4F_6^2 - 543293929F_6^3\\
    \sum_{i=1}^3E_i(^1I)^4&=3888F_0^4 - 864F_0^3F_2 + 75168F_0^3F_4 - 498528F_0^3F_6 + 1384776F_0^2F_2^2 - 1014768F_0^2F_2F_4 - 7069680F_0^2F_2F_6\\& + 4869720F_0^2F_4^2 - 35726832F_0^2F_4F_6 + 784946376F_0^2F_6^2 - 682584F_0F_2^3 + 27486648F_0F_2^2F_4 - 244358280F_0F_2^2F_6 \\&- 1891512F_0F_2F_4^2 + 437324400F_0F_2F_4F_6 + 1447129080F_0F_2F_6^2 + 42996744F_0F_4^3 - 1443281976F_0F_4^2F_6 \\&+ 13717124280F_0F_4F_6^2 - 13039054296F_0F_6^3 + 20581971F_2^4 - 33824916F_2^3F_4 - 181741140F_2^3F_6 + 239171526F_2^2F_4^2 \\&- 2349644220F_2^2F_4F_6 + 29485976850F_2^2F_6^2 - 78404436F_2F_4^3 + 2023712460F_2F_4^2F_6 - 20401921980F_2F_4F_6^2 \\&- 157774130580F_2F_6^3 + 300861297F_4^4 - 8821038996F_4^3F_6 + 159074524422F_4^2F_6^2 - 695032065684F_4F_6^3 \\&+ 5829598075059F_6^4
\end{align*}
\normalsize

The $L_z=5$ sector has 44 states, including additionally four triplets and two singlets. The four triplets are obtained via Eq. \ref{Eq:poly} with
\small
\begin{align*}
    e_1&=24F_0 - 44F_2 - 204F_4 - 2846F_6\\
    e_2&=216F_0^2 - 792F_0F_2 - 3672F_0F_4 - 51228F_0F_6 - 5810F_2^2 - 504F_2F_4 + 239546F_2F_6 - 2793F_4^2 + 550098F_4F_6\\ &+677284F_6^2\\
    e_3&=864F_0^3 - 4752F_0^2F_2 - 22032F_0^2F_4 - 307368F_0^2F_6 - 69720F_0F_2^2 - 6048F_0F_2F_4 + 2874552F_0F_2F_6 - 33516F_0F_4^2\\ &+ 6601176F_0F_4F_6 +
       8127408F_0F_6^2 + 233700F_2^3 + 864420F_2^2F_4 + 3537030F_2^2F_6 + 1016514F_2F_4^2 - 1591044F_2F_4F_6\\ 
       &- 288170112F_2F_6^2 + 1067904F_4^3 +
       1673658F_4^2F_6 - 308443656F_4F_6^2 + 3259826622F_6^3\\
    e_4&=1296F_0^4 - 9504F_0^3F_2 - 44064F_0^3F_4 - 614736F_0^3F_6 - 209160F_0^2F_2^2 - 18144F_0^2F_2F_4 + 8623656F_0^2F_2F_6 - 100548F_0^2F_4^2\\
    &+ 19803528F_0^2F_4F_6 + 24382224F_0^2F_6^2 + 1402200F_0F_2^3 + 5186520F_0F_2^2F_4 + 21222180F_0F_2^2F_6 + 6099084F_0F_2F_4^2\\ 
    &- 9546264F_0F_2F_4F_6 -
       1729020672F_0F_2F_6^2 + 6407424F_0F_4^3 + 10041948F_0F_4^2F_6 - 1850661936F_0F_4F_6^2 + 19558959732F_0F_6^3\\
       & - 1530375F_2^4 - 14296800F_2^3F_4 -
       127202850F_2^3F_6 - 24392025F_2^2F_4^2 - 635166270F_2^2F_4F_6 + 1733499180F_2^2F_6^2 - 8443260F_2F_4^3\\ &
       - 2138633226F_2F_4^2F_6 + 
       2827059732F_2F_4F_6^2 + 87759796098F_2F_6^3 + 20990340F_4^4 - 1526592996F_4^3F_6 + 4272216543F_4^2F_6^2\\ &- 68801310318F_4F_6^3 - 2061418292037F_6^4
\end{align*}
\normalsize
This gives the four triplet energies $E(^3H)$ as
\small
\begin{align*}
   E(^3H) =
   & 6F_0 - 11F_2 - 51F_4 - 711.5F_6 - S_H\pm\frac{1}{2}\sqrt{-4S_H^2-2p_H+\frac{q_H}{S_H}}\\
=& 6F_0 - 11F_2 - 51F_4 - 711.5F_6 + S_H\pm\frac{1}{2}\sqrt{-4S_H^2-2p_H-\frac{q_H}{S_H}} 
\end{align*}
\normalsize
with the parameters $S_H,p_H,q_H$ given in terms of $e_{1,2,3,4}$ in Eq. \ref{Eq:poly}. 
The sum of the squares, cubes, and quartic of these eigenvalues, which is useful for finding other eigenvalues, are 
\small
\begin{align*}
    \sum_{i=1}^4 E_i(^3H)^2&=144F_0^2 - 528F_0F_2 -2448F_0F_4 -34152F_0F_6 + 13556F_2^2 + 18960F_2F_4 -228644F_2F_6 + 47202F_4^2 + 60972F_4F_6\\ &+ 6745148F_6^2\\
    \sum_{i=1}^4 E_i(^3H)^3&=864F_0^3 - 4752F_0^2F_2 - 22032F_0^2F_4 - 307368F_0^2F_6 + 244008F_0F_2^2 + 341280F_0F_2F_4 - 4115592F_0F_2F_6\\ &+ 849636F_0F_4^2 + 1097496F_0F_4F_6 + 121412664F_0F_6^2 - 151004F_2^3 - 2213820F_2^2F_4 - 23904186F_2^2F_6 - 3120894F_2F_4^2\\ 
    &+56864628F_2F_4F_6 + 200972388F_2F_6^2 - 6995268F_4^3 - 37483092F_4^2F_6 - 771122628F_4F_6^2 - 7489661078F_6^3\\
    \sum_{i=1}^4 E_i(^3H)^4 &=5184F_0^4 - 38016F_0^3F_2 - 176256F_0^3F_4 - 2458944F_0^3F_6 + 2928096F_0^2F_2^2 + 4095360F_0^2F_2F_4 - 49387104F_0^2F_2F_4\\ 
    &+ 10195632F_0^2F_4^2 + 13169952F_0^2F_4F_6 + 1456951968F_0^2F_6^2 - 3624096F_0F_2^3 - 53131680F_0F_2^2F_4 - 573700464F_0F_2^2F_6\\
    &- 74901456F_0F_2F_4^2 + 1364751072F_0F_2F_4F_6 + 4823337312F_0F_2F_6^2 - 167886432F_0F_4^3 - 899594208F_0F_4^2F_6\\ &
    - 18506943072F_0F_4F_6^2 - 179751865872F_0F_6^3 + 81243236F_2^4 + 216680640F_2^3F_4 - 3406093768F_2^3F_6 + 787099788F_2^2F_4^2\\ 
    &- 3655990296F_2^2F_4F_6 + 139646432712F_2^2F_6^2 + 800615664F_2F_4^3 - 17501378832F_2F_4^2F_6 - 1598348448F_2F_4F_6^2 \\
    &- 1377678045736F_2F_6^3 + 1257056082F_4^4 + 4485343464F_4^3F_6 + 258385768668F_4^2F_6^2 + 458749330008F_4F_6^3 \\
    &+ 15715401211892F_6^4
\end{align*}
\normalsize
The two singlet energies are 
\small
\begin{align*}
    E(^1H)=6F_0+13F_2-91.5F_4-386F_6
\pm \frac{1}{2}\sqrt{13456F_2^2-7320F_2F_4-252112F_2F_6+60057F_4^2-640164F_4F_6+3307108F_6^2}
\end{align*}
\normalsize

The $L_z=4$ sector has 62 states, gives an additional quintet, three triplets, and four singlets. The quintet energy is
$E(^5G)=6F_0-40F_2-174F_4-2080F_6$. The three triplet energies $E(^3G)$  are roots of the cubic $P_3$ with
\small
\begin{align*}
    e_1 &=18F_0 - 54F_2 - 195F_4 - 3384F_6
    \\ e_2 &= 108F_0^2 - 648F_0F_2 - 2340F_0F_4 - 40608F_0F_6 - 1920F_2^2 + 9070F_2F_4 + 173148F_2F_6 - 5716F_4^2 + 638062F_4F_6 + 2940444F_6^2\\
e_3 &= 
216F_0^3 - 1944F_0^2F_2 - 7020F_0^2F_4 - 121824F_0^2F_6 - 11520F_0F_2^2 + 54420F_0F_2F_4 + 1038888F_0F_2F_6 - 34296F_0F_4^2\\ 
& + 3828372F_0F_4F_6 + 17642664F_0F_6^2 + 35200F_2^3 + 121680F_2^2F_4 + 1403040F_2^2F_6 - 749284F_2F_4^2 - 9249672F_2F_4F_6 \\ &- 136245408F_2F_6^2 + 795036F_4^3 + 3740132F_4^2F_6 - 519024792F_4F_6^2 - 144317888F_6^3
\end{align*}
\normalsize
The power sums are
\small
\begin{align*}
    \sum_{i=1}^3 E_i(^3G)^2&=108F_0^2 - 648F_0F_2 - 2340F_0F_4 - 40608F_0F_6 + 6756F_2^2 + 2920F_2F_4 + 19176F_2F_6 + 49457F_4^2 + 43636F_4F_6\\ &+ 5570568F_6^2\\
    \sum_{i=1}^3 E_i(^3G)^3&=648F_0^3 - 5832F_0^2F_2 - 21060F_0^2F_4 - 365472F_0^2F_6 + 121608F_0F_2^2 + 52560F_0F_2F_4 + 345168F_0F_2F_6\\ 
    &+ 890226F_0F_4^2 + 785448F_0F_4F_6 + 100270224F_0F_6^2 - 362904F_2^3 - 994680F_2^2F_4 - 16835976F_2^2F_6 - 4027944F_2F_4^2\\& + 55186128F_2F_4F_6 - 29721672F_2F_6^2 - 8373627F_4^3 - 59571966F_4^2F_6 - 58410972F_4F_6^2 - 9333293280F_6^3\\
    \sum_{i=1}^3 E_i(^3G)^4&=3888F_0^4 - 46656F_0^3F_2 - 168480F_0^3F_4 - 2923776F_0^3F_6 + 1459296F_0^2F_2^2 + 630720F_0^2F_2F_4 + 4142016F_0^2F_2F_6 \\&+ 10682712F_0^2F_4^2 + 9425376F_0^2F_4F_6 + 1203242688F_0^2F_6^2 - 8709696F_0F_2^3 - 23872320F_0F_2^2F_4 \\& -404063424F_0F_2^2F_6 - 96670656F_0F_2F_4^2 + 1324467072F_0F_2F_4F_6 - 713320128F_0F_2F_6^2 - 200967048F_0F_4^3\\& - 1429727184F_0F_4^2F_6 - 1401863328F_0F_4F_6^2 - 223999038720F_0F_6^3 + 30667536F_2^4 + 55373760F_2^3F_4\\& + 809358912F_2^3F_6 + 535295648F_2^2F_4^2 - 1423398336F_2^2F_4F_6 + 48696842592F_2^2F_6^2 + 908919104F_2F_4^3 \\&- 489240512F_2F_4^2F_6 - 170805966528F_2F_4F_6^2 + 52504939584F_2F_6^3 + 1760521457F_4^4 + 5225950616F_4^3F_6 \\&+ 160108254584F_4^2F_6^2 + 119499829536F_4F_6^3 + 15692292940320F_6^4
\end{align*}
\normalsize
The four singlet energies $E(^1G)$ are roots of the quartic $P_4$ with the coefficients
\small
\begin{align*}
  e_1&=24F_0 + 26F_2 - 78F_4 - 676F_6\\  
  e_2&=216F_0^2 + 468F_0F_2 - 1404F_0F_4 - 12168F_0F_6 - 8820F_2^2 - 2814F_2F_4 + 195348F_2F_6 - 55503F_4^2 + 281256F_4F_6 - 2310288F_6^2\\
  e_3&=864F_0^3 + 2808F_0^2F_2 - 8424F_0^2F_4 - 73008F_0^2F_6 - 105840F_0F_2^2 - 33768F_0F_2F_4 + 2344176F_0F_2F_6 - 666036F_0F_4^2\\& + 3375072F_0F_4F_6 - 27723456F_0F_6^2 + 82200F_2^3 + 846960F_2^2F_4 - 4163040F_2^2F_6 + 426924F_2F_4^2 - 5862024F_2F_4F_6\\& - 54501768F_2F_6^2 + 2165008F_4^3 + 40401060F_4^2F_6 - 7685016F_4F_6^2  + 1080575184F_6^3\\
  e_4&=1296F_0^4 + 5616F_0^3F_2 - 16848F_0^3F_4 - 146016F_0^3F_6 - 317520F_0^2F_2^2 - 101304F_0^2F_2F_4 + 7032528F_0^2F_2F_6 - 1998108F_0^2F_4^2 \\&+ 10125216F_0^2F_4F_6 - 83170368F_0^2F_6^2 + 493200F_0F_2^3 + 5081760F_0F_2^2F_4 - 24978240F_0F_2^2F_6 + 2561544F_0F_2F_4^2 \\&- 35172144F_0F_2F_4F_6 - 327010608F_0F_2F_6^2 + 12990048F_0F_4^3 + 242406360F_0F_4^2F_6 - 46110096F_0F_4F_6^2 + 6483451104F_0F_6^3 \\&+ 696000F_2^4 - 351000F_2^3F_4 - 33745200F_2^3F_6 + 155313780F_2^2F_4^2 - 572900400F_2^2F_4F_6 + 1180362960F_2^2F_6^2 - 298838776F_2F_4^3 \\&- 3328354632F_2F_4^2F_6 + 31536154440F_2F_4F_6^2 - 52829172240F_2F_6^3 + 355683372F_4^4 - 3203526304F_4^3F_6 + 26058554676F_4^2F_6^2 \\&- 320568842880F_4F_6^3 + 624056038320F_6^4
\end{align*}
\normalsize
The moments are
\small
\begin{align*}
    \sum_{i=1}^4 E_i(^1G)^2&=144F_0^2 + 312F_0F_2 - 936F_0F_4 - 8112F_0F_6 + 18316F_2^2 + 1572F_2F_4 - 425848F_2F_6 + 117090F_4^2 - 457056F_4F_6\\&
    + 5077552F_6^2\\
    \sum_{i=1}^4 E_i(^1G)^3&=864F_0^3 + 2808F_0^2F_2 - 8424F_0^2F_4 - 73008F_0^2F_6 + 329688F_0F_2^2 + 28296F_0F_2F_4 - 7665264F_0F_2F_6 \\&+ 2107620F_0F_4^2 - 8227008F_0F_4F_6 + 91395936F_0F_6^2 + 952136F_2^3 + 538308F_2^2F_4 - 46984152F_2^2F_6 + 5426082F_2F_4^2 \\&+ 8706168F_2F_4F_6 + 448507032F_2F_6^2 - 6967230F_4^3 + 62118648F_4^2F_6 - 100207656F_4F_6^2 - 1752454288F_6^3
\\
    \sum_{i=1}^4 E_i(^1G)^4&=5184F_0^4 + 22464F_0^3F_2 - 67392F_0^3F_4 - 584064F_0^3F_6 + 3956256F_0^2F_2^2 + 339552F_0^2F_2F_4 - 91983168F_0^2F_2F_6\\& + 25291440F_0^2F_4^2 - 98724096F_0^2F_4F_6 + 1096751232F_0^2F_6^2 + 22851264F_0F_2^3 + 12919392F_0F_2^2F_4 \\&- 1127619648F_0F_2^2F_6 + 130225968F_0F_2F_4^2 + 208948032F_0F_2F_4F_6 + 10764168768F_0F_2F_6^2 - 167213520F_0F_4^3 \\&+ 1490847552F_0F_4^2F_6  - 2404983744F_0F_4F_6^2 - 42058902912F_0F_6^3 + 185655856F_2^4 + 22149024F_2^3F_4 \\&- 9228030656F_2^3F_6 + 1476622488F_2^2F_4^2 - 5269552992F_2^2F_4F_6 + 210385985568F_2^2F_6^2 + 1030704840F_2F_4^3 \\&- 36436973232F_2F_4^2F_6 + 65372405472F_2F_4F_6^2 - 2048220878528F_2F_6^3 + 5450686098F_4^4 - 50236374144F_4^3F_6 \\&+ 515758820016F_4^2F_6^2 - 1076405569536F_4F_6^3 + 9688573576000F_6^4
\end{align*}
\normalsize

The $L_z=3$ sector has 80 states, which gives an additional quintet, four triplets, and one singlet. The quintet energy is $E(^5F)=6F_0-60F_2-198F_4-1716F_6$. The triplet energies $E(^3F)$ are given by roots of the quartic $P_4$, with coefficients
\small
\begin{align*}
    e_1&=24F_0 - 56F_2 - 294F_4 - 2174F_6\\
    e_2&=216F_0^2 - 1008F_0F_2 - 5292F_0F_4 - 39132F_0F_6 - 2700F_2^2 + 7758F_2F_4 + 148680F_2F_6 - 7941F_4^2 + 768096F_4F_6 - 367452F_6^2\\
    e_3&=864F_0^3 - 6048F_0^2F_2 - 31752F_0^2F_4 - 234792F_0^2F_6 - 33200F_0F_2^2 + 94536F_0F_2F_4 + 1793120F_0F_2F_6 - 95940F_0F_4^2\\& + 9209088F_0F_4F_6 - 4434512F_0F_6^2 + (471200F_2^3)/3 + 1161140F_2^2F_4 - 1576840F_2^2F_6 + 1658682F_2F_4^2 - 26790140F_2F_4F_6 \\&- 46026400F_2F_6^2 + 2256768F_4^3 - 16312770F_4^2F_6 - 117486880F_4F_6^2 + (5266266728F_6^3)/3\\
    e_4&=1296F_0^4 - 12096F_0^3F_2 - 63504F_0^3F_4 - 469584F_0^3F_6 - 102000F_0^2F_2^2 + 287928F_0^2F_2F_4 + 5406240F_0^2F_2F_6 - 289764F_0^2F_4^2 \\&+ 27603072F_0^2F_4F_6 - 13378800F_0^2F_6^2 + 953600F_0F_2^3 + 7005480F_0F_2^2F_4 - 9151680F_0F_2^2F_6 + 9855324F_0F_2F_4^2 \\& - 162069144F_0F_2F_4F_6 - 280676928F_0F_2F_6^2 + 13588236F_0F_4^3 - 96931728F_0F_4^2F_6 - 698694528F_0F_4F_6^2 + 10546168784F_0F_6^3 \\&- (2551600F_2^4)/3 - 29795240F_2^3F_4 - (188361280F_2^3F_6)/3 - 98468952F_2^2F_4^2 + 105872304F_2^2F_4F_6 + 1259789248F_2^2F_6^2 \\& - 88505226F_2F_4^3 - 778192608F_2F_4^2F_6 - 1402209168F_2F_4F_6^2 + (54250849184F_2F_6^3)/3 + 12417057F_4^4 - 482320044F_4^3F_6 \\& - 5678579472F_4^2F_6^2 - 236095123376F_4F_6^3 - (1040198011360F_6^4)/3
\end{align*}
\normalsize
The power sums are
\small
\begin{align*}
    \sum_{i=1}^4 E_i( ^3F)^2 &=
    144F_0^2 - 672F_0F_2 - 3528F_0F_4 - 26088F_0F_6 + 8936F_2^2 + 16692F_2F_4 - 58352F_2F_6 + 102642F_4^2 - 253848F_4F_6 \\&+ 5473724F_6^2\\
    \sum_{i=1}^4 E_i( ^3F)^3 &=864F_0^3 - 6048F_0^2F_2 - 31752F_0^2F_4 - 234792F_0^2F_6 + 160848F_0F_2^2 + 300456F_0F_2F_4 - 1050336F_0F_2F_6 \\&+ 1847556F_0F_4^2 - 4569264F_0F_4F_6 + 98527032F_0F_6^2 - 180416F_2^3 - 437868F_2^2F_4 - 18433392F_2^2F_6 - 3843198F_2F_4^2 \\& + 18303336F_2F_4F_6 - 15097488F_2F_6^2 - 25741098F_4^3 + 11105784F_4^2F_6 + 151939872F_4F_6^2 - 7432449896F_6^3\\
    \sum_{i=1}^4 E_i( ^3F)^4 &=5184F_0^4 - 48384F_0^3F_2 - 254016F_0^3F_4 - 1878336F_0^3F_6 + 1930176F_0^2F_2^2 + 3605472F_0^2F_2F_4 - 12604032F_0^2F_2F_6 \\& + 22170672F_0^2F_4^2 - 54831168F_0^2F_4F_6 + 1182324384F_0^2F_6^2 - 4329984F_0F_2^3 - 10508832F_0F_2^2F_4 - 442401408F_0F_2^2F_6 \\& 
    - 92236752F_0F_2F_4^2 + 439280064F_0F_2F_4F_6 - 362339712F_0F_2F_6^2 - 617786352F_0F_4^3 + 266538816F_0F_4^2F_6 \\&+ 3646556928F_0F_4F_6^2 - 178378797504F_0F_6^3 + 28836896F_2^4 + 61285344F_2^3F_4 - 63672064F_2^3F_6 + 522164376F_2^2F_4^2 \\& - 5215780512F_2^2F_4F_6 + 68624406336F_2^2F_6^2 + 1647645624F_2F_4^3 - 15927271632F_2F_4^2F_6 + 86330620512F_2F_4F_6^2 \\& - 456812427904F_2F_6^3 + 7669804914F_4^4 - 26339649120F_4^3F_6 + 300067609680F_4^2F_6^2 - 1759096645824F_4F_6^3 \\& + 15740119631408F_6^4
\end{align*}
\normalsize

The $L_z=2$ sector has 95 states, giving an additional quintet, two triplets, and four singlets. The quintet energy is
\small
$$E(^5D)=6F_0-5F_2- 132F_4-2717F_6$$
\normalsize
and the two triplet energies are 
\small
\begin{align*}
E(^3D)=6F_0-3F_2-85.5F_4- 1170F_6\pm\frac{1}{2}
\sqrt{17296F_2
^2-25320F_2F_4-228592F_2F_6 + 11289F^2
_4 + 143052F_4F_6 + 828100F^2
_6}.
\end{align*}
\normalsize
The four singlet energies are roots of the quartic $P_4$ with the coefficients
\small
\begin{align*}
    e_1&=24F_0 + 68F_2 + 90F_4 - 2146F_6\\
    e_2&=216F_0^2 + 1224F_0F_2 + 1620F_0F_4 - 38628F_0F_6 - 17794F_2^2 - 5418F_2F_4 + 227758F_2F_6 - 47187F_4^2 + 574164F_4F_6 - 3465904F_6^2\\
    e_3&=864F_0^3 + 7344F_0^2F_2 + 9720F_0^2F_4 - 231768F_0^2F_6 - 213528F_0F_2^2 - 65016F_0F_2F_4 + 2733096F_0F_2F_6 - 566244F_0F_4^2 \\&+ 6889968F_0F_4F_6 - 41590848F_0F_6^2 + 334900F_2^3 + 909078F_2^2F_4 - 4594422F_2^2F_6 - 5095530F_2F_4^2 + 77129976F_2F_4F_6 \\&- 236983032F_2F_6^2 - 1566972F_4^3 + 78401862F_4^2F_6 - 1144910910F_4F_6^2 + 5729478898F_6^3\\
e_4&=1296F_0^4 + 14688F_0^3F_2 + 19440F_0^3F_4 - 463536F_0^3F_6 - 640584F_0^2F_2^2 - 195048F_0^2F_2F_4 + 8199288F_0^2F_2F_6 - 1698732F_0^2F_4^2\\& + 20669904F_0^2F_4F_6  - 124772544F_0^2F_6^2 + 2009400F_0F_2^3 + 5454468F_0F_2^2F_4 - 27566532F_0F_2^2F_6 - 30573180F_0F_2F_4^2 \\&+ 462779856F_0F_2F_4F_6  - 1421898192F_0F_2F_6^2 - 9401832F_0F_4^3 + 470411172F_0F_4^2F_6  - 6869465460F_0F_4F_6^2 +  + 34376873388F_0F_6^3 \\&+ 12278025F_2^4 - 3576390F_2^3F_4 - 391604310F_2^3F_6 + 246956877F_2^2F_4^2 - 4504293180F_2^2F_4F_6  + 24579965592F_2^2F_6^2 \\&+ 628014816F_2F_4^3 - 13451345622F_2F_4^2F_6  + 103686328746F_2F_4F_6^2 - 386246210922F_2F_6^3 + 322609716F_4^4 - 11012900184F_4^3F_6 \\& + 156520032669F_4^2F_6^2 - 830430290976F_4F_6^3 + 1384613875359F_6^4
\end{align*}
\normalsize
This gives the singlet energies $E(^1D)$ as
\small
\begin{align*}
    E(^1D)&=6F_0 + 17F_2 + 22.5F_4 - 536.5F_6- S_D\pm\frac{1}{2}\sqrt{-4S_D^2-2p_D+\frac{q_D}{S_D}}\\
    &=6F_0 + 17F_2 + 22.5F_4 - 536.5F_6+ S_D\pm\frac{1}{2}\sqrt{-4S_D^2-2p_D-\frac{q_D}{S_D}}
\end{align*}
\normalsize

The $L_z=1$ sector has 104 states, giving additionally three triplets. The triplet energies $E(^3P)$ are roots to the cubic $P_3$ with the coefficients:
\small
\begin{align*}
    e_1&=18F_0 + 19F_2 - 29F_4 - 2873F_6\\
    e_2&=108F_0^2 + 228F_0F_2 - 348F_0F_4 - 34476F_0F_6 - 6125F_2^2 + 1638F_2F_4 + 55094F_2F_6 - 30492F_4^2 + 273702F_4F_6 + 388531F_6^2\\
    e_3&=216F_0^3 + 684F_0^2F_2 - 1044F_0^2F_4 - 103428F_0^2F_6 - 36750F_0F_2^2 + 9828F_0F_2F_4 + 330564F_0F_2F_6 - 182952F_0F_4^2 \\& + 1642212F_0F_4F_6 + 2331186F_0F_6^2 + 7425F_2^3 + 87315F_2^2F_4 - 651105F_2^2F_6 - 281160F_2F_4^2 + 15069054F_2F_4F_6  \\& - 16312725F_2F_6^2 - 1285020F_4^3 + 21405384F_4^2F_6 - 44599269F_4F_6^2 + 825423885F_6^3
\end{align*}
\normalsize

Lastly, the $L_z=0$ sector has 111 states, giving an additional quintet and a singlet. The quintet energy is $E(^5S)=6F_0-60F_2-198F_4-1716F_6$ and the singlet energy is $E(^1S)=6F_0+110F_2+111F_4+104F_6$.

I note that some of these results ($^{1,5}S,~^{3,5}D,~^{1,5}F,~^5G,~^1H,~^5I,~^{1,3}J,~^{1,3}K,~^3L$ and $^1M$) were obtained earlier by Reilly in Ref. \cite{Reilly1953}. However $J, K, L, M$ were mislabeled as $K, L, M, N$.

The Hund's rule ground state is $^5I$.

\begin{table}[]
\centering
\resizebox{\columnwidth}{!}{
\begin{tabular}{|c|c|c|c|}
\hline
configuration & term symbol & eigenvalue & degeneracy\\
\hline
 &$^4S$ & $10F_0 - 30F_2 - 99F_4 - 858F_6$ & $4$\\
 &$^2P$ & $10F_0 + 3F_2/2-29F_4-3249F_6/2+ S_P\pm\frac{1}{2}\sqrt{-4S_P^2-2p_P-\frac{q_P}{S_P}}$ & $24$\\
 & & $10F_0 + 3F_2/2 -29F_4 -3249F_6/2- S_P\pm\frac{1}{2}\sqrt{-4S_P^2-2p_P+\frac{q_P}{S_P}}$ & \\
 &$^4P$ & $10F_0 - 47F_2 - 209F_4 - 1859F_6\pm\frac{1}{2}\sqrt{6416F_2^2 - 44880F_2F_4 + 128128F_2F_6 + 101200F_4^2 - 720720F_4F_6 + 1457456F_6^2}$ & $24$\\
 &$^6P$ &$10F_0 - 45F_2 - 264F_4 - 3861F_6$ & $18$\\
 &$^2D$ & 5 doublets & $50$\\
 &$^4D$ & $10F_0 - 109F_2/3 - 136F_4 - 5395F_6/3 - S_D\pm\frac{1}{2}\sqrt{-4S_D^2-2p_D+\frac{q_D}{S_D}}$& $60$\\
& & $10F_0 - 109F_2/3 - 136F_4 - 5395F_6/3 + S_D\pm\frac{1}{2}\sqrt{-4S_D^2-2p_D-\frac{q_D}{S_D}}$& \\
 &$^2F$ & 7 doublets & $98$\\
 &$^4F$ & $10F_0 - 154F_2/4 - 567F_4/4 - 1894F_6 - S_F\pm\frac{1}{2}\sqrt{-4S_F^2-2p_F+\frac{q_F}{S_F}}$& $112$\\
 & & $10F_0 - 154F_2/4 - 567F_4/4 - 1894F_6 + S_F\pm\frac{1}{2}\sqrt{-4S_F^2-2p_F-\frac{q_F}{S_F}}$& \\
 &$^6F$ &$10F_0-100F_2-330F_4-2860F_6$ & $42$\\
 &$^2G$ & 6 doublets & $108$\\
 $f^5/f^{9}$  &$^4G$ & $10F_0 - 186F_2/4 - 723F_4/4 - 1572F_6 - S_G\pm\frac{1}{2}\sqrt{-4S_G^2-2p_G+\frac{q_G}{S_G}}$ & $144$\\
& & $10F_0 - 186F_2/4 - 723F_4/4 - 1572F_6 + S_G\pm\frac{1}{2}\sqrt{-4S_G^2-2p_G-\frac{q_G}{S_G}}$ & \\
 &$^2H$ & 7 doublets & $154$\\
&$^4H$ &$10F_0 - 121F_2/3 - 180F_4 - 6319F_6/3 +\lambda_H \cos(\theta_H /3 + 2\pi k/3)$ & $132$\\
 &$^6H$ &$10F_0 - 115F_2 - 348F_4 - 2587F_6$ & $66$\\
 &$^2I$ & 5 doublets & $130$\\
 &$^4I$ & $10F_0 - 175F_2/3 - 175F_4 - 3967F_6/3 +\lambda_I \cos(\theta_I /3 + 2\pi k/3)$ & $156$\\
 &$^2J$ & 5 doublets & $150$\\
 &$^4J$ &$10F_0 - 62F_2 - 391F_4/2 - 1775F_6\pm \frac{1}{2}\sqrt{5696F_2^2 - 12336F_2F_4 - 51296F_2F_6 + 14017F_4^2 - 32508F_4F_6 + 379652F_6^2}$&$120$\\
 &$^2K$ &$10F_0 - 184F_2/3 - 96F_4 - 1132F_6/3 +\lambda_K \cos(\theta_K /3 + 2\pi k/3)$&$102$\\
 &$^4K$ &$10F_0 - 90F_2 - 276F_4 - 984F_6$& $68$\\
 &$^2L$ &$10F_0 - 53F_2 - 135F_4 - 851F_6\pm\frac{1}{2}\sqrt{6976F_2^2 - 7584F_2F_4 - 107968F_2F_6 + 11124F_4^2 - 50064F_4F_6 + 744016F_6^2}$&$76$\\
 &$^4L$ &$10F_0 - 105F_2 - 231F_4 - 1089F_6$&$76$\\
 &$^2M$ &$10F_0 - 91F_2 - 224F_4 + 101F_6$&$42$\\
 &$^2N$ &$10F_0-80F_2-180F_4-284F_6$&$46$\\
\hline
\end{tabular}}
\vskip 0.4cm
\caption{Table of multiplet energies for $f^{5}/f^{9}$ configurations. The term symbols are to be read as $^{2S+1}L$, and the multiplicity $(2L+1)\times(2S+1)$. Note for $f$-electrons, $F^2=15^2F_2, F^4=33^2F_4$, and $F^6=\frac{429^2}{5^2}F_6$.} 
\label{Table:ATMfb}
\end{table}
\subsubsection{\texorpdfstring{$f^5/f^9$}{TEXT} Configurations}
For the $f^5/f^9$ configurations there are ${14 \choose{5}}= 2002$ states. The allowed term symbols are $^2N$,$~ ^2M$,~ $^{4,2}L,~ ^{4,2}K,~ ^{4,2}J,~ ^{4,2}I,~ ^{6,4,2}H,~ ^{4,2}G,~ ^{6,4,2}F,~ ^{4,2}D,~ ^{6,4,2}P$, and $^4S$. 

The maximum $L_z=11$ occurs for $S_z=\pm \frac{1}{2}$. The doublet eigenvalue $E(^2N)=10F_0-80F_2-180F_4-284F_6.$

$L_z=10$ gives one more doublet having the eigenvalue 
$E(^2M)=10F_0 - 91F_2 - 224F_4 + 101F_6.$

There is one state having $L_z=9, S_z=\frac{3}{2}$, with the quartet eigenvalue $E(^4L)=10F_0 - 105F_2 - 231F_4 - 1089F_6.$

There are five states having $L_z=9, S_z=\frac{1}{2}$, giving two doublets with energies 
\small
$$E(^2L)=10F_0 - 53F_2 - 135F_4 - 851F_6\pm\frac{1}{2}\sqrt{6976F_2^2 - 7584F_2F_4 - 107968F_2F_6 + 11124F_4^2 - 50064F_4F_6 + 744016F_6^2}.$$
\normalsize

There are two states having $L_z=8, S_z=\frac{3}{2}$, giving another quartet with the eigenvalue $E(^4K)=10F_0 - 90F_2 - 276F_4 - 984F_6.$

For $L_z=8, S_z=\frac{1}{2}$ has 9 states giving 3 doublets. The doublets are solutions to the cubic equation with the coefficients
\small
\begin{align*}
    e_1&=30F_0 - 184F_2 - 288F_4 - 1132F_6\\
    e_2&=300F_0^2 - 3680F_0F_2 - 5760F_0F_4 - 22640F_0F_6 + 8100F_2^2 + 32208F_2F_4 + 169752F_2F_6 + 23193F_4^2 + 310668F_4F_6 \\ &
    - 859272F_6^2   \\
    e_3&= 3000F_0^3 - 55200F_0^2F_2 - 86400F_0^2F_4 - 339600F_0^2F_6 + 529680F_0F_2^2 + 1247040F_0F_2F_4 + 2312160F_0F_2F_6 \\ & + 1096740F_0F_4^2 + 920880F_0F_4F_6 + 89999040F_0F_6^2 - 1790704F_2^3 - 5898888F_2^2F_4 - 8157432F_2^2F_6 - 8624232F_2F_4^2\\ & + 13305744F_2F_4F_6 - 501750912F_2F_6^2 - 5250258F_4^3 + 13765788F_4^2F_6 - 794083968F_4F_6^2 - 1549247752F_6^3
\end{align*}
\normalsize
The power sums are
\small
\begin{align*}
    \sum_{i=1}^3 E_i(^2K)^2&=300F_0^2 - 3680F_0F_2 - 5760F_0F_4 - 22640F_0F_6 + 17656F_2^2 + 41568F_2F_4 + 77072F_2F_6 + 36558F_4^2 \\ &+ 30696F_4F_6 + 2999968F_6^2\\
    \sum_{i=1}^3 E_i(^2K)^3&=3000F_0^3 - 55200F_0^2F_2 - 86400F_0^2F_4 - 339600F_0^2F_6 + 529680F_0F_2^2 + 1247040F_0F_2F_4 + 2312160F_0F_2F_6 \\ & + 1096740F_0F_4^2 + 920880F_0F_4F_6 + 89999040F_0F_6^2 - 1790704F_2^3 - 5898888F_2^2F_4 - 8157432F_2^2F_6 \\
    &- 8624232F_2F_4^2 + 13305744F_2F_4F_6 - 501750912F_2F_6^2 - 5250258F_4^3 + 13765788F_4^2F_6 - 794083968F_4F_6^2 \\ & - 1549247752F_6^3\\
    \sum_{i=1}^3 E_i(^2K)^4&=30000F_0^4 - 736000F_0^3F_2 - 1152000F_0^3F_4 - 4528000F_0^3F_6 + 10593600F_0^2F_2^2 + 24940800F_0^2F_2F_4 \\& + 46243200F_0^2F_2F_6 + 21934800F_0^2F_4^2 + 18417600F_0^2F_4F_6 + 1799980800F_0^2F_6^2 - 71628160F_0F_2^3 \\& - 235955520F_0F_2^2F_4 - 326297280F_0F_2^2F_6 - 344969280F_0F_2F_4^2 + 532229760F_0F_2F_4F_6 \\& - 20070036480F_0F_2F_6^2 - 210010320F_0F_4^3 + 550631520F_0F_4^2F_6 - 31763358720F_0F_4F_6^2 - 61969910080F_0F_6^3 \\& + 188463136F_2^4 + 786233856F_2^3F_4 + 801625984F_2^3F_6 + 1590844176F_2^2F_4^2 - 3443758272F_2^2F_4F_6 \\& + 78431796096F_2^2F_6^2 + 1727277696F_2F_4^3 - 8802518688F_2F_4^2F_6 + 196028258688F_2F_4F_6^2 + 198065273344F_2F_6^3 \\& + 798693858F_4^4 - 4595142960F_4^3F_6 + 184891046976F_4^2F_6^2 + 168635648640F_4F_6^3 + 3267678857728F_6^4
\end{align*}
\normalsize

There are four states having $L_z=7, S_z=\frac{3}{2}$, giving two quartets having the eigenvalues 
\small
$$E(^4J)=10F_0 - 62F_2 - 391F_4/2 - 1775F_6\pm \frac{1}{2}\sqrt{5696F_2^2 - 12336F_2F_4 - 51296F_2F_6 + 14017F_4^2 - 32508F_4F_6 + 379652F_6^2}.$$
\normalsize

There are sixteen states having $L_z=7, S_z=\frac{1}{2}$, giving five doublets. The eigenvalues cannot be determined analytically. However, we can determine the power sums:
\small
\begin{align*}
\sum_{i=1}^5 E_i(^2J)&=50F_0 - 148F_2 - 592F_4 - 4192F_6\\
\sum_{i=1}^5 E_i(^2J)^2&=500F_0^2 - 2960F_0F_2 - 11840F_0F_4 - 83840F_0F_6 + 23872F_2^2 + 39328F_2F_4 - 207808F_2F_6 + 156286F_4^2 + \\ & 396904F_4F_6 
+ 9092968F_6^2\\
\sum_{i=1}^5 E_i(^2J)^3&=
5000F_0^3 - 44400F_0^2F_2 - 177600F_0^2F_4 - 1257600F_0^2F_6 + 716160F_0F_2^2 + 1179840F_0F_2F_4 - 6234240F_0F_2F_6 \\& + 4688580F_0F_4^2 + 11907120F_0F_4F_6 + 272789040F_0F_6^2 - 1701568F_2^3 - 8710728F_2^2F_4 - 17224392F_2^2F_6 \\ & - 14810424F_2F_4^2 + 172463568F_2F_4F_6 - 151877424F_2F_6^2 - 39515218F_4^3 - 133682100F_4^2F_6 \\ & - 2149678464F_4F_6^2 - 12189245752F_6^3\\
\sum_{i=1}^5 E_i(^2J)^4&=50000F_0^4 - 592000F_0^3F_2 - 2368000F_0^3F_4 - 16768000F_0^3F_6 + 14323200F_0^2F_2^2 + 23596800F_0^2F_2F_4 \\& - 124684800F_0^2F_2F_6 + 93771600F_0^2F_4^2 + 238142400F_0^2F_4F_6 + 5455780800F_0^2F_6^2 - 68062720F_0F_2^3 \\& - 348429120F_0F_2^2F_4 - 688975680F_0F_2^2F_6 - 592416960F_0F_2F_4^2 + 6898542720F_0F_2F_4F_6 - 6075096960F_0F_2F_6^2 \\& - 1580608720F_0F_4^3 - 5347284000F_0F_4^2F_6 - 85987138560F_0F_4F_6^2 - 487569830080F_0F_6^3 + 202992192F_2^4 \\& + 926763776F_2^3F_4 - 2635547136F_2^3F_6 + 3683995664F_2^2F_4^2 - 12638236608F_2^2F_4F_6 + 198355662144F_2^2F_6^2 \\& + 5353379264F_2F_4^3 - 69791055392F_2F_4^2F_6 - 30618342528F_2F_4F_6^2 - 875031606144F_2F_6^3 + 10586373730F_4^4 \\& + 29399227664F_4^3F_6 + 881048921024F_4^2F_6^2 + 2583766692992F_4F_6^3 + 22985943306528F_6^4
\end{align*}
\normalsize

There are seven states having $L_z=6, S_z=\frac{3}{2}$, giving 3 quartets. The quartets $E(^4I)$ are roots of $P_3$ with the coefficients
\small
\begin{align*}
    e_1&=30F_0 - 175F_2 - 525F_4 - 3967F_6\\
    e_2&=300F_0^2 - 3500F_0F_2 - 10500F_0F_4 - 79340F_0F_6 + 6875F_2^2 + 55350F_2F_4 + 519190F_2F_6 + 78423F_4^2 + 1474326F_4F_6 \\& + 4248971F_6^2
\\ 
    e_3&=1000F_0^3 - 17500F_0^2F_2 - 52500F_0^2F_4 - 396700F_0^2F_6 + 68750F_0F_2^2 + 553500F_0F_2F_4 + 5191900F_0F_2F_6 + 784230F_0F_4^2 \\& + 14743260F_0F_4F_6 + 42489710F_0F_6^2 + 17875F_2^3 - 693225F_2^2F_4 - 12516375F_2^2F_6 - 3317535F_2F_4^2 - 83627730F_2F_4F_6 \\& - 351687735F_2F_6^2 - 3331611F_4^3 - 115171407F_4^2F_6 - 876591513F_4F_6^2 - 660177869F_6^3 
\end{align*}
\normalsize
The power sums are
\small
\begin{align*}
    \sum_{i=1}^3 E_i(^4I)^2&=300F_0^2 - 3500F_0F_2 - 10500F_0F_4 - 79340F_0F_6 + 16875F_2^2 + 73050F_2F_4 + 350070F_2F_6 + 118779F_4^2 \\& + 1216698F_4F_6 + 7239147F_6^2\\
    \sum_{i=1}^3 E_i(^4I)^3&=3000F_0^3 - 52500F_0^2F_2 - 157500F_0^2F_4 - 1190100F_0^2F_6 + 506250F_0F_2^2 + 2191500F_0F_2F_4 + 10502100F_0F_2F_6 \\& + 3563370F_0F_4^2 + 36500940F_0F_4F_6 + 217174410F_0F_6^2 - 1696375F_2^3 - 10427175F_2^2F_4 - 47623125F_2^2F_6 \\&- 26307405F_2F_4^2 - 187226190F_2F_4F_6 - 907444965F_2F_6^2 - 31181733F_4^3 - 370351773F_4^2F_6 - 3177606663F_4F_6^2 \\& - 13842561799F_6^3\\
    \sum_{i=1}^3 E_i(^4I)^4&=30000F_0^4 - 700000F_0^3F_2 - 2100000F_0^3F_4 - 15868000F_0^3F_6 + 10125000F_0^2F_2^2 + 43830000F_0^2F_2F_4 \\& + 210042000F_0^2F_2F_6 + 71267400F_0^2F_4^2 + 730018800F_0^2F_4F_6 + 4343488200F_0^2F_6^2 - 67855000F_0F_2^3 \\& - 417087000F_0F_2^2F_4 - 1904925000F_0F_2^2F_6 - 1052296200F_0F_2F_4^2 - 7489047600F_0F_2F_4F_6 - 36297798600F_0F_2F_6^2 \\& - 1247269320F_0F_4^3 - 14814070920F_0F_4^2F_6 - 127104266520F_0F_4F_6^2 - 553702471960F_0F_6^3 + 177721875F_2^4 \\& + 1391032500F_2^3F_4 + 6014959500F_2^3F_6 + 4839263250F_2^2F_4^2 + 32540046300F_2^2F_4F_6 + 155698254450F_2^2F_6^2 \\& + 9289710900F_2F_4^3 + 80520844260F_2F_4^2F_6 + 586117993500F_2F_4F_6^2 + 2287028853900F_2F_6^3 + 8804500083F_4^4 \\& + 121278029940F_4^3F_6 + 1188310850322F_4^2F_6^2 + 7854365332212F_4F_6^3 + 26773442595219F_6^4
\end{align*}
\normalsize

There are 24 states having $L_z=6, S_z=\frac{1}{2}$, giving 5 doublets. This again has no analytic solution (quintic), but the power sums are:
\small
\begin{align*}
\sum_{i=1}^5 E_i(^2I)&=50F_0 - 155F_2 - 522F_4 - 4535F_6\\  
\sum_{i=1}^5 E_i(^2I)^2&=500F_0^2 - 3100F_0F_2 - 10440F_0F_4 - 90700F_0F_6 + 12917F_2^2 + 15780F_2F_4 + 205906F_2F_6 + 125556F_4^2 \\& + 529428F_4F_6 + 7732973F_6^2\\    
\sum_{i=1}^5 E_i(^2I)^3&=5000F_0^3 - 46500F_0^2F_2 - 156600F_0^2F_4 - 1360500F_0^2F_6 + 387510F_0F_2^2 + 473400F_0F_2F_4 + 6177180F_0F_2F_6 \\& +3766680F_0F_4^2 + 15882840F_0F_4F_6 + 231989190F_0F_6^2 - 929195F_2^3 - 2491326F_2^2F_4 - 29564853F_2^2F_6 - 5104980F_2F_4^2 \\& + 38112372F_2F_4F_6 - 630754809F_2F_6^2 - 33480000F_4^3 - 169483572F_4^2F_6 - 1153033038F_4F_6^2 - 11942516591F_6^3\\    
\sum_{i=1}^5 E_i(^2I)^4&=50000F_0^4 - 620000F_0^3F_2 - 2088000F_0^3F_4 - 18140000F_0^3F_6 + 7750200F_0^2F_2^2 + 9468000F_0^2F_2F_4 + 123543600F_0^2F_2F_6 \\&+ 75333600F_0^2F_4^2 + 317656800F_0^2F_4F_6 + 4639783800F_0^2F_6^2 - 37167800F_0F_2^3 - 99653040F_0F_2^2F_4 - 1182594120F_0F_2^2F_6 \\& - 204199200F_0F_2F_4^2 + 1524494880F_0F_2F_4F_6 - 25230192360F_0F_2F_6^2 - 1339200000F_0F_4^3 - 6779342880F_0F_4^2F_6 \\& - 46121321520F_0F_4F_6^2 - 477700663640F_0F_6^3 + 74739557F_2^4 + 213153480F_2^3F_4 + 2625384932F_2^3F_6 + 865366872F_2^2F_4^2 \\& + 206881272F_2^2F_4F_6 + 104195921070F_2^2F_6^2 + 1729756800F_2F_4^3 - 10762031184F_2F_4^2F_6 -60657410664F_2F_4F_6^2 \\& + 1214907840836F_2F_6^3 + 9657369360F_4^4 + 46857441024F_4^3F_6 + 433584146904F_4^2F_6^2 + 1835677668456F_4F_6^3 \\& + 20163761045525F_6^4    
\end{align*}
\normalsize

There is one state having $L_z=5, S_z=\frac{5}{2}$. The sextet eigenvalue is 
$E(^6H)=10F_0 - 115F_2 - 348F_4 - 2587F_6.$

There are 11 states having $L_z=5, S_z=\frac{3}{2}$, giving 3 quartets. The quartets $E(^4H)$ are roots of $P_3$ with the coefficients
\small
\begin{align*}
    e_1&=30F_0 - 121F_2 - 540F_4 - 6319F_6\\
    e_2&=300F_0^2 - 2420F_0F_2 - 10800F_0F_4 - 126380F_0F_6 + 2555F_2^2 + 46956F_2F_4 + 540514F_2F_6 + 87801F_4^2 + 2350272F_4F_6 \\&  + 12914327F_6^2
\\ 
    e_3&=1000F_0^3 - 12100F_0^2F_2 - 54000F_0^2F_4 - 631900F_0^2F_6 + 25550F_0F_2^2 + 469560F_0F_2F_4 + 5405140F_0F_2F_6 + 878010F_0F_4^2 \\& + 23502720F_0F_4F_6 + 129143270F_0F_6^2 + 25525F_2^3 - 641640F_2^2F_4 - 6267135F_2^2F_6 - 4102899F_2F_4^2 - 107099412F_2F_4F_6 \\& - 590796189F_2F_6^2 - 4078890F_4^3 - 200165721F_4^2F_6 - 2475702588F_4F_6^2 - 8508677801F_6^3
\end{align*}
\normalsize
The power sums are
\small
\begin{align*}
    \sum_{i=1}^3 E_i(^4H)^2&=300F_0^2 - 2420F_0F_2 - 10800F_0F_4 - 126380F_0F_6 + 9531F_2^2 + 36768F_2F_4 + 448170F_2F_6 + 115998F_4^2 + \\& 2123976F_4F_6  + 14101107F_6^2\\
    \sum_{i=1}^3 E_i(^4H)^3&=3000F_0^3 - 36300F_0^2F_2 - 162000F_0^2F_4 - 1895700F_0^2F_6 + 285930F_0F_2^2 + 1103040F_0F_2F_4 + 13445100F_0F_2F_6 \\& + 3479940F_0F_4^2 + 63719280F_0F_4F_6 + 423033210F_0F_6^2 - 767521F_2^3 - 4459212F_2^2F_4 - 51709125F_2^2F_6 \\& - 10219014F_2F_4^2 - 179672688F_2F_4F_6 - 1332467211F_2F_6^2 - 27463050F_4^3 - 656474166F_4^2F_6 - 6638004540F_4F_6^2\\&  - 33025296223F_6^3
\\
    \sum_{i=1}^3 E_i(^4H)^4&=30000F_0^4 - 484000F_0^3F_2 - 2160000F_0^3F_4 - 25276000F_0^3F_6 + 5718600F_0^2F_2^2 + 22060800F_0^2F_2F_4 \\& + 268902000F_0^2F_2F_6 + 69598800F_0^2F_4^2 + 1274385600F_0^2F_4F_6 + 8460664200F_0^2F_6^2 - 30700840F_0F_2^3 \\& - 178368480F_0F_2^2F_4 - 2068365000F_0F_2^2F_6 - 408760560F_0F_2F_4^2 - 7186907520F_0F_2F_4F_6 \\& - 53298688440F_0F_2F_6^2 - 1098522000F_0F_4^3 - 26258966640F_0F_4^2F_6 - 265520181600F_0F_4F_6^2 - 1321011848920F_0F_6^3 \\& + 65429811F_2^4 + 476401056F_2^3F_4 + 5407086900F_2^3F_6 + 1627727124F_2^2F_4^2 + 29493797904F_2^2F_4F_6 \\& + 197709919938F_2^2F_6^2 + 2875338504F_2F_4^3 + 60813993480F_2F_4^2F_6 + 615107977056F_2F_4F_6^2 + 3769052600916F_2F_6^3 \\& + 6847907202F_4^4 + 202784989608F_4^3F_6 + 2606460574548F_4^2F_6^2 + 19446703759536F_4F_6^3 + 80346874997667F_6^4
\end{align*}
\normalsize

There are 35 states having $L_z=5, S_z=\frac{1}{2}$, giving 7 doublets. The power sums are:
\small
\begin{align*}
    \sum_{i=1}^7 E_i(^2H)&=70F_0 - 143F_2 - 558F_4 - 7091F_6\\
    \sum_{i=1}^7 E_i(^2H)^2&=700F_0^2 - 2860F_0F_2 - 11160F_0F_4 - 141820F_0F_6 + 33183F_2^2 + 32196F_2F_4 - 229674F_2F_6 + 169098F_4^2 \\ & + 43356F_4F_6 + 15529983F_6^2\\
    \sum_{i=1}^7 E_i(^2H)^3&=7000F_0^3 - 42900F_0^2F_2 - 167400F_0^2F_4 - 2127300F_0^2F_6 + 995490F_0F_2^2 + 965880F_0F_2F_4- 6890220F_0F_2F_6 \\& + 5072940F_0F_4^2 + 1300680F_0F_4F_6 + 465899490F_0F_6^2 - 351623F_2^3 - 6724314F_2^2F_4 - 113304129F_2^2F_6 \\& - 15576030F_2F_4^2 + 176781492F_2F_4F_6 + 642836379F_2F_6^2 - 35389710F_4^3 -180877482F_4^2F_6 - 1204427898F_4F_6^2 \\& - 28751587067F_6^3\\
    \sum_{i=1}^7 E_i(^2H)^4&=70000F_0^4 - 572000F_0^3F_2 - 2232000F_0^3F_4 - 28364000F_0^3F_6 + 19909800F_0^2F_2^2 + 19317600F_0^2F_2F_4 \\& - 137804400F_0^2F_2F_6 + 101458800F_0^2F_4^2 + 26013600F_0^2F_4F_6 + 9317989800F_0^2F_6^2 - 14064920F_0F_2^3 \\& - 268972560F_0F_2^2F_4 - 4532165160F_0F_2^2F_6 - 623041200F_0F_2F_4^2 + 7071259680F_0F_2F_4F_6 + 25713455160F_0F_2F_6^2 \\& - 1415588400F_0F_4^3 - 7235099280F_0F_4^2F_6 - 48177115920F_0F_4F_6^2 - 1150063482680F_0F_6^3 + 267209463F_2^4 \\& + 367117512F_2^3F_4 - 7372135668F_2^3F_6 + 3457473132F_2^2F_4^2 - 5915973528F_2^2F_4F_6 + 530988455754F_2^2F_6^2 \\& + 3893630760F_2F_4^3 - 41130800184F_2F_4^2F_6 - 305303817384F_2F_4F_6^2 - 3621007732596F_2F_6^3 + 9447421842F_4^4 \\& + 18432390744F_4^3F_6 + 1137017658828F_4^2F_6^2 - 151496076936F_4F_6^3 + 65553988428663F_6^4
\end{align*}
\normalsize

There are 15 states having $L_z=4, S_z=\frac{3}{2}$, giving 4 quartets. The quartets $E(^4G)$ are roots of $P_4$ with the coefficients:
\small
\begin{align*}
    e_1&=40F_0 - 186F_2 - 723F_4 - 6288F_6\\
    e_2&= 600F_0^2 - 5580F_0F_2 - 21690F_0F_4 - 188640F_0F_6 + 5720F_2^2 + 91694F_2F_4 + 1061248F_2F_6 + 167897F_4^2 + 3696920F_4F_6 \\& + 12293216F_6^2
\\ 
    e_3&=4000F_0^3 - 55800F_0^2F_2 - 216900F_0^2F_4 - 1886400F_0^2F_6 + 114400F_0F_2^2 + 1833880F_0F_2F_4 + 21224960F_0F_2F_6 \\& + 3357940F_0F_4^2 + 73938400F_0F_4F_6 + 245864320F_0F_6^2 + 440000F_2^3 - 867960F_2^2F_4 - 38614920F_2^2F_6 - 12018238F_2F_4^2 \\& - 371342784F_2F_4F_6  - 1794647088F_2F_6^2 - 15091425F_4^3 - 604138600F_4^2F_6 - 5553074280F_4F_6^2 - 6222803080F_6^3\\
    e_4&=10000F_0^4 - 186000F_0^3F_2 - 723000F_0^3F_4 - 6288000F_0^3F_6 + 572000F_0^2F_2^2 + 9169400F_0^2F_2F_4 + 106124800F_0^2F_2F_6 \\& + 16789700F_0^2F_4^2 + 369692000F_0^2F_4F_6 + 1229321600F_0^2F_6^2 + 4400000F_0F_2^3 - 8679600F_0F_2^2F_4 - 386149200F_0F_2^2F_6 \\& - 120182380F_0F_2F_4^2 - 3713427840F_0F_2F_4F_6 - 17946470880F_0F_2F_6^2 - 150914250F_0F_4^3 - 6041386000F_0F_4^2F_6 \\& - 55530742800F_0F_4F_6^2 - 62228030800F_0F_6^3 - 17248000F_2^4 - 107430400F_2^3F_4 - 336001600F_2^3F_6 - 79993760F_2^2F_4^2 \\& + 4658357880F_2^2F_4F_6 + 50807808480F_2^2F_6^2 + 375178650F_2F_4^3 + 25752501632F_2F_4^2F_6 + 340643770896F_2F_4F_6^2 \\& + 824506297472F_2F_6^3 + 431219250F_4^4 + 28459503600F_4^3F_6 + 493325023048F_4^2F_6^2 + 2124929161784F_4F_6^3 \\& - 2243550598431.93F_6^4
\end{align*}
\normalsize
The power sums are:
\small
\begin{align*}
    \sum_{i=1}^4 E_i(^4G)^2&=600F_0^2 - 5580F_0F_2 - 21690F_0F_4 - 188640F_0F_6 + 5720F_2^2 + 91694F_2F_4 + 1061248F_2F_6 + 167897F_4^2 \\& + 3696920F_4F_6  + 12293216F_6^2\\
    \sum_{i=1}^4 E_i(^4G)^3&=4000F_0^3 - 55800F_0^2F_2 - 216900F_0^2F_4 - 1886400F_0^2F_6 + 694680F_0F_2^2 + 2567040F_0F_2F_4 + 6499200F_0F_2F_6 \\& + 5608050F_0F_4^2 + 50958240F_0F_4F_6 + 448575360F_0F_6^2 - 1923096F_2^3 - 14070672F_2^2F_4 - 68385240F_2^2F_6 \\& - 35166684F_2F_4^2 - 93170448F_2F_4F_6 - 567675216F_2F_6^2 - 59038749F_4^3 - 487347168F_4^2F_6 - 6016507992F_4F_6^2\\&  - 35390062488F_6^3
\\
    \sum_{i=1}^4 E_i(^4G)^4&=40000F_0^4 - 744000F_0^3F_2 - 2892000F_0^3F_4 - 25152000F_0^3F_6 + 13893600F_0^2F_2^2 + 51340800F_0^2F_2F_4 \\& + 129984000F_0^2F_2F_6 + 112161000F_0^2F_4^2 + 1019164800F_0^2F_4F_6 + 8971507200F_0^2F_6^2 - 76923840F_0F_2^3 \\& - 562826880F_0F_2^2F_4 - 2735409600F_0F_2^2F_6 - 1406667360F_0F_2F_4^2 - 3726817920F_0F_2F_4F_6 - 22707008640F_0F_2F_6^2 \\& - 2361549960F_0F_4^3 - 19493886720F_0F_4^2F_6 - 240660319680F_0F_4F_6^2 - 1415602499520F_0F_6^3 + 212395536F_2^4 \\& + 1667870336F_2^3F_4 + 4758304320F_2^3F_6 + 7093838144F_2^2F_4^2 + 33065889024F_2^2F_4F_6 + 308878875648F_2^2F_6^2 \\& + 14894767984F_2F_4^3 + 25700389120F_2F_4^2F_6 + 391678814016F_2F_4F_6^2 + 764724684096F_2F_6^3 + 20485413107F_4^4 \\& + 165161802400F_4^3F_6 + 2166638272064F_4^2F_6^2 + 18176322778496F_4F_6^3 + 86821441326720F_6^4
\end{align*}
\normalsize

There are 45 states having $L_z=4, S_z=\frac{1}{2}$, giving 6 doublets. The power sums are:
\small
\begin{align*}
    \sum_{i=1}^6 E_i(^2G)&=60F_0 - 48F_2 - 346F_4 - 7164F_6\\    
    \sum_{i=1}^6 E_i(^2G)^2&=600F_0^2 - 960F_0F_2 - 6920F_0F_4 - 143280F_0F_6 + 26808F_2^2 - 1864F_2F_4 - 403824F_2F_6 + 151714F_4^2 \\& - 176992F_4F_6 + 17377344F_6^2\\    
    \sum_{i=1}^6 E_i(^2G)^3&=6000F_0^3 - 14400F_0^2F_2 - 103800F_0^2F_4 - 2149200F_0^2F_6 + 804240F_0F_2^2 - 55920F_0F_2F_4 - 12114720F_0F_2F_6 \\& + 4551420F_0F_4^2 - 5309760F_0F_4F_6 + 521320320F_0F_6^2 - 136848F_2^3 - 6581544F_2^2F_4 - 92191824F_2^2F_6 - 3038520F_2F_4^2 \\& + 235129632F_2F_4F_6 + 1081149264F_2F_6^2 - 28134286F_4^3 - 296278284F_4^2F_6 - 300105600F_4F_6^2 - 36871899168F_6^3\\    
    \sum_{i=1}^6 E_i(^2G)^4&=60000F_0^4 - 192000F_0^3F_2 - 1384000F_0^3F_4 - 28656000F_0^3F_6 + 16084800F_0^2F_2^2 - 1118400F_0^2F_2F_4 \\& - 242294400F_0^2F_2F_6 + 91028400F_0^2F_4^2 - 106195200F_0^2F_4F_6 + 10426406400F_0^2F_6^2 - 5473920F_0F_2^3 \\& - 263261760F_0F_2^2F_4 - 3687672960F_0F_2^2F_6 - 121540800F_0F_2F_4^2 + 9405185280F_0F_2F_4F_6 + 43245970560F_0F_2F_6^2 \\& - 1125371440F_0F_4^3 -11851131360F_0F_4^2F_6 - 12004224000F_0F_4F_6^2 - 1474875966720F_0F_6^3 + 230969888F_2^4 \\& - 121574848F_2^3F_4 - 8128859008F_2^3F_6 + 3391836752F_2^2F_4^2 + 16174399872F_2^2F_4F_6 + 435741504384F_2^2F_6^2 \\& + 371646640F_2F_4^3 - 77713940384F_2F_4^2F_6 - 699473447424F_2F_4F_6^2 - 3832425273856F_2F_6^3 + 9095111026F_4^4 \\& + 23542246528F_4^3F_6 + 1522215822464F_4^2F_6^2 - 1383724872448F_4F_6^3 + 85486078465280F_6^4    
\end{align*}
\normalsize

There are 2 states having $L_z=3, S_z=\frac{5}{2}$. The sextet eigenvalue is  
$E(^6F)=10F_0-100F_2-330F_4-2860F_6.$

There are 20 states having $L_z=3, S_z=\frac{3}{2}$, having 4 quartets. The quartets $E(^4F)$ are roots of the quartic $P_4$ with the coefficients:
\small
\begin{align*}
    e_1&=40F_0 - 154F_2 - 567F_4 - 7576F_6\\
    e_2&=600F_0^2 - 4620F_0F_2 - 17010F_0F_4 - 227280F_0F_6 + 1520F_2^2 + 69042F_2F_4 + 1000216F_2F_6 + 97449F_4^2 + 3380916F_4F_6 \\& + 19399952F_6^2
\\ 
    e_3&=4000F_0^3 - 46200F_0^2F_2 - 170100F_0^2F_4 - 2272800F_0^2F_6 + 30400F_0F_2^2 + 1380840F_0F_2F_4 + 20004320F_0F_2F_6 + 1948980F_0F_4^2 \\& + 67618320F_0F_4F_6 + 387999040F_0F_6^2 + 378400F_2^3 - 997320F_2^2F_4 - 17030280F_2^2F_6 - 8424438F_2F_4^2 - 288275856F_2F_4F_6 \\& - 2022196176F_2F_6^2 - 4618053F_4^3 - 430635480F_4^2F_6 - 6064882824F_4F_6^2 - 18498671048F_6^3\\
    e_4&=10000F_0^4 - 154000F_0^3F_2 - 567000F_0^3F_4 - 7576000F_0^3F_6 + 152000F_0^2F_2^2 + 6904200F_0^2F_2F_4 + 100021600F_0^2F_2F_6 \\& + 9744900F_0^2F_4^2 + 338091600F_0^2F_4F_6 + 1939995200F_0^2F_6^2 + 3784000F_0F_2^3 - 9973200F_0F_2^2F_4 - 170302800F_0F_2^2F_6 \\& - 84244380F_0F_2F_4^2 - 2882758560F_0F_2F_4F_6 - 20221961760F_0F_2F_6^2 - 46180530F_0F_4^3 - 4306354800F_0F_4^2F_6 \\&- 60648828240F_0F_4F_6^2 - 184986710480F_0F_6^3 - 5280000F_2^4 - 25344000F_2^3F_4 - 552156000F_2^3F_6 + 91179000F_2^2F_4^2 \\& + 2257140600F_2^2F_4F_6 + 28431374400F_2^2F_6^2 + 288764190F_2F_4^3 + 16116871320F_2F_4^2F_6 + 294178856400F_2F_4F_6^2 \\& + 1227080094240F_2F_6^3 - 18719910F_4^4 + 13569876540F_4^3F_6 + 431554626360F_4^2F_6^2 + 3018353378040F_4F_6^3 \\& + 4240461688320F_6^4
\end{align*}
\normalsize
The power sums are:
\small
\begin{align*}
    \sum_{i=1}^4 E_i(^4F)^2&=400F_0^2 - 3080F_0F_2 - 11340F_0F_4 - 151520F_0F_6 + 20676F_2^2 + 36552F_2F_4 + 332976F_2F_6 + 126591F_4^2 \\&+ 1829352F_4F_6 + 18595872F_6^2\\
    \sum_{i=1}^4 E_i(^4F)^3&=4000F_0^3 - 46200F_0^2F_2 - 170100F_0^2F_4 - 2272800F_0^2F_6 + 620280F_0F_2^2 + 1096560F_0F_2F_4 + 9989280F_0F_2F_6 \\& + 3797730F_0F_4^2 + 54880560F_0F_4F_6 + 557876160F_0F_6^2 - 1814824F_2^3 - 8849952F_2^2F_4 - 93461736F_2^2F_6 \\& - 11339352F_2F_4^2 - 1417392F_2F_4F_6 - 887749968F_2F_6^2 - 30377673F_4^3 - 632949444F_4^2F_6 - 5984086248F_4F_6^2 \\& - 49404303064F_6^3\\
    \sum_{i=1}^4 E_i(^4F)^4&=40000F_0^4 - 616000F_0^3F_2 - 2268000F_0^3F_4 - 30304000F_0^3F_6 + 12405600F_0^2F_2^2 + 21931200F_0^2F_2F_4 \\&+ 199785600F_0^2F_2F_6 + 75954600F_0^2F_4^2 + 1097611200F_0^2F_4F_6 + 11157523200F_0^2F_6^2 - 72592960F_0F_2^3 \\& - 353998080F_0F_2^2F_4 - 3738469440F_0F_2^2F_6 - 453574080F_0F_2F_4^2 - 56695680F_0F_2F_4F_6 - 35509998720F_0F_2F_6^2 \\& - 1215106920F_0F_4^3 - 25317977760F_0F_4^2F_6 - 239363449920F_0F_4F_6^2 - 1976172122560F_0F_6^3 + 210901776F_2^4 \\& + 949236864F_2^3F_4 + 8920153152F_2^3F_6 + 3531413856F_2^2F_4^2 + 40602445056F_2^2F_4F_6 + 409066665984F_2^2F_6^2 \\&+ 3138302304F_2F_4^3 + 4363253376F_2F_4^2F_6 - 425032944192F_2F_4F_6^2 + 2534882497728F_2F_6^3 + 7581289923F_4^4 \\&+ 207638705520F_4^3F_6 + 2710372442016F_4^2F_6^2 + 19350140328960F_4F_6^3 + 136712060921088F_6^4
\end{align*}
\normalsize

There are 57 states having $L_z=3, S_z=\frac{1}{2}$, giving 7 doublets. The power sums are:
\small
\begin{align*}
\sum_{i=1}^7 E_i(^2F)&=70F_0 + 40F_2 - 372F_4 - 7448F_6\\    
\sum_{i=1}^7 E_i(^2F)^2&=700F_0^2 + 800F_0F_2 - 7440F_0F_4 - 148960F_0F_6 + 49752F_2^2 + 22872F_2F_4 - 949296F_2F_6 + 207150F_4^2 \\& - 125376F_4F_6 + 27187440F_6^2\\    
\sum_{i=1}^7 E_i(^2F)^3&=7000F_0^3 + 12000F_0^2F_2 - 111600F_0^2F_4 - 2234400F_0^2F_6 + 1492560F_0F_2^2 + 686160F_0F_2F_4 - 28478880F_0F_2F_6 \\& + 6214500F_0F_4^2 - 3761280F_0F_4F_6 + 815623200F_0F_6^2 + 2320720F_2^3 - 3113928F_2^2F_4 - 215145456F_2^2F_6 \\& + 5874768F_2F_4^2 + 334512576F_2F_4F_6 + 2394791184F_2F_6^2 - 34977366F_4^3 - 337938372F_4^2F_6 - 525955968F_4F_6^2 \\& - 45813279008F_6^3\\    
\sum_{i=1}^7 E_i(^2F)^4&=70000F_0^4 + 160000F_0^3F_2 - 1488000F_0^3F_4 - 29792000F_0^3F_6 + 29851200F_0^2F_2^2 + 13723200F_0^2F_2F_4 \\& - 569577600F_0^2F_2F_6 + 124290000F_0^2F_4^2 - 75225600F_0^2F_4F_6 + 16312464000F_0^2F_6^2 + 92828800F_0F_2^3 \\& - 124557120F_0F_2^2F_4 - 8605818240F_0F_2^2F_6 + 234990720F_0F_2F_4^2 + 13380503040F_0F_2F_4F_6 + 95791647360F_0F_2F_6^2 \\& - 1399094640F_0F_4^3 - 13517534880F_0F_4^2F_6 - 21038238720F_0F_4F_6^2 - 1832531160320F_0F_6^3 + 757496352F_2^4 \\& + 759758784F_2^3F_4 - 39023190912F_2^3F_6 + 7424889744F_2^2F_4^2 - 28912879872F_2^2F_4F_6 + 1475033956992F_2^2F_6^2 \\& + 1662796368F_2F_4^3 - 171799505760F_2F_4^2F_6 - 100657022976F_2F_4F_6^2 - 16337535661824F_2F_6^3 + 14204363490F_4^4 \\& - 2277991008F_4^3F_6 + 3059434685568F_4^2F_6^2 - 8732853090048F_4F_6^3 + 178407443848703F_6^4
\end{align*}
\normalsize

There are 23 states having $L_z=2, S_z=\frac{3}{2}$, having 3 quartets. The quartets $E(^4D)$ are roots of the cubic $P_3$ with the coefficients:
\small
\begin{align*}
    e_1&=30F_0 - 109F_2 - 408F_4 - 5395F_6\\
    e_2&=300F_0^2 - 2180F_0F_2 - 8160F_0F_4 - 107900F_0F_6 + 995F_2^2 + 29268F_2F_4 + 447850F_2F_6 + 42009F_4^2 + 1564992F_4F_6 \\& + 8233511F_6^2
\\ 
    e_3&=1000F_0^3 - 10900F_0^2F_2 - 40800F_0^2F_4 - 539500F_0^2F_6 + 9950F_0F_2^2 + 292680F_0F_2F_4 + 4478500F_0F_2F_6 + 420090F_0F_4^2 \\& + 15649920F_0F_4F_6 + 82335110F_0F_6^2 + 3025F_2^3 - 148500F_2^2F_4 - 2820675F_2^2F_6 - 1189287F_2F_4^2 - 60102900F_2F_4F_6 \\& - 422340633F_2F_6^2 - 882090F_4^3 - 98773389F_4^2F_6 - 1186696368F_4F_6^2 - 3086101733F_6^3
\end{align*}
\normalsize
The power sums are:
\small
\begin{align*}
    \sum_{i=1}^3 E_i(^4D)^2&=300F_0^2 - 2180F_0F_2 - 8160F_0F_4 - 107900F_0F_6 + 9891F_2^2 + 30408F_2F_4 + 280410F_2F_6 + 82446F_4^2 \\&+ 1272336F_4F_6  + 12639003F_6^2\\
    \sum_{i=1}^3 E_i(^4D)^3&=3000F_0^3 - 32700F_0^2F_2 - 122400F_0^2F_4 - 1618500F_0^2F_6 + 296730F_0F_2^2 + 912240F_0F_2F_4 + 8412300F_0F_2F_6 \\& + 2473380F_0F_4^2 + 38170080F_0F_4F_6 + 379170090F_0F_6^2 - 960589F_2^3 - 4199328F_2^2F_4 - 38204985F_2^2F_6 \\& - 8440614F_2F_4^2 - 86243976F_2F_4F_6 - 843881727F_2F_6^2 - 19144566F_4^3 - 395074134F_4^2F_6 - 3778650720F_4F_6^2 \\& - 33025934539F_6^3\\
    \sum_{i=1}^3 E_i(^4D)^4&=30000F_0^4 - 436000F_0^3F_2 - 1632000F_0^3F_4 - 21580000F_0^3F_6 + 5934600F_0^2F_2^2 + 18244800F_0^2F_2F_4 \\& + 168246000F_0^2F_2F_6 + 49467600F_0^2F_4^2 + 763401600F_0^2F_4F_6 + 7583401800F_0^2F_6^2 - 38423560F_0F_2^3 \\& - 167973120F_0F_2^2F_4 - 1528199400F_0F_2^2F_6 - 337624560F_0F_2F_4^2 - 3449759040F_0F_2F_4F_6 - 33755269080F_0F_2F_6^2 \\& - 765782640F_0F_4^3 - 15802965360F_0F_4^2F_6 - 151146028800F_0F_4F_6^2 - 1321037381560F_0F_6^3 + 94532931F_2^4 \\& + 544853616F_2^3F_4 + 4929162420F_2^3F_6 + 1436046900F_2^2F_4^2 + 17576237952F_2^2F_4F_6 + 139756589154F_2^2F_6^2 \\&+ 2421465912F_2F_4^3 + 31962029736F_2F_4^2F_6 + 218444924880F_2F_4F_6^2 + 2798345272788F_2F_6^3 + 4707401634F_4^4 \\& + 127056705048F_4^3F_6 + 1489201412244F_4^2F_6^2 + 11266027304064F_4F_6^3 + 90761065457907F_6^4
\end{align*}
\normalsize

There are 65 states having $L_z=2, S_z=\frac{1}{2}$, giving 5 doublets. The power sums are:
\small
\begin{align*}
\sum_{i=1}^5 E_i(^2D)&=50F_0 + 27F_2 - 354F_4 - 5697F_6\\    
\sum_{i=1}^5 E_i(^2D)^2&=500F_0^2 + 540F_0F_2 - 7080F_0F_4 - 113940F_0F_6 + 47797F_2^2 + 25068F_2F_4 - 1079182F_2F_6 + 174498F_4^2 \\& - 659004F_4F_6 + 17525413F_6^2\\    
\sum_{i=1}^5 E_i(^2D)^3&=5000F_0^3 + 8100F_0^2F_2 - 106200F_0^2F_4 - 1709100F_0^2F_6 + 1433910F_0F_2^2 + 752040F_0F_2F_4 - 32375460F_0F_2F_6 \\& + 5234940F_0F_4^2 - 19770120F_0F_4F_6 + 525762390F_0F_6^2 + 3743307F_2^3 - 5836518F_2^2F_4 - 271313523F_2^2F_6 \\& - 5340798F_2F_4^2 + 73365084F_2F_4F_6 + 4574467521F_2F_6^2 - 42381198F_4^3 - 82823562F_4^2F_6 + 344507706F_4F_6^2 \\& - 44916861393F_6^3\\    
\sum_{i=1}^5 E_i(^2D)^4&=50000F_0^4 + 108000F_0^3F_2 - 1416000F_0^3F_4 - 22788000F_0^3F_6 + 28678200F_0^2F_2^2 + 15040800F_0^2F_2F_4 \\& - 647509200F_0^2F_2F_6 + 104698800F_0^2F_4^2 - 395402400F_0^2F_4F_6 + 10515247800F_0^2F_6^2 + 149732280F_0F_2^3 \\& - 233460720F_0F_2^2F_4 - 10852540920F_0F_2^2F_6 - 213631920F_0F_2F_4^2 + 2934603360F_0F_2F_4F_6 + 182978700840F_0F_2F_6^2 \\& - 1695247920F_0F_4^3 - 3312942480F_0F_4^2F_6 + 13780308240F_0F_4F_6^2 - 1796674455720F_0F_6^3 + 986998117F_2^4\\& - 70605864F_2^3F_4 - 58263278684F_2^3F_6 + 3440147676F_2^2F_4^2 + 4017573144F_2^2F_4F_6 + 1680406271934F_2^2F_6^2 \\& + 3427247880F_2F_4^3 - 112373296152F_2F_4^2F_6 + 257652342792F_2F_4F_6^2 - 20669678072732F_2F_6^3 + 16457153202F_4^4 \\& - 72124561224F_4^3F_6 + 1582741949820F_4^2F_6^2 - 5398341939576F_4F_6^3 + 134929358189319F_6^4
\end{align*}
\normalsize

There are 3 states having $L_z=1, S_z=\frac{5}{2}$. The sextet eigenvalue is  $E(^6P)=10F_0 - 45F_2 - 264F_4 - 3861F_6$.

There are 26 states having $L_z=1, S_z=\frac{3}{2}$, yielding 2 quartet eigenvalues
\small
\begin{align*}
    E(^4P)=10F_0 - 47F_2 - 209F_4 - 1859F_6\pm\frac{1}{2}\sqrt{6416F_2^2 - 44880F_2F_4 + 128128F_2F_6 + 101200F_4^2 - 720720F_4F_6 + 1457456F_6^2}
\end{align*}
\normalsize

There are 72 states having $L_z=1, S_z=\frac{1}{2}$, giving 4 doublets. The four eigenvalues $E(^2P)$ are roots of the quartic with the coefficients:
\small
\begin{align*}
e_1&=40F_0 + 6F_2 - 116F_4 - 6498F_6\\    
e_2&=600F_0^2 + 180F_0F_2 - 3480F_0F_4 - 194940F_0F_6 - 13440F_2^2 + 5964F_2F_4 + 175152F_2F_6 - 39424F_4^2 + 933372F_4F_6 \\& + 12897456F_6^2\\    
e_3&=4000F_0^3 + 1800F_0^2F_2 - 34800F_0^2F_4 - 1949400F_0^2F_6 - 268800F_0F_2^2 + 119280F_0F_2F_4 + 3503040F_0F_2F_6 - 788480F_0F_4^2 \\& + 18667440F_0F_4F_6 + 257949120F_0F_6^2 - 187350F_2^3 + 2151300F_2^2F_4 + 41775510F_2^2F_6 - 1883992F_2F_4^2 - 3786552F_2F_4F_6 \\& - 754790322F_2F_6^2 + 3825008F_4^3 + 78395240F_4^2F_6 - 1931630844F_4F_6^2 - 6409255710F_6^3
\\    
e_4&=10000F_0^4 + 6000F_0^3F_2 - 116000F_0^3F_4 - 6498000F_0^3F_6 - 1344000F_0^2F_2^2 + 596400F_0^2F_2F_4 + 17515200F_0^2F_2F_6 \\& - 3942400F_0^2F_4^2 + 93337200F_0^2F_4F_6 + 1289745600F_0^2F_6^2 - 1873500F_0F_2^3 + 21513000F_0F_2^2F_4 + 417755100F_0F_2^2F_6 \\& - 18839920F_0F_2F_4^2 - 37865520F_0F_2F_4F_6 - 7547903220F_0F_2F_6^2 + 38250080F_0F_4^3 + 783952400F_0F_4^2F_6 \\& - 19316308440F_0F_4F_6^2 - 64092557100F_0F_6^3 + 27031375F_2^4 + 94500F_2^3F_4 - 792726500F_2^3F_6 - 170351720F_2^2F_4^2 \\& - 1122642180F_2^2F_4F_6 - 21773101350F_2^2F_6^2 + 108064880F_2F_4^3 + 409799984F_2F_4^2F_6 - 23960084148F_2F_4F_6^2 \\& + 720285559708F_2F_6^3 + 199214400F_4^4 - 7909336336F_4^3F_6 + 8652443800F_4^2F_6^2 + 1068996259188F_4F_6^3 - 2711117643377F_6^4    
\end{align*}
\normalsize
The power sums are:
\small
\begin{align*}
\sum_{i=1}^4 E_i(^2P)^2&=400F_0^2 + 120F_0F_2 - 2320F_0F_4 - 129960F_0F_6 + 26916F_2^2 - 13320F_2F_4 - 428280F_2F_6 + 92304F_4^2 \\& - 359208F_4F_6  + 16429092F_6^2\\    
\sum_{i=1}^4 E_i(^2P)^3&=4000F_0^3 + 1800F_0^2F_2 - 34800F_0^2F_4 - 1949400F_0^2F_6 + 807480F_0F_2^2 - 399600F_0F_2F_4 - 12848400F_0F_2F_6 \\& + 2769120F_0F_4^2 - 10776240F_0F_4F_6 + 492872760F_0F_6^2 - 319914F_2^3 + 1656900F_2^2F_4 - 140527350F_2^2F_6 \\& - 2624664F_2F_4^2 + 176190408F_2F_4F_6 + 1677919986F_2F_6^2 - 3805424F_4^3 - 470843544F_4^2F_6 + 2194622532F_4F_6^2 \\& - 42176337858F_6^3\\    
\sum_{i=1}^4 E_i(^2P)^4&=40000F_0^4 + 24000F_0^3F_2 - 464000F_0^3F_4 - 25992000F_0^3F_6 + 16149600F_0^2F_2^2 - 7992000F_0^2F_2F_4 \\&- 256968000F_0^2F_2F_6 + 55382400F_0^2F_4^2 - 215524800F_0^2F_4F_6 + 9857455200F_0^2F_6^2 - 12796560F_0F_2^3 \\& + 66276000F_0F_2^2F_4 - 5621094000F_0F_2^2F_6 - 104986560F_0F_2F_4^2 + 7047616320F_0F_2F_4F_6 + 67116799440F_0F_2F_6^2 \\& - 152216960F_0F_4^3 - 18833741760F_0F_4^2F_6 + 87784901280F_0F_4F_6^2 - 1687053514320F_0F_6^3 + 250581956F_2^4 \\& - 258234000F_2^3F_4 - 4595878000F_2^3F_6 + 2593746368F_2^2F_4^2 - 32828588016F_2^2F_4F_6 + 682993809048F_2^2F_6^2 \\& - 984766656F_2F_4^3 - 13172399744F_2F_4^2F_6 - 593473441584F_2F_4F_6^2 - 6525018223216F_2F_6^3 + 2839863552F_4^4 \\& - 23281293504F_4^3F_6 + 2277500702144F_4^2F_6^2 - 11050561254864F_4F_6^3 + 114660166388420F_6^4    
\end{align*}
\normalsize

There are 27 states having $L_z=0, S_z=\frac{3}{2}$, yielding 1 quartet $E(^4S)=10F_0 - 30F_2 - 99F_4 - 858F_6.$

This completes the $f^5/f^9$ multiplets. The Hund's rule ground state is $^6H$.

\begin{table}[]
\centering
\resizebox{\columnwidth}{!}{
\begin{tabular}{|c|c|c|c|}
\hline
configuration & term symbol & eigenvalue & degeneracy\\
\hline\hline
 &$^1S$ & $15F_0 - 9F_2 - 57F_4 - 846F_6 - S_S\pm\frac{1}{2}\sqrt{-4S_S^2-2p_S+\frac{q_S}{S_S}}$ & $4$\\
 & & $15F_0 - 9F_2 - 57F_4 - 846F_6+ S_S\pm\frac{1}{2}\sqrt{-4S_S^2-2p_S-\frac{q_S}{S_S}}$ & \\
 &$^5S$ & $15F_0 - 80F_2 - 264F_4 - 2288F_6$ & $5$\\
 &$^1P$ & $15F_0 - F_2 - 193F_4 - 2197F_6$ & $3$\\
 &$^3P$ & 6 triplets & $54$\\
 &$^5P$ & $15F_0 - 85F_2 - 319F_4 - 3289F_6$ & $15$\\
 &$^1D$ & 6 singlets & $30$\\
 &$^3D$ & 5 triplets & $75$\\
 &$^5D$ & $15F_0 - 87F_2 - 299F_4 - 2925F_6 +\lambda_D\cos(\theta_D/3+2\pi k/3)$& $75$\\
 &$^1F$ & $15F_0 - 65F_2/2 - 144F_4 - 1798F_6 - S_F\pm\frac{1}{2}\sqrt{-4S_F^2-2p_F+\frac{q_F}{S_F}}$& $28$\\
& & $15F_0 - 65F_2/2 - 144F_4 - 1798F_6 + S_F\pm\frac{1}{2}\sqrt{-4S_F^2-2p_F-\frac{q_F}{S_F}}$& \\
 &$^3F$ & 9 triplets & $189$\\
 &$^5F$ & $15F_0 - 90F_2 - 255F_4 - 2925F_6\pm \frac{1}{2}\sqrt{4800F_2^2 + 9840F_2F_4 - 109200F_2F_6 + 6660F_4^2 - 238056F_4F_6 + 3080532F_6^2}$ & $70$\\
 &$^7F$ & $15F_0 - 150F_2 - 495F_4 - 4290F_6$ & $49$\\
 &$^1G$ & 8 singlets & $72$\\
$f^6/f^{8}$ &$^3G$ & 7 triplets & $189$\\
 &$^5G$ & $15F_0 - 254F_2/3 - 967F_4/3 - 8432F_6/3+\lambda_G\cos(\theta_G/3+2\pi k/3)$ & $135$\\
 &$^1H$ & $15F_0 - 22F_2 - 429F_4/4 - 3575F_6/2- S_H\pm\frac{1}{2}\sqrt{-4S_H^2-2p_H+\frac{q_H}{S_H}}$ & $44$\\
& & $15F_0 - 22F_2 - 429F_4/4 - 3575F_6/2+ S_H\pm\frac{1}{2}\sqrt{-4S_H^2-2p_H-\frac{q_H}{S_H}}$ & \\
&$^3H$ & 9 triplets & $297$\\
 &$^5H$ & $15F_0 - 102F_2 - 669F_4/2 - 3009F_6 \pm \frac{1}{2}\sqrt{3396F_2^2 - 18780F_2F_4 + 37968F_2F_6 + 29745F_4^2 - 150360F_4F_6 + 242256F_6^2}$
 & $110$\\
 &$^1I$ & 7 singlets & $91$\\
  &$^3I$ & 6 triplets & $234$\\
 &$^5I$ & $15F_0 - 90F_2 - 615F_4/2 - 2610F_6 \pm \frac{1}{2}\sqrt{2500F_2^2 + 11460F_2F_4 - 3640F_2F_6 + 14121F_4^2 + 23268F_4F_6 + 254212F_6^2}$ & $130$\\
 &$^1J$ & $15F_0 - 44F_2 - 521F_4/3 - 1378F_6 +\lambda_J\cos(\theta_J/3+2\pi k/3)$& $45$\\
 &$^3J$ & 6 triplets & $270$\\
 &$^5J$ & $15F_0 - 100F_2 - 400F_4 - 2638F_6$ & $75$\\
 &$^1K$ &$15F_0 - 48F_2 - 621F_4/4 - 4854F_6/4- S_K\pm\frac{1}{2}\sqrt{-4S_K^2-2p_K+\frac{q_K}{S_K}}$&$68$\\
 & &$15F_0 - 48F_2 - 621F_4/4 - 4854F_6/4+ S_K\pm\frac{1}{2}\sqrt{-4S_K^2-2p_K-\frac{q_K}{S_K}}$&\\
 &$^3K$ &$15F_0 - 72F_2 - 260F_4 - 1476F_6+\lambda_K\cos(\theta_K/3+2\pi k/3)$ &$153$\\
 &$^5K$ &$15F_0 - 140F_2 - 336F_4 - 2582F_6$& $85$\\
 &$^1L$ &$15F_0 - 102F_2-156F_4- 1308F_6\pm \frac{1}{2}\sqrt{676F_2^2 -4728F_2F_4 + 13496F_2F_6 + 15156F_4^2 -129864F_4F_6 + 315364F_6^2}$&$38$\\
 &$^3L$ &$15F_0 - 93F_2- 260F_4- 1245F_6+\lambda_L\cos(\theta_L/3+2\pi k/3)$ &$171$\\
 &$^1M$ &$15F_0 - 82F_2 - 481F_4/2 - 328F_6 \pm \frac{1}{2}\sqrt{10596F_2^2 + 22020F_2F_4 - 418824F_2F_6 + 18945F_4^2 - 289380F_4F_6 + 4846884F_6^2}$ & $42$\\
 &$^3M$ &$ 15F_0 - 119F_2 - 259F_4 - 1427F_6$&$63$\\
 &$^3N$ &$15F_0 - 130F_2 - 303F_4 - 1042F_6$&$69$\\
 &$^1O$ &$15F_0 - 126F_2 - 273F_4 - 342F_6$ & $25$\\
\hline
\end{tabular}}
\vskip 0.4cm
\caption{Table of multiplet energies for $f^{6}/f^{8}$ configurations. The term symbols are to be read as $^{2S+1}L$, and the multiplicity $(2L+1)\times(2S+1)$. Note for $f$-electrons, $F^2=15^2F_2, F^4=33^2F_4$, and $F^6=\frac{429^2}{5^2}F_6$.} 
\label{Table:ATMfb}
\end{table}

\subsubsection{\texorpdfstring{$f^6/f^8$}{TEXT}Configurations}
For the $f^6/f^8$ configurations there are ${14 \choose{6}}= 3003$ states. The allowed term symbols are $^1O,~ ^3N,~ ^{3,1}M,~ ^{3,1}L,~ ^{5,3,1}K,~ ^{5,3,1}J,~ ^{5,3,1}I,~ ^{5,3,1}H,~ ^{5,3,1}G,~ ^{7,5,3,1}D,~ ^{5,3,1}D,~ ^{5,3,1}P$, and $~ ^{5,3,1}S$. 

The maximum $L_z=12$ occurs for $S_z=0$. The singlet eigenvalue is $E(^1O)=15F_0 - 126F_2 - 273F_4 - 342F_6$.

There is one state for $L_z=11, S_z=1$, giving the triplet eigenvalue  $E(^3N)=15F_0 - 130F_2 - 303F_4 - 1042F_6$.

There are two states for $L_z=10, S_z=1$, giving the triplet eigenvalue  $E(^3M)=15F_0 - 119F_2 - 259F_4 - 1427F_6$.

There are five states for $L_z=10, S_z=0$, giving two singlet eigenvalues:
\small
\begin{align*}
    E(^1M)=15F_0 - 82F_2 - 481F_4/2 - 328F_6 \pm \frac{1}{2}\sqrt{10596F_2^2 + 22020F_2F_4 - 418824F_2F_6 + 18945F_4^2 - 289380F_4F_6 + 4846884F_6^2}
\end{align*}
\normalsize

There are five states for $L_z=9, S_z=1$, giving three additional triplet eigenvalues. The three eigenvalues are roots of the cubic polynomial with the coefficients:
\small
\begin{align*}
    e_1&=45F_0 - 279F_2 - 780F_4 - 3735F_6\\
    e_2&=675F_0^2 - 8370F_0F_2 - 23400F_0F_4 - 112050F_0F_6 + 23615F_2^2 + 142734F_2F_4 + 744382F_2F_6 + 195399F_4^2 + 2006922F_4F_6\\ & + 3778463F_6^2\\
    e_3&=3375F_0^3 - 62775F_0^2F_2 - 175500F_0^2F_4 - 840375F_0^2F_6 + 354225F_0F_2^2 + 2141010F_0F_2F_4 + 11165730F_0F_2F_6 + 2930985F_0F_4^2 + \\ & 30103830F_0F_4F_6 + 56676945F_0F_6^2 - 566025F_2^3 - 5813070F_2^2F_4 - 34923675F_2^2F_6 - 17394225F_2F_4^2 - 196240152F_2F_4F_6\\ & - 416665563F_2F_6^2 - 15736320F_4^3 - 261962637F_4^2F_6 - 1048120314F_4F_6^2 - 704367369F_6^3
\end{align*}
\normalsize
The power sums are:
\small
\begin{align*}
    \sum_{i=1}^3 E_i(^3L)^2&=675F_0^2 - 8370F_0F_2 - 23400F_0F_4 - 112050F_0F_6 + 30611F_2^2 + 149772F_2F_4 + 595366F_2F_6 + 217602F_4^2 \\ & + 1812756F_4F_6 + 6393299F_6^2\\
    \sum_{i=1}^3 E_i(^3L)^3&=10125F_0^3 - 188325F_0^2F_2 - 526500F_0^2F_4 - 2521125F_0^2F_6 + 1377495F_0F_2^2 + 6739740F_0F_2F_4 + 26791470F_0F_2F_6 \\ & + 9792090F_0F_4^2 + 81574020F_0F_4F_6 + 287698455F_0F_6^2 - 3649959F_2^3 - 24859692F_2^2F_4 - 89325621F_2^2F_6 \\ & - 63866952F_2F_4^2 - 444602592F_2F_4F_6 - 1422961173F_2F_6^2 - 64527300F_4^3 - 717366636F_4^2F_6 - 4458723012F_4F_6^2 \\ & - 11879514567F_6^3\\
    \sum_{i=1}^3 E_i(^3L)^4&=151875F_0^4 - 3766500F_0^3F_2 - 10530000F_0^3F_4 - 50422500F_0^3F_6 + 41324850F_0^2F_2^2 + 202192200F_0^2F_2F_4 \\ & + 803744100F_0^2F_2F_6 + 293762700F_0^2F_4^2 + 2447220600F_0^2F_4F_6 + 8630953650F_0^2F_6^2 - 218997540F_0F_2^3 \\ & - 1491581520F_0F_2^2F_4 - 5359537260F_0F_2^2F_6 - 3832017120F_0F_2F_4^2 - 26676155520F_0F_2F_4F_6 \\ & - 85377670380F_0F_2F_6^2 - 3871638000F_0F_4^3 - 43041998160F_0F_4^2F_6 - 267523380720F_0F_4F_6^2 - 712770874020F_0F_6^3 \\ & + 453380771F_2^4 + 3940071864F_2^3F_4 + 13566408332F_2^3F_6 + 14099036076F_2^2F_4^2 + 89563668696F_2^2F_4F_6 \\ &+ 267506959314F_2^2F_6^2 + 25452765144F_2F_4^3 + 238965133896F_2F_4^2F_6 + 1342180630728F_2F_4F_6^2 \\ & + 3373283820428F_2F_6^3 + 20086410402F_4^4 + 272740502952F_4^3F_6 + 2243627394876F_4^2F_6^2 + 10703203842984F_4F_6^3 \\ & + 22843955311523F_6^4
\end{align*}
\normalsize
There are ten states for $L_z=9, S_z=0$. This gives 2 singlets
\small
\begin{align*}
    E(^1L)=15F_0 - 102F_2-156F_4- 1308F_6\pm \frac{1}{2}\sqrt{676F_2^2 -4728F_2F_4 + 13496F_2F_6 + 15156F_4^2 -129864F_4F_6 + 315364F_6^2}
\end{align*}
\normalsize

There is one state for $L_z=8, S_z=2$, giving the quintet $E(^5K)=15F_0 - 140F_2 - 336F_4 - 2582F_6$.

There are nine states for $L_z=8, S_z=1$.  Three triplets eigenvalues are roots of the cubic equation with the coefficients:
\small
\begin{align*}
e_1&= 45F_0 - 216F_2 - 780F_4 - 4428F_6\\
e_2&=675F_0^2 - 6480F_0F_2 - 23400F_0F_4 - 132840F_0F_6 + 14000F_2^2 + 111720F_2F_4 + 681844F_2F_6 + 191079F_4^2 + 2304282F_4F_6 \\& + 6048080F_6^2 \\
e_3&=3375F_0^3 - 48600F_0^2F_2 - 175500F_0^2F_4 - 996300F_0^2F_6 + 210000F_0F_2^2 + 1675800F_0F_2F_4 + 10227660F_0F_2F_6 + 2866185F_0F_4^2 \\ & + 34564230F_0F_4F_6 + 90721200F_0F_6^2 - 240000F_2^3 - 3582240F_2^2F_4 - 24492720F_2^2F_6 - 13423524F_2F_4^2 - 177155832F_2F_4F_6 \\& - 501820584F_2F_6^2 - 14409540F_4^3 - 279891906F_4^2F_6 - 1569981948F_4F_6^2 - 2425253256F_6^3
\end{align*}
\normalsize
The power sums are:
\small
\begin{align*}
\sum_{i=1}^3 E_i(^3K)^2&=675F_0^2 - 6480F_0F_2 - 23400F_0F_4 - 132840F_0F_6 + 18656F_2^2 + 113520F_2F_4 + 549208F_2F_6 + 226242F_4^2 \\& + 2299116F_4F_6 + 7511024F_6^2\\
\sum_{i=1}^3 E_i(^3K)^3&=10125F_0^3 - 145800F_0^2F_2 - 526500F_0^2F_4 - 2988900F_0^2F_6 + 839520F_0F_2^2 + 5108400F_0F_2F_4 + 24714360F_0F_2F_6 \\ & + 10180890F_0F_4^2 + 103460220F_0F_4F_6 + 337996080F_0F_6^2 - 1725696F_2^3 - 14767200F_2^2F_4 - 65445552F_2^2F_6 \\& - 49269780F_2F_4^2 - 434865960F_2F_4F_6 - 1234145448F_2F_6^2 - 70655760F_4^3 - 991348002F_4^2F_6 - 5828167116F_4F_6^2 \\& - 13753675800F_6^3
\\
\sum_{i=1}^3 E_i(^3K)^4&=
151875F_0^4 - 2916000F_0^3F_2 - 10530000F_0^3F_4 - 59778000F_0^3F_6 + 25185600F_0^2F_2^2 + 153252000F_0^2F_2F_4 \\& + 741430800F_0^2F_2F_6 + 305426700F_0^2F_4^2 + 3103806600F_0^2F_4F_6 + 10139882400F_0^2F_6^2 - 103541760F_0F_2^3 \\& - 886032000F_0F_2^2F_4 - 3926733120F_0F_2^2F_6 - 2956186800F_0F_2F_4^2 - 26091957600F_0F_2F_4F_6 - 74048726880F_0F_2F_6^2 \\& - 4239345600F_0F_4^3 - 59480880120F_0F_4^2F_6 - 349690026960F_0F_4F_6^2 - 825220548000F_0F_6^3 + 163406336F_2^4 \\& + 1823193600F_2^3F_4 + 7721374976F_2^3F_6 + 8439704640F_2^2F_4^2 + 69663121920F_2^2F_4F_6 + 180753835296F_2^2F_6^2 \\& + 20307837600F_2F_4^3 + 251927570160F_2F_4^2F_6 + 1303207996320F_2F_4F_6^2 + 2738505696896F_2F_6^3 + 23120838882F_4^4 \\& + 407598122232F_4^3F_6 + 3298266297576F_4^2F_6^2 + 14165813817024F_4F_6^3 + 26213023826048F_6^4
\end{align*}
\normalsize

There are 18 states for $L_z=8, S_z=0$.  The four singlets $E(^1K)$ are solutions to the quartic equation with the coefficients:
\small
\begin{align*}
e_1&=60F_0 - 192F_2 - 621F_4 - 4854F_6 \\    
e_2&= 1350F_0^2 - 8640F_0F_2 - 27945F_0F_4 - 218430F_0F_6 + 5776F_2^2 + 88152F_2F_4 + 869804F_2F_6 + 110007F_4^2 + 2571540F_4F_6 \\& + 6434812F_6^2\\ 
e_3&=13500F_0^3 - 129600F_0^2F_2 - 419175F_0^2F_4 - 3276450F_0^2F_6 + 173280F_0F_2^2 + 2644560F_0F_2F_4 + 26094120F_0F_2F_6 \\& + 3300210F_0F_4^2 + 77146200F_0F_4F_6 + 193044360F_0F_6^2 + 387840F_2^3 - 1632816F_2^2F_4 - 32383440F_2^2F_6 - 9792900F_2F_4^2 \\& - 288443412F_2F_4F_6 - 1078789392F_2F_6^2 - 5025267F_4^3 - 361342782F_4^2F_6 - 2837297772F_4F_6^2 - 771631584F_6^3
 \\    
e_4&=50625F_0^4 - 648000F_0^3F_2 - 2095875F_0^3F_4 - 16382250F_0^3F_6 + 1299600F_0^2F_2^2 + 19834200F_0^2F_2F_4 + 195705900F_0^2F_2F_6 \\& + 24751575F_0^2F_4^2 + 578596500F_0^2F_4F_6 + 1447832700F_0^2F_6^2 + 5817600F_0F_2^3 - 24492240F_0F_2^2F_4 - 485751600F_0F_2^2F_6 \\& - 146893500F_0F_2F_4^2 - 4326651180F_0F_2F_4F_6 - 16181840880F_0F_2F_6^2 - 75379005F_0F_4^3 - 5420141730F_0F_4^2F_6 \\& - 42559466580F_0F_4F_6^2 - 11574473760F_0F_6^3 - 15142400F_2^4 - 73691520F_2^3F_4 - 4567040F_2^3F_6 + 142601760F_2^2F_4^2 \\& + 4188784464F_2^2F_4F_6 + 31076854656F_2^2F_6^2 + 199882620F_2F_4^3 + 17127493560F_2F_4^2F_6 + 218228241456F_2F_4F_6^2 \\& + 252139241344F_2F_6^3 + 37898280F_4^4 + 8119497888F_4^3F_6 + 283506802560F_4^2F_6^2 + 446788554816F_4F_6^3 - 1312199692544F_6^4 
\end{align*}
\normalsize
The power sums are:
\small
\begin{align*}
\sum_{i=1}^4 E_i(^1K)^2&=900F_0^2 - 5760F_0F_2 - 18630F_0F_4 - 145620F_0F_6 + 25312F_2^2 + 62160F_2F_4 + 124328F_2F_6 + 165627F_4^2 \\& + 885588F_4F_6  + 10691692F_6^2 \\   
\sum_{i=1}^4 E_i(^1K)^3&= 13500F_0^3 - 129600F_0^2F_2 - 419175F_0^2F_4 - 3276450F_0^2F_6 + 1139040F_0F_2^2 + 2797200F_0F_2F_4 + 5594760F_0F_2F_6 \\& + 7453215F_0F_4^2 + 39851460F_0F_4F_6 + 481126140F_0F_6^2 - 2587392F_2^3 - 12039840F_2^2F_4 - 48846672F_2^2F_6 \\& - 23916708F_2F_4^2 + 47478312F_2F_4F_6 - 435148632F_2F_6^2 - 49615821F_4^3 - 307031634F_4^2F_6 - 2971804788F_4F_6^2 \\& - 22977790272F_6^3
\\   
\sum_{i=1}^4 E_i(^1K)^4&=202500F_0^4 - 2592000F_0^3F_2 - 8383500F_0^3F_4 - 65529000F_0^3F_6 + 34171200F_0^2F_2^2 + 83916000F_0^2F_2F_4 \\& + 167842800F_0^2F_2F_6 + 223596450F_0^2F_4^2 + 1195543800F_0^2F_4F_6 + 14433784200F_0^2F_6^2 - 155243520F_0F_2^3 \\& - 722390400F_0F_2^2F_4 - 2930800320F_0F_2^2F_6 - 1435002480F_0F_2F_4^2 + 2848698720F_0F_2F_4F_6 - 26108917920F_0F_2F_6^2 \\& - 2976949260F_0F_4^3 - 18421898040F_0F_4^2F_6 - 178308287280F_0F_4F_6^2 - 1378667416320F_0F_6^3 + 336681472F_2^4 \\& + 1695498240F_2^3F_4 + 3556477696F_2^3F_6 + 5171870016F_2^2F_4^2 + 11088377856F_2^2F_4F_6 + 227885479584F_2^2F_6^2 \\& + 9186838560F_2F_4^3 - 22569579216F_2F_4^2F_6 - 80249522976F_2F_4F_6^2 + 800283424768F_2F_6^3 + 15560393139F_4^4 \\& + 124473027240F_4^3F_6 + 1198450360296F_4^2F_6^2 + 7965914644416F_4F_6^3 + 51729464477296F_6^4 
\end{align*}
\normalsize

There are two states for $L_z=7, S_z=2$, giving the quintet $E(^5J)=15F_0 - 100F_2 - 400F_4 - 2638F_6$.

There are 16 states for $L_z=7, S_z=1$, giving  6 triplets. The power sums are:
\small
\begin{align*}
    \sum_{i=1}^6 E_i(^3J)&=90F_0 - 428F_2 - 1194F_4 - 11096F_6\\
    \sum_{i=1}^6 E_i(^3J)^2&=1350F_0^2 - 12840F_0F_2 - 35820F_0F_4 - 332880F_0F_6 + 49488F_2^2 + 166416F_2F_4 + 1309656F_2F_6 + 311524F_4^2 \\& + 3735948F_4F_6 + 25474560F_6^2\\
    \sum_{i=1}^6 E_i(^3J)^3&=20250F_0^3 - 288900F_0^2F_2 - 805950F_0^2F_4 - 7489800F_0^2F_6 + 2226960F_0F_2^2 + 7488720F_0F_2F_4 + 58934520F_0F_2F_6 \\&+ 14018580F_0F_4^2 + 168117660F_0F_4F_6 + 1146355200F_0F_6^2 - 5855168F_2^3 - 29171904F_2^2F_4 - 215979504F_2^2F_6 \\&- 64466412F_2F_4^2 - 537228264F_2F_4F_6 - 4387714152F_2F_6^2 - 95646870F_4^3 - 1249987002F_4^2F_6 - 12504513516F_4F_6^2 \\&- 59751966296F_6^3\\
    \sum_{i=1}^6 E_i(^3J)^4&=303750F_0^4 - 5778000F_0^3F_2 - 16119000F_0^3F_4 - 149796000F_0^3F_6 + 66808800F_0^2F_2^2 + 224661600F_0^2F_2F_4 \\&+ 1768035600F_0^2F_2F_6 + 420557400F_0^2F_4^2 + 5043529800F_0^2F_4F_6 + 34390656000F_0^2F_6^2 - 351310080F_0F_2^3 \\&- 1750314240F_0F_2^2F_4 - 12958770240F_0F_2^2F_6 - 3867984720F_0F_2F_4^2 - 32233695840F_0F_2F_4F_6 \\&- 263262849120F_0F_2F_6^2 - 5738812200F_0F_4^3 - 74999220120F_0F_4^2F_6 - 750270810960F_0F_4F_6^2 - 3585117977760F_0F_6^3 \\&+ 730760448F_2^4 + 4646277632F_2^3F_4 + 31591548672F_2^3F_6 + 14626435072F_2^2F_4^2 + 118105708032F_2^2F_4F_6 \\&+ 962279380512F_2^2F_6^2 + 26601510816F_2F_4^3 + 192223794800F_2F_4^2F_6 + 2289701409312F_2F_4F_6^2 \\&+ 13302047263872F_2F_6^3 + 32092753636F_4^4 + 455755540728F_4^3F_6 + 5362282214440F_4^2F_6^2 + 38057249438528F_4F_6^3 \\&+ 143146275810432F_6^4
\end{align*}
\normalsize

There are 28 states for $L_z=7, S_z=0$, giving 3 singlets. The eigenvalues $E(^1J)$ are solutions to the cubic equation with the coefficients:
\small
\begin{align*}
e_1&=45F_0 - 132F_2 - 521F_4 - 4134F_6\\
e_2&=675F_0^2 - 3960F_0F_2 - 15630F_0F_4 - 124020F_0F_6 + 2976F_2^2 + 43828F_2F_4 + 450228F_2F_6 + 72767F_4^2 + 1531132F_4F_6 \\& + 4647660F_6^2\\
e_3&=3375F_0^3 - 29700F_0^2F_2 - 117225F_0^2F_4 - 930150F_0^2F_6 + 44640F_0F_2^2 + 657420F_0F_2F_4 + 6753420F_0F_2F_6 + 1091505F_0F_4^2 \\& + 22966980F_0F_4F_6 + 69714900F_0F_6^2 + 68480F_2^3 - 488192F_2^2F_4 - 7610784F_2^2F_6 - 2743376F_2F_4^2 - 75679540F_2F_4F_6 \\& - 365583168F_2F_6^2 - 2676535F_4^3 - 106150742F_4^2F_6 - 996316940F_4F_6^2 - 769026208F_6^3
\end{align*}
\normalsize
The power sums are:
\small
\begin{align*}
\sum_{i=1}^e E(^1J)^2&=675F_0^2 - 3960F_0F_2 - 15630F_0F_4 - 124020F_0F_6 + 11472F_2^2 + 49888F_2F_4 + 190920F_2F_6 + 125907F_4^2 \\& + 1245364F_4F_6 + 7794636F_6^2\\
\sum_{i=1}^e E(^1J)^3&=10125F_0^3 - 89100F_0^2F_2 - 351675F_0^2F_4 - 2790450F_0^2F_6 + 516240F_0F_2^2 + 2244960F_0F_2F_4 + 8591400F_0F_2F_6 \\& + 5665815F_0F_4^2 + 56041380F_0F_4F_6 + 350758620F_0F_6^2 - 916032F_2^3 - 6690912F_2^2F_4 - 23726160F_2^2F_6 \\&- 18401868F_2F_4^2 - 79269816F_2F_4F_6 - 440171064F_2F_6^2 - 35715545F_4^3 - 389247858F_4^2F_6 - 3447160404F_4F_6^2 \\& - 15316677408F_6^3\\
\sum_{i=1}^e E(^1J)^4&=151875F_0^4 - 1782000F_0^3F_2 - 7033500F_0^3F_4 - 55809000F_0^3F_6 + 15487200F_0^2F_2^2 + 67348800F_0^2F_2F_4 \\& + 257742000F_0^2F_2F_6 + 169974450F_0^2F_4^2 + 1681241400F_0^2F_4F_6 + 10522758600F_0^2F_6^2 - 54961920F_0F_2^3 \\&- 401454720F_0F_2^2F_4 - 1423569600F_0F_2^2F_6 - 1104112080F_0F_2F_4^2 - 4756188960F_0F_2F_4F_6 - 26410263840F_0F_2F_6^2 \\& - 2142932700F_0F_4^3 - 23354871480F_0F_4^2F_6 - 206829624240F_0F_4F_6^2 - 919000644480F_0F_6^3 + 77736192F_2^4 \\& + 737954816F_2^3F_4 + 1907063040F_2^3F_6 + 3135511872F_2^2F_4^2 + 14358312960F_2^2F_4F_6 + 73434163104F_2^2F_6^2 \\& + 6935974592F_2F_4^3 + 31989209328F_2F_4^2F_6 + 220390368864F_2F_4F_6^2 + 1057606248192F_2F_6^3 + 10840399011F_4^4 \\& + 133413890408F_4^3F_6 + 1303847631144F_4^2F_6^2 + 9027341867840F_4F_6^3 + 30271480796784F_6^4
\end{align*}
\normalsize

There are four states for $L_z=6, S_z=2$. This gives two quintet eigenvalues:
\small
\begin{align*}
    E(^5I)=15F_0 - 90F_2 - 615F_4/2 - 2610F_6 \pm \frac{1}{2}\sqrt{2500F_2^2 + 11460F_2F_4 - 3640F_2F_6 + 14121F_4^2 + 23268F_4F_6 + 254212F_6^2}
\end{align*}
\normalsize

There are 24 states for $L_z=6, S_z=1$, giving  6 triplets. The power sums are:
\small
\begin{align*}
\sum_{i=1}^6 E_i(^3I)&=90F_0 - 400F_2 - 1278F_4 - 10900F_6\\
\sum_{i=1}^6 E_i(^3I)^2&=1350F_0^2 - 12000F_0F_2 - 38340F_0F_4 - 327000F_0F_6 + 44182F_2^2 + 161208F_2F_4 + 1113908F_2F_6 + 355008F_4^2 \\& + 3895128F_4F_6 + 26093038F_6^2\\
\sum_{i=1}^6 E_i(^3I)^3&=20250F_0^3 - 270000F_0^2F_2 - 862650F_0^2F_4 - 7357500F_0^2F_6 + 1988190F_0F_2^2 + 7254360F_0F_2F_4 + 50125860F_0F_2F_6 \\&+ 15975360F_0F_4^2 + 175280760F_0F_4F_6 + 1174186710F_0F_6^2 - 4855120F_2^3 - 26325018F_2^2F_4 - 187043772F_2^2F_6 \\& - 66156282F_2F_4^2 - 451160100F_2F_4F_6 - 3555996888F_2F_6^2 - 111348864F_4^3 - 1454002506F_4^2F_6 - 13019562666F_4F_6^2 \\& - 66984809452F_6^3\\
\sum_{i=1}^6 E_i(^3I)^4&=303750F_0^4 - 5400000F_0^3F_2 - 17253000F_0^3F_4 - 147150000F_0^3F_6 + 59645700F_0^2F_2^2 + 217630800F_0^2F_2F_4 \\& + 1503775800F_0^2F_2F_6 + 479260800F_0^2F_4^2 + 5258422800F_0^2F_4F_6 + 35225601300F_0^2F_6^2 - 291307200F_0F_2^3 \\& - 1579501080F_0F_2^2F_4 - 11222626320F_0F_2^2F_6 - 3969376920F_0F_2F_4^2 - 27069606000F_0F_2F_4F_6 \\& - 213359813280F_0F_2F_6^2 - 6680931840F_0F_4^3 - 87240150360F_0F_4^2F_6 - 781173759960F_0F_4F_6^2 - 4019088567120F_0F_6^3 \\& + 569498662F_2^4 + 3814036655F_2^3F_4 + 26120148136F_2^3F_6 + 13704105216F_2^2F_4^2 + 102589413840F_2^2F_4F_6 \\& + 805614945395F_2^2F_6^2 + 27725317416F_2F_4^3 + 194811669600F_2F_4^2F_6 + 1553609034000F_2F_4F_6^2 \\& + 11289224029576F_2F_6^3 + 37108680804F_4^4 + 568799538984F_4^3F_6 + 6116365261920F_4^2F_6^2 + 42113993426160F_4F_6^3 \\& + 178297505829526F_6^4
\end{align*}
\normalsize

There are 43 states for $L_z=6, S_z=0$, giving 7 singlets. The power sums are:
\small
\begin{align*}
\sum_{i=1}^7 E_i(^1I)&=105F_0 - 301F_2 - 987F_4 - 7861F_6\\
\sum_{i=1}^7 E_i(^1I)^2&=1575F_0^2 - 9030F_0F_2 - 29610F_0F_4 - 235830F_0F_6 + 56823F_2^2 + 110730F_2F_4 - 366786F_2F_6 + 307413F_4^2 \\& + 961338F_4F_6 + 26285511F_6^2\\
\sum_{i=1}^7 E_i(^1I)^3&=23625F_0^3 - 203175F_0^2F_2 - 666225F_0^2F_4 - 5306175F_0^2F_6 + 2557035F_0F_2^2 + 4982850F_0F_2F_4 - 16505370F_0F_2F_6 \\&+ 13833585F_0F_4^2 + 43260210F_0F_4F_6 + 1182847995F_0F_6^2 - 4375501F_2^3 - 27721665F_2^2F_4 - 93710703F_2^2F_6 \\&- 44865873F_2F_4^2 + 426256542F_2F_4F_6 - 461391735F_2F_6^2 - 100604403F_4^3 - 418547385F_4^2F_6 - 7908723585F_4F_6^2 \\&- 51786797461F_6^3\\
\sum_{i=1}^7 E_i(^1I)^4&=354375F_0^4 - 4063500F_0^3F_2 - 13324500F_0^3F_4 - 106123500F_0^3F_6 + 76711050F_0^2F_2^2 + 149485500F_0^2F_2F_4 \\&- 495161100F_0^2F_2F_6 + 415007550F_0^2F_4^2 + 1297806300F_0^2F_4F_6 + 35485439850F_0^2F_6^2 - 262530060F_0F_2^3 \\&- 1663299900F_0F_2^2F_4 - 5622642180F_0F_2^2F_6 - 2691952380F_0F_2F_4^2 + 25575392520F_0F_2F_4F_6 \\&- 27683504100F_0F_2F_6^2 - 6036264180F_0F_4^3 - 25112843100F_0F_4^2F_6 - 474523415100F_0F_4F_6^2 - 3107207847660F_0F_6^3 \\&+ 754960743F_2^4 + 3755937300F_2^3F_4 - 14082065412F_2^3F_6 + 14675101326F_2^2F_4^2 - 30936976164F_2^2F_4F_6 \\&+ 1089065898378F_2^2F_6^2 + 18225466020F_2F_4^3 - 258612557796F_2F_4^2F_6 + 402902284284F_2F_4F_6^2 - 7780406872644F_2F_6^3 \\&+ 35465832261F_4^4 + 163915624836F_4^3F_6 + 4034621151630F_4^2F_6^2 + 12364393909812F_4F_6^3 + 165973578507976F_6^4
\end{align*}
\normalsize

There are six states for $L_z=5, S_z=2$.  This gives two quintets:
\small
\begin{align*}
    E(^5H)=15F_0 - 102F_2 - 669F_4/2 - 3009F_6 \pm \frac{1}{2}\sqrt{3396F_2^2 - 18780F_2F_4 + 37968F_2F_6 + 29745F_4^2 - 150360F_4F_6 + 242256F_6^2}
\end{align*}
\normalsize

There are 35 states for $L_z=5, S_z=1$, giving  9 triplets. The power sums are:
\small
\begin{align*}
\sum_{i=1}^9 E_i(^3H)&=135F_0 - 427F_2 - 1638F_4 - 16231F_6\\
\sum_{i=1}^9 E_i(^3H)^2&=2025F_0^2 - 12810F_0F_2 - 49140F_0F_4 - 486930F_0F_6 + 62793F_2^2 + 195912F_2F_4 + 647514F_2F_6 + 463098F_4^2 \\& + 4808916F_4F_6 + 44497161F_6^2\\
\sum_{i=1}^9 E_i(^3H)^3&=30375F_0^3 - 288225F_0^2F_2 - 1105650F_0^2F_4 - 10955925F_0^2F_6 + 2825685F_0F_2^2 + 8816040F_0F_2F_4 + 29138130F_0F_2F_6 \\&+ 20839410F_0F_4^2 + 216401220F_0F_4F_6 + 2002372245F_0F_6^2 - 5894587F_2^3 - 38862882F_2^2F_4 - 227109909F_2^2F_6 \\&- 82998504F_2F_4^2 - 235154232F_2F_4F_6 - 2134490673F_2F_6^2 - 151542684F_4^3 - 1749769038F_4^2F_6 - 19040335146F_4F_6^2 \\&- 123073360471F_6^3
\\
\sum_{i=1}^9 E_i(^3H)^4&=455625F_0^4 - 5764500F_0^3F_2 - 22113000F_0^3F_4 - 219118500F_0^3F_6 + 84770550F_0^2F_2^2 + 264481200F_0^2F_2F_4 \\&+ 874143900F_0^2F_2F_6 + 625182300F_0^2F_4^2 + 6492036600F_0^2F_4F_6 + 60071167350F_0^2F_6^2 - 353675220F_0F_2^3 \\&- 2331772920F_0F_2^2F_4 - 13626594540F_0F_2^2F_6 - 4979910240F_0F_2F_4^2 -14109253920F_0F_2F_4F_6 \\&- 128069440380F_0F_2F_6^2 - 9092561040F_0F_4^3 - 104986142280F_0F_4^2F_6 - 1142420108760F_0F_4F_6^2 \\&- 7384401628260F_0F_6^3 + 817818633F_2^4 + 5299392144F_2^3F_4 + 16465278708F_2^3F_6 + 21319372140F_2^2F_4^2 \\&+ 116447605560F_2^2F_4F_6 + 1206773348022F_2^2F_6^2 + 36046391256F_2F_4^3 + 63300302544F_2F_4^2F_6 \\&+ 935442990864F_2F_4F_6^2 + 4267774067508F_2F_6^3 + 54436553046F_4^4 + 677841640704F_4^3F_6 + 8800385825148F_4^2F_6^2 \\& + 65237341217160F_4F_6^3 + 362799335864457F_6^4
\end{align*}
\normalsize

There are 58 states for $L_z=5, S_z=0$, giving  4 singlets. The eigenvalues $E(^1H)$ are solutions to the quartic equation ith the coefficients:
\small
\begin{align*}
    e_1&=60F_0 - 88F_2 - 429F_4 - 7150F_6\\
    e_2&=1350F_0^2 - 3960F_0F_2 - 19305F_0F_4 - 321750F_0F_6 - 10034F_2^2 + 40779F_2F_4 + 584726F_2F_6 + 23535F_4^2 + 2781927F_4F_6 \\&+ 16460968F_6^2\\
    e_3&=13500F_0^3 - 59400F_0^2F_2 - 289575F_0^2F_4 - 4826250F_0^2F_6 - 301020F_0F_2^2 + 1223370F_0F_2F_4 + 17541780F_0F_2F_6 \\& + 706050F_0F_4^2 + 83457810F_0F_4F_6 + 493829040F_0F_6^2 + 785880F_2^3 + 2323173F_2^2F_4 + 17372286F_2^2F_6 - 6554538F_2F_4^2 \\&- 157624158F_2F_4F_6 - 1117745676F_2F_6^2 + 4319109F_4^3 - 163383120F_4^2F_6 - 5566404003F_4F_6^2 - 12452152338F_6^3\\
    e_4&=50625F_0^4 - 297000F_0^3F_2 - 1447875F_0^3F_4 - 24131250F_0^3F_6 - 2257650F_0^2F_2^2 + 9175275F_0^2F_2F_4 + 131563350F_0^2F_2F_6 \\& + 5295375F_0^2F_4^2 + 625933575F_0^2F_4F_6 + 3703717800F_0^2F_6^2 + 11788200F_0F_2^3 + 34847595F_0F_2^2F_4 + 260584290F_0F_2^2F_6 \\& - 98318070F_0F_2F_4^2 - 2364362370F_0F_2F_4F_6 - 16766185140F_0F_2F_6^2 + 64786635F_0F_4^3 - 2450746800F_0F_4^2F_6 \\& - 83496060045F_0F_4F_6^2 - 186782285070F_0F_6^3 - 10224975F_2^4 - 62852835F_2^3F_4 - 1161444630F_2^3F_6 + 92311299F_2^2F_4^2 \\& - 4504791789F_2^2F_4F_6 + 3599456760F_2^2F_6^2 + 92273175F_2F_4^3 + 9336689316F_2F_4^2F_6 + 185519273391F_2F_4F_6^2 \\&+ 507187534038F_2F_6^3 - 117993672F_4^4 - 6160789449F_4^3F_6 + 234667639809F_4^2F_6^2 + 3190682805945F_4F_6^3 \\& + 1199938956567F_6^4
\end{align*}
\normalsize
The power sums are:
\small
\begin{align*}
    \sum_{i=1}^4 E_i(^1H)^2&=900F_0^2 - 2640F_0F_2 - 12870F_0F_4 - 214500F_0F_6 + 27812F_2^2 - 6054F_2F_4 + 88948F_2F_6 + 136971F_4^2 \\& + 570846F_4F_6 +18200564F_6^2\\
    \sum_{i=1}^4 E_i(^1H)^3&=13500F_0^3 - 59400F_0^2F_2 - 289575F_0^2F_4 - 4826250F_0^2F_6 + 1251540F_0F_2^2 - 272430F_0F_2F_4 + 4002660F_0F_2F_6 \\&+ 6163695F_0F_4^2 + 25688070F_0F_4F_6 + 819025380F_0F_6^2 - 972808F_2^3 - 5145111F_2^2F_4 - 174853578F_2^2F_6 \\& - 9554625F_2F_4^2 + 269247366F_2F_4F_6 + 38491224F_2F_6^2 - 35706717F_4^3 - 352663011F_4^2F_6 - 1636269543F_4F_6^2 \\& - 49794568414F_6^3\\
    \sum_{i=1}^4 E_i(^1H)^4&=202500F_0^4 - 1188000F_0^3F_2 - 5791500F_0^3F_4 - 96525000F_0^3F_6 + 37546200F_0^2F_2^2 - 8172900F_0^2F_2F_4 \\&+ 120079800F_0^2F_2F_6 + 184910850F_0^2F_4^2 + 770642100F_0^2F_4F_6 + 24570761400F_0^2F_6^2 - 58368480F_0F_2^3 \\& - 308706660F_0F_2^2F_4 - 10491214680F_0F_2^2F_6 - 573277500F_0F_2F_4^2 + 16154841960F_0F_2F_4F_6 + 2309473440F_0F_2F_6^2 \\& - 2142403020F_0F_4^3 - 21159780660F_0F_4^2F_6 - 98176172580F_0F_4F_6^2 - 2987674104840F_0F_6^3 + 336415172F_2^4 \\&- 614957388F_2^3F_4 + 4470772136F_2^3F_6 + 3225660210F_2^2F_4^2 + 24202291572F_2^2F_4F_6 + 879369611448F_2^2F_6^2 \\&+ 3860788212F_2F_4^3 - 13261177620F_2F_4^2F_6 - 1667129038884F_2F_4F_6^2 - 940882545304F_2F_6^3 + 10713646035F_4^4 \\&+ 75970161468F_4^3F_6 + 1569928806738F_4^2F_6^2 + 5410909686972F_4F_6^3 + 140665395964579F_6^4
\end{align*}
\normalsize

There are 9 states for $L_z=4, S_z=2$.  The 3 quintets eigenvalues are solutions to the cubic equation with the coefficients:
\small
\begin{align*}
e_1&=45F_0 - 254F_2 - 967F_4 - 8432F_6\\
e_2&=675F_0^2 - 7620F_0F_2 - 29010F_0F_4 - 252960F_0F_6 + 16780F_2^2 + 161960F_2F_4 + 1545260F_2F_6 + 296550F_4^2 + 5567462F_4F_6 \\
&+ 22212424F_6^2 \\
e_3&=3375F_0^3 - 57150F_0^2F_2 - 217575F_0^2F_4 - 1897200F_0^2F_6 + 251700F_0F_2^2 + 2429400F_0F_2F_4 + 23178900F_0F_2F_6 \\& + 4448250F_0F_4^2 + 83511930F_0F_4F_6 + 333186360F_0F_6^2 - 143400F_2^3 - 5211700F_2^2F_4 - 59046840F_2^2F_6 - 24169560F_2F_4^2 \\ & - 500013968F_2F_4F_6 - 2236694616F_2F_6^2 - 29026800F_4^3 - 875970744F_4^2F_6 - 7551521692F_4F_6^2 - 17623007688F_6^3
\end{align*}
\normalsize
The power sums are:
\small
\begin{align*}
\sum_{i=1}^3 E_i(^5G)^2&=675F_0^2 - 7620F_0F_2 - 29010F_0F_4 - 252960F_0F_6 + 30956F_2^2 + 167316F_2F_4 + 1192936F_2F_6 + 341989F_4^2 \\ & + 5172564F_4F_6 + 26673776F_6^2\\
\sum_{i=1}^3 E_i(^5G)^3&=
10125F_0^3 - 171450F_0^2F_2 - 652725F_0^2F_4 - 5691600F_0^2F_6 + 1393020F_0F_2^2 + 7529220F_0F_2F_4 + 53682120F_0F_2F_6 \\ &+ 15389505F_0F_4^2 + 232765380F_0F_4F_6 + 1200319920F_0F_6^2 - 4030904F_2^3 - 30703716F_2^2F_4 - 207182256F_2^2F_6 \\ & - 89229438F_2F_4^2 - 1104202296F_2F_4F_6 - 4872471288F_2F_6^2 - 131019913F_4^3 - 2629187514F_4^2F_6 \\& - 23638912524F_4F_6^2 - 90487143128F_6^3\\
\sum_{i=1}^3 E_i(^5G)^4&=
253125F_0^4 - 6183000F_0^3F_2 - 22086000F_0^3F_4 - 195075000F_0^3F_6 + 72173700F_0^2F_2^2 + 397442700F_0^2F_2F_4 \\ & + 3293449200F_0^2F_2F_6 + 783866700F_0^2F_4^2 + 12316625100F_0^2F_4F_6 + 60619139100F_0^2F_6^2 - 400374480F_0F_2^3 \\ & - 3024908820F_0F_2^2F_4 - 24969181320F_0F_2^2F_6 - 9170063460F_0F_2F_4^2 - 134847178920F_0F_2F_4F_6 \\ & - 637320155760F_0F_2F_6^2 - 13247942640F_0F_4^3 - 282483900900F_0F_4^2F_6 - 2475201310740F_0F_4F_6^2 \\ & - 8764081502520F_0F_6^3 + 864755522F_2^4 + 8270767164F_2^3F_4 + 67983839264F_2^3F_6 + 33778899558F_2^2F_4^2 \\& + 496877053236F_2^2F_4F_6 + 2372474258508F_2^2F_6^2 + 75488933916F_2F_4^3 + 1494768089628F_2F_4^2F_6 \\ & + 12555836682708F_2F_4F_6^2 + 43399141228256F_2F_6^3 + 88482311698F_4^4 + 2330059355580F_4^3F_6 \\& + 27554691651894F_4^2F_6^2 + 174413836877148F_4F_6^3 + 489635887333010F_6^4
\end{align*}
\normalsize

There are 45 states for $L_z=4, S_z=1$, giving 7 triplets. The power sums are:
\small
\begin{align*}
    \sum_{i=1}^7 E_i(^3G)&=105F_0 - 312F_2 - 1234F_4 - 15036F_6\\
    \sum_{i=1}^7 E_i(^3G)^2&=1575F_0^2 - 9360F_0F_2 - 37020F_0F_4 - 451080F_0F_6 + 43368F_2^2 + 95252F_2F_4 + 853824F_2F_6 + 345680F_4^2 \\& + 4324568F_4F_6 + 39597936F_6^2\\
    \sum_{i=1}^7 E_i(^3G)^3&=23625F_0^3 - 210600F_0^2F_2 - 832950F_0^2F_4 - 10149300F_0^2F_6 + 1951560F_0F_2^2 + 4286340F_0F_2F_4 + 38422080F_0F_2F_6 \\&+ 15555600F_0F_4^2 + 194605560F_0F_4F_6 + 1781907120F_0F_6^2 - 4140912F_2^3 - 20108412F_2^2F_4 - 227099304F_2^2F_6 \\&- 37620654F_2F_4^2 - 182793672F_2F_4F_6 - 2410861248F_2F_6^2 - 106667308F_4^3 - 1743133920F_4^2F_6 \\&- 14768156304F_4F_6^2 - 113811724488F_6^3\\
    \sum_{i=1}^7 E_i(^3G)^4&=253125F_0^4 - 1458000F_0^3F_2 - 7627500F_0^3F_4 - 121743000F_0^3F_6 + 28163700F_0^2F_2^2 - 42975900F_0^2F_2F_4 \\&- 5303232000F_0^2F_2F_6 + 144486450F_0^2F_4^2 + 504503100F_0^2F_4F_6 + 28847672100F_0^2F_6^2 - 89934480F_0F_2^3 \\&- 23818860F_0F_2^2F_4 - 1087712280F_0F_2^2F_6 + 1559057940F_0F_2F_4^2 + 57627420840F_0F_2F_4F_6 \\&+ 200320203600F_0F_2F_6^2 - 1013290620F_0F_4^3 + 20144614860F_0F_4^2F_6 + 170777181060F_0F_4F_6^2 - 3493850554440F_0F_6^3 \\&+ 189913422F_2^4 - 520545468F_2^3F_4 - 8201058144F_2^3F_6 - 2749229894F_2^2F_4^2 - 135807839988F_2^2F_4F_6 \\&- 236519432172F_2^2F_6^2 - 16195752332F_2F_4^3 - 650409740380F_2F_4^2F_6 - 6964346572500F_2F_4F_6^2 \\&- 16442258341920F_2F_6^3 + 335412227F_4^4 - 323441195300F_4^3F_6 - 4722070178942F_4^2F_6^2 - 18855745265916F_4F_6^3 \\&+ 169978935625515F_6^4\\
\end{align*}
\normalsize

There are 76 states for $L_z=4, S_z=0$, giving  8 singlets. The power sums are:
\small
\begin{align*}
\sum_{i=1}^8 E_i(^1G)&=120F_0 - 74F_2 - 800F_4 - 13460F_6\\    
\sum_{i=1}^8 E_i(^1G)^2&=1800F_0^2 - 2220F_0F_2 - 24000F_0F_4 - 403800F_0F_6 + 64748F_2^2 + 9420F_2F_4 - 883016F_2F_6 + 364206F_4^2 \\& + 877320F_4F_6 + 40071536F_6^2\\    
\sum_{i=1}^8 E_i(^1G)^3&=27000F_0^3 - 49950F_0^2F_2 - 540000F_0^2F_4 - 9085500F_0^2F_6 + 2913660F_0F_2^2 + 423900F_0F_2F_4 - 39735720F_0F_2F_6 \\&+16389270F_0F_4^2 + 39479400F_0F_4F_6 + 1803219120F_0F_6^2 + 306376F_2^3 - 20775132F_2^2F_4 - 366467352F_2^2F_6 \\&- 9801162F_2F_4^2 + 638700840F_2F_4F_6 + 3744859512F_2F_6^2 - 107672684F_4^3 - 910951704F_4^2F_6 - 4353684768F_4F_6^2 \\&- 114073037840F_6^3\\    
\sum_{i=1}^8 E_i(^1G)^4&=405000F_0^4 - 999000F_0^3F_2 - 10800000F_0^3F_4 - 181710000F_0^3F_6 + 87409800F_0^2F_2^2 + 12717000F_0^2F_2F_4 \\&+ 3580837200F_0^2F_2F_6 + 491678100F_0^2F_4^2 + 1184382000F_0^2F_4F_6 + 54096573600F_0^2F_6^2 + 18382560F_0F_2^3 \\&- 1246507920F_0F_2^2F_4 - 21988041120F_0F_2^2F_6 - 588069720F_0F_2F_4^2 + 38322050400F_0F_2F_4F_6 \\&+ 224691570720F_0F_2F_6^2 - 6460361040F_0F_4^3 - 54657102240F_0F_4^2F_6 - 261221086080F_0F_4F_6^2 \\&- 6844382270400F_0F_6^3 + 967449968F_2^4 - 143612640F_2^3F_4 - 39577596352F_2^3F_6 + 14078665464F_2^2F_4^2 \\&+ 54646486944F_2^2F_4F_6 + 2475427344672F_2^2F_6^2 + 4701817464F_2F_4^3 - 324116406096F_2F_4^2F_6 \\&- 2330216586528F_2F_4F_6^2 - 21970247878336F_2F_6^3 + 43424820342F_4^4 + 218625940848F_4^3F_6 \\&+ 5653679910288F_4^2F_6^2 + 5980901040960F_4F_6^3 + 368451537345538F_6^4    
\end{align*}
\normalsize

There is one state for $L_z=3, S_z=3$ giving the septet $E(^7F)=15F_0 - 150F_2 - 495F_4 - 4290F_6$.

There are 12 states for $L_z=3, S_z=2$.  This gives 2 quintets:
\small
\begin{align*}
    E(^5F)=15F_0 - 90F_2 - 255F_4 - 2925F_6\pm \frac{1}{2}\sqrt{4800F_2^2 + 9840F_2F_4 - 109200F_2F_6 + 6660F_4^2 - 238056F_4F_6 + 3080532F_6^2}
\end{align*}
\normalsize

There are 57 states for $L_z=3, S_z=1$, giving 9 triplets. The power sums are:
\small
\begin{align*}
\sum_{i=1}^9 E_i(^3F)&=135F_0 - 256F_2 - 1458F_4 - 19192F_6\\    
\sum_{i=1}^9 E_i(^3F)^2&=2025F_0^2 - 7680F_0F_2 - 43740F_0F_4 - 575760F_0F_6 + 67488F_2^2 + 84708F_2F_4 + 57504F_2F_6 + 441786F_4^2 \\& + 4909824F_4F_6 + 57920424F_6^2\\    
\sum_{i=1}^9 E_i(^3F)^3&=30375F_0^3 - 172800F_0^2F_2 - 984150F_0^2F_4 - 12954600F_0^2F_6 + 3036960F_0F_2^2 + 3811860F_0F_2F_4 + 2587680F_0F_2F_6 \\&+ 19880370F_0F_4^2 + 220942080F_0F_4F_6 + 2606419080F_0F_6^2 - 3126016F_2^3 - 31181868F_2^2F_4 - 405004296F_2^2F_6 \\&- 34977546F_2F_4^2 + 224404344F_2F_4F_6 + 1258528560F_2F_6^2 - 146903328F_4^3 - 2147097744F_4^2F_6 \\&- 19045507536F_4F_6^2 - 184147133896F_6^3\\    
\sum_{i=1}^9 E_i(^3F)^4&=455625F_0^4 - 3456000F_0^3F_2 - 19683000F_0^3F_4 - 259092000F_0^3F_6 + 91108800F_0^2F_2^2 + 114355800F_0^2F_2F_4 \\&+ 4850539200F_0^2F_2F_6 + 596411100F_0^2F_4^2 + 6628262400F_0^2F_4F_6 + 78192572400F_0^2F_6^2 - 187560960F_0F_2^3 \\&- 1870912080F_0F_2^2F_4 - 24300257760F_0F_2^2F_6 - 2098652760F_0F_2F_4^2 + 13464260640F_0F_2F_4F_6 \\&+ 75511713600F_0F_2F_6^2 - 8814199680F_0F_4^3 - 128825864640F_0F_4^2F_6 - 1142730452160F_0F_4F_6^2 \\&- 11048828033760F_0F_6^3 + 911480448F_2^4 + 1943131296F_2^3F_4 - 3523581312F_2^3F_6 + 14916036120F_2^2F_4^2 \\&+ 179786196000F_2^2F_4F_6 + 2548132289856F_2^2F_6^2 + 13513570632F_2F_4^3 - 86430893424F_2F_4^2F_6 \\&- 2325685070304F_2F_4F_6^2 - 13228791512064F_2F_6^3 + 53372194038F_4^4 + 911196676944F_4^3F_6 \\&+ 10234450827312F_4^2F_6^2 + 69434747483520F_4F_6^3 + 629133140720353F_6^4    
\end{align*}
\normalsize

There are 92 states for $L_z=3, S_z=0$, giving  4 singlets. The eigenvalues $E(^1F)$ are roots of the quartic equation  with the coefficients:
\small
\begin{align*}
e_1&=60F_0 - 130F_2 - 576F_4 - 7192F_6\\    
e_2&=1350F_0^2 - 5850F_0F_2 - 25920F_0F_4 - 323640F_0F_6 - 1556F_2^2 + 63642F_2F_4 + 868100F_2F_6 + 94734F_4^2 + 3251928F_4F_6 \\&+ 17500720F_6^2\\    
e_3&=13500F_0^3 - 87750F_0^2F_2 - 388800F_0^2F_4 - 4854600F_0^2F_6 - 46680F_0F_2^2 + 1909260F_0F_2F_4 + 26043000F_0F_2F_6 \\&+ 2842020F_0F_4^2 + 97557840F_0F_4F_6 + 525021600F_0F_6^2 + 499560F_2^3 - 533592F_2^2F_4 - 5519952F_2^2F_6 - 6740046F_2F_4^2 \\&- 283577256F_2F_4F_6 - 1810906200F_2F_6^2 - 4635360F_4^3 - 378396936F_4^2F_6 - 5564641536F_4F_6^2 - 15729680928F_6^3\\    
e_4&=50625F_0^4 - 438750F_0^3F_2 - 1944000F_0^3F_4 - 24273000F_0^3F_6 - 350100F_0^2F_2^2 + 14319450F_0^2F_2F_4 + 1388549700F_0^2F_2F_6 \\& + 21315150F_0^2F_4^2 + 731683800F_0^2F_4F_6 + 3937662000F_0^2F_6^2 + 7493400F_0F_2^3 - 8003880F_0F_2^2F_4 - 82799280F_0F_2^2F_6 \\&- 101100690F_0F_2F_4^2 - 4253658840F_0F_2F_4F_6 - 27163593000F_0F_2F_6^2 - 69530400F_0F_4^3 - 5675954040F_0F_4^2F_6 \\&- 83469623040F_0F_4F_6^2 - 235945213920F_0F_6^3 - 5347200F_2^4 - 48624600F_2^3F_4 - 1007301360F_2^3F_6 + 12893940F_2^2F_4^2 \\&+ 2984516784F_2^2F_4F_6 + 19875268656F_2^2F_6^2 + 70726230F_2F_4^3 + 16168522932F_2F_4^2F_6 + 292144129128F_2F_4F_6^2 \\&+ 1166920488240F_2F_6^3 + 20417265F_4^4 + 10796555880F_4^3F_6 + 338590322136F_4^2F_6^2 + 2728722673248F_4F_6^3 \\&+ 3081498001294F_6^4    
\end{align*}
\normalsize
The power sums are:
\small
\begin{align*}
\sum_{i=1}^4 E_i(^1F)^2&=900F_0^2 - 3900F_0F_2 - 17280F_0F_4 - 215760F_0F_6 + 20012F_2^2 + 22476F_2F_4 + 133720F_2F_6 + 142308F_4^2 \\&+ 1781328F_4F_6 + 16723424F_6^2\\    
\sum_{i=1}^4 E_i(^1F)^3&=13500F_0^3 - 87750F_0^2F_2 - 388800F_0^2F_4 - 4854600F_0^2F_6 + 900540F_0F_2^2 + 1011420F_0F_2F_4 + 6017400F_0F_2F_6 \\& + 6403860F_0F_4^2 + 80159760F_0F_4F_6 + 752554080F_0F_6^2 - 1305160F_2^3 - 8672364F_2^2F_4 - 76207512F_2^2F_6 \\&- 2693142F_2F_4^2 + 59514984F_2F_4F_6 - 50009160F_2F_6^2 - 41308704F_4^3 - 630277416F_4^2F_6 - 5669646912F_4F_6^2 \\&- 41598729952F_6^3\\    
\sum_{i=1}^4 E_i(^1F)^4&= 202500F_0^4 - 1755000F_0^3F_2 - 7776000F_0^3F_4 - 97092000F_0^3F_6 + 27016200F_0^2F_2^2 + 30342600F_0^2F_2F_4 \\&- 4592386800F_0^2F_2F_6 + 192115800F_0^2F_4^2 + 2404792800F_0^2F_4F_6 + 22576622400F_0^2F_6^2 - 78309600F_0F_2^3 \\&- 520341840F_0F_2^2F_4 - 4572450720F_0F_2^2F_6 - 161588520F_0F_2F_4^2 + 3570899040F_0F_2F_4F_6 - 3000549600F_0F_2F_6^2 \\&- 2478522240F_0F_4^3 - 37816644960F_0F_4^2F_6 - 340178814720F_0F_4F_6^2 - 2495923797120F_0F_6^3 + 157255472F_2^4 \\&+ 616667232F_2^3F_4 + 3283302080F_2^3F_6 + 3372566184F_2^2F_4^2 + 40146822432F_2^2F_4F_6 + 309916750368F_2^2F_6^2 \\&- 62667432F_2F_4^3 - 59306084784F_2F_4^2F_6 - 463652892576F_2F_4F_6^2 - 689085922240F_2F_6^3 + 12900705732F_4^4 \\&+ 236712214944F_4^3F_6 + 2503455085728F_4^2F_6^2 + 17345383174272F_4F_6^3 + 107307978178498F_6^4   
\end{align*}
\normalsize

There are 15 states for $L_z=2, S_z=2$.  The 3 quintets are solutions of the cubic equation with the coefficients:
\small
\begin{align*}
    e_1&=45F_0 - 261F_2 - 897F_4 - 8775F_6\\
    e_2&=  675F_0^2 - 7830F_0F_2 - 26910F_0F_4 - 263250F_0F_6 + 16535F_2^2 + 158334F_2F_4 + 1664806F_2F_6 + 242748F_4^2 + 5510076F_4F_6 \\& + 23447567F_6^2\\
    e_3&=4500F_0^3 - 92475F_0^2F_2 - 313200F_0^2F_4 - 2939625F_0^2F_6 + 585525F_0F_2^2 + 4602510F_0F_2F_4 + 44277090F_0F_2F_6 \\ &+ 7316595F_0F_4^2 + 146357640F_0F_4F_6 + 627775005F_0F_6^2 - 1179675F_2^3 - 16425465F_2^2F_4 - 157977495F_2^2F_6 - 58340898F_2F_4^2 \\&-1148017728F_2F_4F_6 - 5297285565F_2F_6^2 - 59939649F_4^3 - 1849951818F_4^2F_6 - 16904104503F_4F_6^2 - 44091990657F_6^3
\end{align*}
\normalsize
The power sums are:
\small
\begin{align*}
\sum_{i=1}^3 E_i(^5D)^2&= 675F_0^2 - 7830F_0F_2 - 26910F_0F_4 - 263250F_0F_6 + 35051F_2^2 + 151566F_2F_4 + 1250938F_2F_6 + 319113F_4^2 \\&+ 4722198F_4F_6 + 30105491F_6^2\\   
\sum_{i=1}^3 E_i(^5D)^3&= 13500F_0^3 - 277425F_0^2F_2 - 939600F_0^2F_4 - 8818875F_0^2F_6 + 2589795F_0F_2^2 + 13502970F_0F_2F_4 \\&+ 114207210F_0F_2F_6 + 25386210F_0F_4^2 + 403618410F_0F_4F_6 + 2182931595F_0F_6^2 - 8371701F_2^3 - 64118799F_2^2F_4 \\&- 528390837F_2^2F_6  - 188883063F_2F_4^2 - 2807788230F_2F_4F_6 - 13997883159F_2F_6^2 - 248318352F_4^3 \\&- 5513231763F_4^2F_6  - 49770841887F_4F_6^2 - 190699255071F_6^3\\   \sum_{i=1}^3 E_i(^5D)^4&=202500F_0^4 - 5548500F_0^3F_2 - 18792000F_0^3F_4 - 176377500F_0^3F_6 + 77693850F_0^2F_2^2 + 405089100F_0^2F_2F_4 \\& + 3426216300F_0^2F_2F_6 + 761586300F_0^2F_4^2 + 12108552300F_0^2F_4F_6 + 65487947850F_0^2F_6^2 - 502302060F_0F_2^3 \\& - 3847127940F_0F_2^2F_4 - 31703450220F_0F_2^2F_6 - 11332983780F_0F_2F_4^2 - 168467293800F_0F_2F_4F_6 \\& - 839872989540F_0F_2F_6^2 - 14899101120F_0F_4^3 - 330793905780F_0F_4^2F_6 - 2986250513220F_0F_4F_6^2 \\& - 11441955304260F_0F_6^3 + 1245090851F_2^4 + 12552178332F_2^3F_4 + 101573484116F_2^3F_6 + 53736623982F_2^2F_4^2 \\& + 804309134148F_2^2F_4F_6 + 3871141037634F_2^2F_6^2 + 113588413668F_2F_4^3 + 2440619540892F_2F_4^2F_6 \\& + 20968355049876F_2F_4F_6^2 + 74143083998516F_2F_6^3 + 114020934498F_4^4 + 3295775482740F_4^3F_6 \\& + 41083604084718F_4^2F_6^2 + 261210382896684F_4F_6^3 + 769346569482804F_6^4   
\end{align*}
\normalsize

There are 65 states for $L_z=2, S_z=1$, giving 5 triplets. The power sums are:
\small
\begin{align*}
    \sum_{i=1}^5 E_i(^3D)&=75F_0 - 177F_2 - 813F_4 - 11853F_6\\
    \sum_{i=1}^5 E_i(^3D)^2&=1125F_0^2 - 5310F_0F_2 - 24390F_0F_4 - 355590F_0F_6 + 26997F_2^2 + 58938F_2F_4 + 480546F_2F_6 + 237891F_4^2 \\&+ 2676966F_4F_6 + 35010309F_6^2\\
    \sum_{i=1}^5 E_i(^3D)^3&=13500F_0^3 - 18225F_0^2F_2 - 214650F_0^2F_4 - 5105025F_0^2F_6 + 202365F_0F_2^2 - 4030290F_0F_2F_4 - 36290430F_0F_2F_6 \\&-321030F_0F_4^2 - 70656030F_0F_4F_6 + 747279405F_0F_6^2 + 1247943F_2^3 + 22338693F_2^2F_4 + 111798801F_2^2F_6 \\&+ 81142029F_2F_4^2 + 1846864854F_2F_4F_6 + 7065627957F_2F_6^2 + 49254426F_4^3 + 2048598171F_4^2F_6 \\&+ 17956760589F_4F_6^2 - 33775757781F_6^3
\\
    \sum_{i=1}^5 E_i(^3D)^4&=202500F_0^4 - 364500F_0^3F_2 - 4293000F_0^3F_4 - 102100500F_0^3F_6 + 6070950F_0^2F_2^2 - 120908700F_0^2F_2F_4 \\& - 1088712900F_0^2F_2F_6 - 9630900F_0^2F_4^2 - 2119680900F_0^2F_4F_6 + 22418382150F_0^2F_6^2 + 74876580F_0F_2^3 \\& + 1340321580F_0F_2^2F_4 + 6707928060F_0F_2^2F_6 + 4868521740F_0F_2F_4^2 + 110811891240F_0F_2F_4F_6 \\& + 423937677420F_0F_2F_6^2 + 2955265560F_0F_4^3 + 122915890260F_0F_4^2F_6 + 1077405635340F_0F_4F_6^2 \\& - 2026545466860F_0F_6^3 - 227853483F_2^4 - 5601530124F_2^3F_4 - 42242306748F_2^3F_6 - 27227124294F_2^2F_4^2 \\& - 517841931996F_2^2F_4F_6 - 1481326181730F_2^2F_6^2 - 60284979540F_2F_4^3 - 1817087985876F_2F_4^2F_6 \\& - 16813292056452F_2F_4F_6^2 - 45054358697148F_2F_6^3 - 37496178414F_4^4 - 1661068805436F_4^3F_6 \\& - 22152925237974F_4^2F_6^2 - 122620134004884F_4F_6^3 + 42681126748343F_6^4
\end{align*}
\normalsize

There are 106 states for $L_z=2, S_z=0$, giving 6 singlets. The power sums are:
\small
\begin{align*}
    \sum_{i=1}^6 E_i(^1D)&=90F_0 + 32F_2 - 390F_4 - 10354F_6\\
    \sum_{i=1}^6 E_i(^1D)^2&=1350F_0^2 + 960F_0F_2 - 11700F_0F_4 - 310620F_0F_6 + 48558F_2^2 + 20712F_2F_4 - 836016F_2F_6 + 189684F_4^2 \\& - 197340F_4F_6 + 30839094F_6^2\\
    \sum_{i=1}^6 E_i(^1D)^3&=20250F_0^3 + 21600F_0^2F_2 - 263250F_0^2F_4 - 6988950F_0^2F_6 + 2185110F_0F_2^2 + 932040F_0F_2F_4 - 37620720F_0F_2F_6 \\& +8535780F_0F_4^2 - 8880300F_0F_4F_6 + 1387759230F_0F_6^2 + 890552F_2^3 - 6759378F_2^2F_4 - 240777246F_2^2F_6 \\& - 18693666F_2F_4^2 + 269152776F_2F_4F_6 + 3602411664F_2F_6^2 - 39506238F_4^3 - 534715992F_4^2F_6 \\& - 198860922F_4F_6^2 - 87680103442F_6^3\\
\sum_{i=1}^6 E_i(^1D)^4&=303750F_0^4 + 432000F_0^3F_2 - 5265000F_0^3F_4 - 139779000F_0^3F_6 + 65553300F_0^2F_2^2 + 27961200F_0^2F_2F_4 \\& - 1128621600F_0^2F_2F_6 + 256073400F_0^2F_4^2 - 266409000F_0^2F_4F_6 + 41632776900F_0^2F_6^2 + 53433120F_0F_2^3 \\&- 405562680F_0F_2^2F_4 - 14446634760F_0F_2^2F_6 - 1121619960F_0F_2F_4^2 + 16149166560F_0F_2F_4F_6 \\&+ 216144699840F_0F_2F_6^2 - 2370374280F_0F_4^3 - 32082959520F_0F_4^2F_6 - 11931655320F_0F_4F_6^2 \\& - 5260806206520F_0F_6^3 + 735641238F_2^4 + 851190864F_2^3F_4 - 28659528192F_2^3F_6 + 6543746280F_2^2F_4^2 \\&- 16717100328F_2^2F_4F_6 + 1414585814580F_2^2F_6^2 + 7047925344F_2F_4^3 - 103083991128F_2F_4^2F_6 \\&- 640041932592F_2F_4F_6^2 - 17808246884832F_2F_6^3 + 15504170148F_4^4 + 5367696984F_4^3F_6 + 3720831961464F_4^2F_6^2 \\&- 6049603124761F_4F_6^3 + 279204858739307F_6^4
\end{align*}
\normalsize

There are 16 states for $L_z=1, S_z=2$.  This gives the one quintet eigenvalue $E(^5P)=15F_0 - 85F_2 - 319F_4 - 3289F_6$.

There are 72 states for $L_z=1, S_z=1$, giving 6 triplets. The power sums are:
\small
\begin{align*}
    \sum_{i=1}^6 E_i(^3P)&=90F_0 - 162F_2 - 956F_4 - 11754F_6\\
    \sum_{i=1}^6 E_i(^3P)^2&=1350F_0^2 - 4860F_0F_2 - 28680F_0F_4 - 352620F_0F_6 + 60062F_2^2 + 72200F_2F_4 - 695396F_2F_6 + 351516F_4^2 \\& + 2203256F_4F_6 + 42174110F_6^2\\
    \sum_{i=1}^6 E_i(^3P)^3&=20250F_0^3 - 109350F_0^2F_2 - 645300F_0^2F_4 - 7933950F_0^2F_6 + 2702790F_0F_2^2 + 3249000F_0F_2F_4 - 31292820F_0F_2F_6 \\& + 15818220F_0F_4^2 + 99146520F_0F_4F_6 + 1897834950F_0F_6^2 - 2295642F_2^3 - 35288220F_2^2F_4 - 281773206F_2^2F_6 \\& - 37509984F_2F_4^2 + 668384184F_2F_4F_6 + 3117732930F_2F_6^2 - 142009496F_4^3 - 1005675624F_4^2F_6 \\& - 14043288684F_4F_6^2 - 135697523970F_6^3\\
    \sum_{i=1}^6 E_i(^3P)^4&=303750F_0^4 - 2187000F_0^3F_2 - 12906000F_0^3F_4 - 158679000F_0^3F_6 + 81083700F_0^2F_2^2 + 97470000F_0^2F_2F_4 \\& - 938784600F_0^2F_2F_6 + 474546600F_0^2F_4^2 + 2974395600F_0^2F_4F_6 + 56935048500F_0^2F_6^2 - 137738520F_0F_2^3 \\& - 2117293200F_0F_2^2F_4 - 16906392360F_0F_2^2F_6 - 2250599040F_0F_2F_4^2 + 40103051040F_0F_2F_4F_6 \\& + 187063975800F_0F_2F_6^2 - 8520569760F_0F_4^3 - 60340537440F_0F_4^2F_6 - 842597321040F_0F_4F_6^2 \\& - 8141851438200F_0F_6^3 + 1096079222F_2^4 + 2508695280F_2^3F_4 - 39591775912F_2^3F_6 + 22024356424F_2^2F_4^2 \\&+ 123785790960F_2^2F_4F_6 + 2450116529796F_2^2F_6^2 + 22148983696F_2F_4^3 - 492114407824F_2F_4^2F_6 \\&- 2607422463984F_2F_4F_6^2 - 25286753036584F_2F_6^3 + 64414373268F_4^4 + 398422669936F_4^3F_6 \\&+ 8653386533032F_4^2F_6^2 + 51157124830993F_4F_6^3 + 497329781921848F_6^4
\end{align*}
\normalsize

There are 114 states for $L_z=1, S_z=0$, giving  1 singlet: $E(^1P)=15F_0 - F_2 - 193F_4 - 2197F_6$.

There are 17 states for $L_z=0, S_z=2$.  This gives 1 quintet eigenvalue $E(^5S)=15F_0 - 80F_2 - 264F_4 - 2288F_6$.

There are 119 states for $L_z=0, S_z=0$. giving 4 singlets. The eigenvalues $E(^1S)$ are given as roots to the cubic equation with the coefficients:
\small
\begin{align*}
    e_1&=60F_0 - 36F_2 - 228F_4 - 3384F_6\\
    e_2&=1350F_0^2 - 1620F_0F_2 - 10260F_0F_4 - 152280F_0F_6 - 29920F_2^2 - 63744F_2F_4 + 689056F_2F_6 - 92943F_4^2 + 1199928F_4F_6 \\& - 9996064F_6^2\\
    e_3&=13500F_0^3 - 24300F_0^2F_2 - 153900F_0^2F_4 - 2284200F_0^2F_6 - 897600f0F_2^2 - 1912320F_0F_2F_4 + 20671680F_0F_2F_6 \\& - 2788290F_0F_4^2 + 35997840F_0F_4F_6 - 299881920F_0F_6^2 + 1651200F_2^3 + 8474400F_2^2F_4 + 31331520F_2^2F_6 + 13189500F_2F_4^2 \\& + 196546752F_2F_4F_6 - 1392550848F_2F_6^2 + 12778722F_4^3 + 278158320F_4^2F_6 + 582753600F_4F_6^2 + 34163720448F_6^3\\
    e_4&=50625F_0^4 - 121500F_0^3F_2 - 769500F_0^3F_4 - 11421000F_0^3F_6 - 6732000F_0^2F_2^2 - 14342400F_0^2F_2F_4 + 155037600F_0^2F_2F_6 \\& - 20912175F_0^2F_4^2 + 269983800F_0^2F_4F_6 - 2249114400F_0^2F_6^2 + 24768000F_0F_2^3 + 127116000F_0F_2^2F_4 + 469972800F_0F_2^2F_6 \\& + 197842500F_0F_2F_4^2 + 2948201280F_0F_2F_4F_6 - 20888262720F_0F_2F_6^2 + 191680830F_0F_4^3 + 4172374800F_0F_4^2F_6 \\& + 8741304000F_0F_4F_6^2 + 512455806720F_0F_6^3 + 27648000F_2^4 - 345600F_2^3F_4 - 2354918400F_2^3F_6 + 136455840F_2^2F_4^2 \\& - 10713409920F_2^2F_4F_6 + 2192935680F_2^2F_6^2 + 588102768F_2F_4^3 - 56316061296F_2F_4^2F_6 - 132442801920F_2F_4F_6^2 \\& + 1034968416768F_2F_6^3 + 1568708856F_4^4 - 50174925768F_4^3F_6 - 255704640048F_4^2F_6^2 - 2327997919104F_4F_6^3 \\& - 20063284091136F_6^4
\end{align*}
\normalsize
The power sums are:
\small
\begin{align*}
    \sum_{i=1}^4 E_i(^1S)^2&=900F_0^2 - 1080F_0F_2 - 6840F_0F_4 - 101520F_0F_6 + 61136F_2^2 + 143904F_2F_4 - 1134464F_2F_6 + 237870F_4^2 \\&- 856752F_4F_6 + 31443584F_6^2\\
    \sum_{i=1}^4 E_i(^1S)^3&=13500F_0^3 - 24300F_0^2F_2 - 153900F_0^2F_4 - 2284200F_0^2F_6 + 2751120F_0F_2^2 + 6475680F_0F_2F_4 - 51050880F_0F_2F_6 \\&+ 10704150F_0F_4^2 - 38553840F_0F_4F_6 + 1414961280F_0F_6^2 + 1675584F_2^3 - 2812896F_2^2F_4 - 148492224F_2^2F_6 \\&- 19684512F_2F_4^2 + 376762464F_2F_4F_6 + 501311808F_2F_6^2 - 37089198F_4^3 + 183926808F_4^2F_6 - 740173824F_4F_6^2 \\&- 37740607488F_6^3\\
    \sum_{i=1}^4 E_i(^1S)^4&=202500F_0^4 - 486000F_0^3F_2 - 3078000F_0^3F_4 - 45684000F_0^3F_6 + 82533600F_0^2F_2^2 + 194270400F_0^2F_2F_4 \\& - 1531526400F_0^2F_2F_6 + 321124500F_0^2F_4^2 - 1156615200F_0^2F_4F_6 + 42448838400F_0^2F_6^2 + 100535040F_0F_2^3 \\& - 168773760F_0F_2^2F_4 - 8909533440F_0F_2^2F_6 - 1181070720F_0F_2F_4^2 + 22605747840F_0F_2F_4F_6 \\& + 30078708480F_0F_2F_6^2 - 2225351880F_0F_4^3 + 11035608480F_0F_4^2F_6 - 44410429440F_0F_4F_6^2 - 2264436449280F_0F_6^3 \\& + 1598832896F_2^4 + 7241722368F_2^3F_4 - 73689668608F_2^3F_6 + 20369424384F_2^2F_4^2 - 240697387008F_2^2F_4F_6 \\& + 2753405313024F_2^2F_6^2 + 28541283552F_2F_4^3 - 396739676736F_2F_4^2F_6 + 4193001086976F_2F_4F_6^2 \\& - 34001717788672F_2F_6^3 + 21376304514F_4^4 - 187445029824F_4^3F_6 + 5823279783360F_4^2F_6^2 - 35633952726527F_4F_6^3 \\& + 406669400161276F_6^4
\end{align*}
\normalsize
This complete the $f^6/f^8$ configurations. The Hund's rule ground state is $^7F$.

\begin{table}[]
\centering
\resizebox{\columnwidth}{!}{
\begin{tabular}{|c|c|c|c|}
\hline
configuration & term symbol & eigenvalue & degeneracy\\
\hline
 &$^2S$ & $21F_0 - 42F_2 - 336F_4 - 3066F_6\pm\frac{1}{2}\sqrt{16896F_2^2 - 49632F_2F_4 - 73920F_2F_6 + 60336F_4^2 - 178080F_4F_6 - 845917680F_6^2}$  & $4$\\
 &$^4S$ & $21F_0 - 90F_2 - 213F_4 - 3276F_6\pm\frac{1}{2}\sqrt{19200F_2^2 + 39360F_2F_4 - 436800F_2F_6 + 26640F_4^2 -952224F_4F_6 + 12322128F_6^2} $ & $8$\\
 &$^8S$ & $21F_0-210F_2-693F_4-6006F_6$ & $8$\\
 &$^2P$ & 5 doublets & $30$\\
 &$^4P$ & $21F_0 - 27F_2 - 269F_4 - 4095F_6\pm\frac{1}{2}\sqrt{36096F_2^2 - 33344F_2F_4 - 594048F_2F_6 + 72304F_4^2 - 500864F_4F_6 + 4769856F_6^2}  $ & $24$\\
 &$^6P$ & $21F_0 - 195F_2 - 528F_4 - 3003F_6$ & $18$\\
 &$^2D$ & 7 doublets & $70$\\
 &$^4D$ & 6 quartets & $120$\\
 &$^6D$ & $21F_0 - 169F_2 - 396F_4 - 5005F_6$ & $30$\\
 &$^2F$ & 10 doublets & $140$\\
 &$^4F$ & 5 quartets & $140$\\
 &$^6F$ & $21F_0 - 140F_2 - 462F_4 - 4004F_6$ & $42$\\
 &$^2G$ & 10 doublets & $180$\\
 &$^4G$ & 7 quartets & $252$\\
 &$^6G$ & $21F_0 - 120F_2 - 592F_4 - 4368F_6$ & $54$\\
$f^7$ &$^2H$ & 9 doublets & $198$\\
 &$^4H$ & 5 quartets& $220$\\
 &$^6H$ & $21F_0 - 125F_2 - 444F_4 - 4277F_6$ & $66$\\
  &$^2I$ & 9 doublets & $234$\\
 &$^4I$ & 5 quartets & $260$\\
 &$^6I$ & $21F_0 - 175F_2 - 504F_4 - 4291F_6$ & $78$\\
 &$^2J$ & 7 doublets & $210$\\
 &$^4J$ & $21F_0-122F_2-1177F_4/3- 2772F_6+\lambda_J\cos(\theta_J/3+2\pi k/3)$ & $180$\\
 &$^2K$ & 5 doublets &$170$\\
 &$^4K$ &$21F_0-118F_2-395F_4- 2954F_6+\lambda_K\cos(\theta_K/3+2\pi k/3)$ &$204$\\
 &$^2L$ &$21F_0-118F_2-1440F_4/4 -1624F_6- S_L\pm\frac{1}{2}\sqrt{-4S_L^2-2p_L+\frac{q_L}{S_L}}$& $152$\\
& &$21F_0-118F_2-1440F_4/4 -1624F_6 + S_L\pm\frac{1}{2}\sqrt{-4S_L^2-2p_L-\frac{q_L}{S_L}}$& \\
 &$^4L$ &$21F_0 - 135F_2 - 351F_4 - 2415F_6$&$76$\\
 &$^2M$ & $21F_0 - 101F_2 - 285F_4 - 2261F_6 \pm \frac{1}{2}\sqrt{6400F_2^2 + 4160F_2F_4 - 17920F_2F_6 + 676F_4^2 - 5824F_4F_6 + 12544F_6^2}$ &$84$\\
 &$^4M$ &$21F_0 - 165F_2 - 415F_4 - 2625F_6$&$84$\\
 &$^2N$ &$21F_0 - 160F_2 - 360F_4 - 700F_6$&$46$\\
 &$^2O$ &$21F_0 - 164F_2 - 390F_4 - 1400F_6$ & $50$\\
\hline
\end{tabular}}
\vskip 0.4cm
\caption{Table of multiplet energies for $f^7$ configurations. The term symbols are to be read as $^{2S+1}L$, and the multiplicity $(2L+1)\times(2S+1)$. Note for $f$-electrons, $F^2=15^2F_2, F^4=33^2F_4$, and $F^6=\frac{429^2}{5^2}F_6$.} 
\label{Table:ATMfc}
\end{table}

\subsubsection{\texorpdfstring{$f^7$}{TEXT}Configurations}
For the $f^7$ configuration there are ${14 \choose{7}}= 3432$ states. The allowed term symbols are $^2O,~ ^2N,~ ^{4,2}M,~ ^{4,2}L,~ ^{4,2}K,~ ^{4,2}J,~ ^{6,4,2}I,~ ^{6,4,2}H,~ ^{6,4,2}G,~ ^{6,4,2}F,~ ^{6,4,2}D,~ ^{6,4,2}P$, and $^{8,4,2}S$. 

The maximum $L_z=12$ occurs for $S_z=\pm 1/2$. The doublet eigenvalue is $E(^2O)=
21F_0 - 164F_2 - 390F_4 - 1400F_6$.

There is an additional state for $L_z=11, S_z=\pm 1/2$, giving doublet eigenvalue  $E(^2N)=
21F_0 - 160F_2 - 360F_4 - 700F_6$.

For $L_z=10, S_z=\pm 3/2$ there is one state, having the quartet eigenvalue $E(^4M)=21F_0 - 165F_2 - 415F_4 - 2625F_6$.

For $L_z=10, S_z=\pm 1/2$ there are 5 states. This gives two doublets: 
\small
\begin{align*}
    E(^2M)=21F_0 - 101F_2 - 285F_4 - 2261F_6 \pm \frac{1}{2}\sqrt{6400F_2^2 + 4160F_2F_4 - 17920F_2F_6 + 676F_4^2 - 5824F_4F_6 + 12544F_6^2}
\end{align*}
\normalsize

For $L_z=9, S_z=\pm 3/2$ there are 2 states, giving one quartet eigenvalue $E(^4L)=21F_0 - 135F_2 - 351F_4 - 2415F_6$.

For $L_z=9, S_z=\pm 1/2$ there are 10 states, giving 4 doublets. The eigenvalues $E(^2L)$ are solutions to the quartic equation with the coefficients:
\small
\begin{align*}
e_1&= 84F_0 - 472F_2 - 1440F_4 - 6496F_6\\    
e_2&= 2646F_0^2 - 29736F_0F_2 - 90720F_0F_4 - 409248F_0F_6 + 81366F_2^2 + 511740F_2F_4 + 2372580F_2F_6 + 762156F_4^2 + 7107156F_4F_6 \\& + 14734398F_6^2\\    
e_3&= 37044F_0^3 - 624456F_0^2F_2 - 1905120F_0^2F_4 - 8594208F_0^2F_6 + 3417372F_0F_2^2 + 21493080F_0F_2F_4 + 99648360F_0F_2F_6 \\ & + 32010552F_0F_4^2 + 298500552F_0F_4F_6 + 618844716F_0F_6^2 - 6061320F_2^3 - 59253840F_2^2F_4 - 280106064F_2^2F_6 - 181217520F_2F_4^2 \\& - 1727898480F_2F_4F_6 - 3770236008F_2F_6^2 - 174858480F_4^3 - 2546165664F_4^2F_6 - 10995534144F_4F_6^2 - 13007383200F_6^3\\    
e_4&= 194481F_0^4 - 4371192F_0^3F_2 - 13335840F_0^3F_4 - 60159456F_0^3F_6 + 35882406F_0^2F_2^2 + 225677340F_0^2F_2F_4 + 1046307780F_0^2F_2F_6 \\& + 336110796F_0^2F_4^2 + 3134255796F_0^2F_4F_6 + 6497869518F_0^2F_6^2 - 127287720F_0F_2^3 - 1244330640F_0F_2^2F_4 - 5882227344F_0F_2^2F_6 \\& - 3805567920F_0F_2F_4^2 - 36285868080F_0F_2F_4F_6 - 79174956168F_0F_2F_6^2 - 3672028080F_0F_4^3 - 53469478944F_0F_4^2F_6 \\& - 230906217024F_0F_4F_6^2 - 273155047200F_0F_6^3 + 164010825F_2^4 + 2234625300F_2^3F_4 + 10695990060F_2^3F_6 + 10562300100F_2^2F_4^2 \\& + 102235199220F_2^2F_4F_6 + 231494425470F_2^2F_6^2 + 20860502400F_2F_4^3 + 309130872120F_2F_4^2F_6 + 1390993018380F_2F_4F_6^2 \\& + 1828688728860F_2F_6^3 + 14566878000F_4^4 + 297128109600F_4^3F_6 + 2021764438260F_4^2F_6^2 + 5087559799980.01F_4F_6^3 \\& + 3045071671424.99F_6^4   
\end{align*}
\normalsize
The power sums are:
\small
\begin{align*}
\sum_{i=1}^4 E_i(^2L)^2&=1764F_0^2 - 19824F_0F_2 - 60480F_0F_4 - 272832F_0F_6 + 60052F_2^2 + 335880F_2F_4 + 1387064F_2F_6 + 549288F_4^2 \\& + 4494168F_4F_6 + 12729220F_6^2\\    
\sum_{i=1}^4 E_i(^2L)^3&=37044F_0^3 - 624456F_0^2F_2 - 1905120F_0^2F_4 - 8594208F_0^2F_6 + 3783276F_0F_2^2 + 21160440F_0F_2F_4 + 87385032F_0F_2F_6 \\& + 34605144F_0F_4^2 + 283132584F_0F_4F_6 + 801940860F_0F_6^2 - 8123752F_2^3 - 64063440F_2^2F_4 - 236698896F_2^2F_6 \\& - 189940464F_2F_4^2 - 1388835504F_2F_4F_6 - 3962352072F_2F_6^2 - 218045520F_4^3 - 2493003744F_4^2F_6 \\& - 13125176064F_4F_6^2 - 25996513312F_6^3\\    
\sum_{i=1}^4 E_i(^2L)^4&=777924F_0^4 - 17484768F_0^3F_2 - 53343360F_0^3F_4 - 240637824F_0^3F_6 + 158897592F_0^2F_2^2 + 888738480F_0^2F_2F_4 \\& + 3670171344F_0^2F_2F_6 + 1453416048F_0^2F_4^2 + 11891568528F_0^2F_4F_6 + 33681516120F_0^2F_6^2 - 682395168F_0F_2^3 \\& - 5381328960F_0F_2^2F_4 - 19882707264F_0F_2^2F_6 - 15954998976F_0F_2F_4^2 - 116662182336F_0F_2F_4F_6 \\& - 332837574048F_0F_2F_6^2 - 18315823680F_0F_4^3 - 209412314496F_0F_4^2F_6 - 1102514789376F_0F_4F_6^2 \\& - 2183707118208F_0F_6^3 + 1153119652F_2^4 + 11633536080F_2^3F_4 + 37956185008F_2^3F_6 + 48168660528F_2^2F_4^2 \\& + 308236129488F_2^2F_4F_6 + 869492803704F_2^2F_6^2 + 99390576960F_2F_4^3 + 993717710496F_2F_4^2F_6 \\ & + 5218339953456F_2F_4F_6^2 + 10687330590832F_2F_6^3 + 88871103072F_4^4 + 1291063281984F_4^3F_6 \\& + 9645376947696F_4^2F_6^2 + 35816092999344F_4F_6^3 + 53631631346692F_6^4    
\end{align*}
\normalsize

For $L_z=8, S_z=\pm 3/2$ there are 5 states, giving 3 quartets. The eigenvalues $E(^4K)$ are solutions to the cubic equation with the coefficients:
\small
\begin{align*}
    e_1&=63F_0 - 354F_2 - 1185F_4 - 8862F_6\\
    e_2&=1323F_0^2 - 14868F_0F_2 - 49770F_0F_4 - 372204F_0F_6 + 40380F_2^2 + 282300F_2F_4 + 2080344F_2F_6 + 461070F_4^2 + 7023660F_4F_6 \\& + 26126016F_6^2\\
    e_3&=9261F_0^3 - 156114F_0^2F_2 - 522585F_0^2F_4 - 3908142F_0^2F_6 + 847980F_0F_2^2 + 5928300F_0F_2F_4 + 43687224F_0F_2F_6 + 9682470F_0F_4^2 \\& + 147496860F_0F_4F_6 + 548646336F_0F_6^2 - 1495800F_2^3 - 16154100F_2^2F_4 - 118190520F_2^2F_6 - 55449900F_2F_4^2 - 831128760F_2F_4F_6 \\& - 3051861120F_2F_6^2 - 59065200F_4^3 - 1370491920F_4^2F_6 - 10385373600F_4F_6^2 - 25626545280F_6^3
\end{align*}
\normalsize
The power sums are:
\small
\begin{align*}
\sum_{i=1}^3 E_i(^4K)^2&=1323F_0^2 - 14868F_0F_2 - 49770F_0F_4 - 372204F_0F_6 + 44556F_2^2 + 274380F_2F_4 + 2113608F_2F_6 + 482085F_4^2 \\&+ 6955620F_4F_6 + 26283012F_6^2 \\   
\sum_{i=1}^3 E_i(^4K)^3&=27783F_0^3 - 468342F_0^2F_2 - 1567755F_0^2F_4 - 11724426F_0^2F_6 + 2807028F_0F_2^2 + 17285940F_0F_2F_4 + 133157304F_0F_2F_6 \\&+ 30371355F_0F_4^2 + 438204060F_0F_4F_6 + 1655829756F_0F_6^2 - 5965704F_2^3 - 50607180F_2^2F_4 - 403354728F_2^2F_6 \\& - 164403810F_2F_4^2 - 2438530920F_2F_4F_6 - 9505945512F_2F_6^2 - 202098375F_4^3 - 4217083290F_4^2F_6 - 30739190580F_4F_6^2\\& - 78270934392F_6^3\\   \sum_{i=1}^3 E_i(^4K)^4&=583443F_0^4 - 13113576F_0^3F_2 - 43897140F_0^3F_4 - 328283928F_0^3F_6 + 117895176F_0^2F_2^2 + 726009480F_0^2F_2F_4 \\&+ 5592606768F_0^2F_2F_6 + 1275596910F_0^2F_4^2 + 18404570520F_0^2F_4F_6 + 69544849752F_0^2F_6^2 - 501119136F_0F_2^3 \\& - 4251003120F_0F_2^2F_4 - 33881797152F_0F_2^2F_6 - 13809920040F_0F_2F_4^2 - 204836597280F_0F_2F_4F_6 \\& - 798499423008F_0F_2F_6^2 - 16976263500F_0F_4^3 - 354234996360F_0F_4^2F_6 - 2582092008720F_0F_4F_6^2 \\& - 6574758488928F_0F_6^3 + 842201136F_2^4 + 8817752160F_2^3F_4 + 72711567936F_2^3F_6 + 39472828920F_2^2F_4^2 \\& + 605838653280F_2^2F_4F_6 + 2444987020896F_2^2F_6^2 + 90377569800F_2F_4^3 + 1932744688560F_2F_4^2F_6 \\& + 14511323806560F_2F_4F_6^2 + 38169127434816F_2F_6^3 + 87203905425F_4^4 + 2342679381000F_4^3F_6 + 24682521328920F_4^2F_6^2 \\& + 121238821700640F_4F_6^3 + 234069072813072F_6^4
\end{align*}
\normalsize

For $L_z=8, S_z=\pm 1/2$ there are 18 states, giving 5 doublets. The power sums are:
\small
\begin{align*}
\sum_{i=1}^5 E_i(^2K)&=105F_0 - 478F_2 - 1254F_4 - 10318F_6\\
\sum_{i=1}^5 E_i(^2K)^2&=2205F_0^2 - 20076F_0F_2 - 52668F_0F_4 - 433356F_0F_6 + 65332F_2^2 + 258504F_2F_4 + 1491224F_2F_6 + 385506F_4^2 \\& + 4316928F_4F_6 + 30694972F_6^2\\
\sum_{i=1}^5 E_i(^2K)^3&=46305F_0^3 - 632394F_0^2F_2 - 1659042F_0^2F_4 - 13650714F_0^2F_6 + 4115916F_0F_2^2 + 16285752F_0F_2F_4 \\& + 93947112F_0F_2F_6 + 24286878F_0F_4^2 + 271966464F_0F_4F_6 + 1933783236F_0F_6^2 - 9049528F_2^3 - 55291464F_2^2F_4 \\& - 281482824F_2^2F_6 - 127114920F_2F_4^2 - 841539888F_2F_4F_6 - 6393793224F_2F_6^2 - 139172850F_4^3 - 1657334196F_4^2F_6 \\& - 19429853184F_4F_6^2 - 90561019528F_6^3\\
\sum_{i=1}^5 E_i(^2K)^4&=972405F_0^4 - 17707032F_0^3F_2 - 46453176F_0^3F_4 - 382219992F_0^3F_6 + 172868472F_0^2F_2^2 + 684001584F_0^2F_2F_4 \\& + 3945778704F_0^2F_2F_6 + 1020048876F_0^2F_4^2 + 11422591488F_0^2F_4F_6 + 81218895912F_0^2F_6^2 - 760160352F_0F_2^3 \\& - 4644482976F_0F_2^2F_4 - 23644557216F_0F_2^2F_6 - 10677653280F_0F_2F_4^2 - 70689350592F_0F_2F_4F_6 \\& - 537078630816F_0F_2F_6^2 - 11690519400F_0F_4^3 - 139216072464F_0F_4^2F_6 - 1632107667456F_0F_4F_6^2 \\& - 7607125640352F_0F_6^3 + 1305573712F_2^4 + 10567522368F_2^3F_4 + 46681851328F_2^3F_6 + 36716896800F_2^2F_4^2 \\& + 215017334784F_2^2F_4F_6 + 1609405076832F_2^2F_6^2 + 63811684224F_2F_4^3 + 388659700800F_2F_4^2F_6 \\& + 4756795654464F_2F_4F_6^2 + 23415824974528F_2F_6^3 + 55790559378F_4^4 + 691171890192F_4^3F_6 \\& + 9921145198608F_4^2F_6^2 + 75322567067328F_4F_6^3 + 275443158211312F_6^4
\end{align*}
\normalsize

For $L_z=7, S_z=\pm 3/2$ there are 8 states, giving 3 quartets. The eigenvalues $E(^4J)$ are solutions to the cubic equation with the coefficients:
\small
\begin{align*}
    e_1&=63F_0 - 378F_2 - 1169F_4 - 7770F_6\\
    e_2&=1323F_0^2 - 15876F_0F_2 - 49098F_0F_4 - 326340F_0F_6 + 46236F_2^2 + 294092F_2F_4 + 2002728F_2F_6 + 444334F_4^2 + 6146756F_4F_6 \\& + 19404000F_6^2\\
    e_3&=9261F_0^3 - 166698F_0^2F_2 - 515529F_0^2F_4 - 3426570F_0^2F_6 + 970956F_0F_2^2 + 6175932F_0F_2F_4 + 42057288F_0F_2F_6 \\& + 9331014F_0F_4^2 + 129081876F_0F_4F_6 + 407484000F_0F_6^2 - 1839960F_2^3 - 17909844F_2^2F_4 - 124429032F_2^2F_6 - 55802492F_2F_4^2 \\& - 790012104F_2F_4F_6 - 2581903296F_2F_6^2 - 54366000F_4^3 - 1189906592F_4^2F_6 - 7829168256F_4F_6^2 - 15277538304F_6^3
\end{align*}
\normalsize
The power sums are:
\small
\begin{align*}
\sum_{i=1}^3 E_i(^4J)^2&= 1323F_0^2 - 15876F_0F_2 - 49098F_0F_4 - 326340F_0F_6 + 50412F_2^2 + 295580F_2F_4 + 1868664F_2F_6 + 477893F_4^2 \\& + 5872748F_4F_6 + 21564900F_6^2\\
\sum_{i=1}^3 E_i(^4J)^3&=27783F_0^3 - 500094F_0^2F_2 - 1546587F_0^2F_4 - 10279710F_0^2F_6 + 3175956F_0F_2^2 + 18621540F_0F_2F_4 \\& + 117725832F_0F_2F_6 + 30107259F_0F_4^2 + 369983124F_0F_4F_6 + 1358588700F_0F_6^2 - 7098408F_2^3 - 59173740F_2^2F_4 \\& - 355058424F_2^2F_6 - 181832250F_2F_4^2 - 2121302232F_2F_4F_6 - 7520852808F_2F_6^2 - 202328471F_4^3 \\& - 3510157854F_4^2F_6 - 23884554708F_4F_6^2 - 62622807912F_6^3\\
\sum_{i=1}^3 E_i(^4J)^4&=583443F_0^4 - 14002632F_0^3F_2 - 43304436F_0^3F_4 - 287831880F_0^3F_6 + 133390152F_0^2F_2^2 + 782104680F_0^2F_2F_4 \\& + 4944484944F_0^2F_2F_6 + 1264504878F_0^2F_4^2 + 15539291208F_0^2F_4F_6 + 57060725400F_0^2F_6^2 - 596266272F_0F_2^3 \\& - 4970594160F_0F_2^2F_4 - 29824907616F_0F_2^2F_6 - 15273909000F_0F_2F_4^2 - 178189387488F_0F_2F_4F_6 \\& - 631751635872F_0F_2F_6^2 - 16995591564F_0F_4^3 - 294853259736F_0F_4^2F_6 - 2006302595472F_0F_4F_6^2 \\& - 5260315864608F_0F_6^3 + 1047853872F_2^4 + 11094344160F_2^3F_4 + 63336305088F_2^3F_6 + 48513302456F_2^2F_4^2 \\& + 537008987424F_2^2F_4F_6 + 1826764460640F_2^2F_6^2 + 102944771560F_2F_4^3 + 1694985779536F_2F_4^2F_6 \\& + 11093584155360F_2F_4F_6^2 + 28496560494528F_2F_6^3 + 87731728337F_4^4 + 1941938102104F_4^3F_6 \\& + 18639439836248F_4^2F_6^2 + 90972151500384F_4F_6^3 + 186840370498320F_6^4
\end{align*}
\normalsize

For $L_z=7, S_z=\pm 1/2$ there are 28 states, giving 7 doublets. The power sums are:
\small
\begin{align*}
\sum_{i=1}^7 E_i(^2J)&=147F_0 - 562F_2 - 1786F_4 - 15442F_6\\
\sum_{i=1}^7 E_i(^2J)^2&=3087F_0^2 - 23604F_0F_2 - 75012F_0F_4 - 648564F_0F_6 + 67868F_2^2 + 292008F_2F_4 + 2084488F_2F_6 + 559698F_4^2 \\& + 7121184F_4F_6 + 41243300F_6^2 \\   
\sum_{i=1}^7 E_i(^2J)^3&=64827F_0^3 - 743526F_0^2F_2 - 2362878F_0^2F_4 - 20429766F_0^2F_6 + 4275684F_0F_2^2 + 18396504F_0F_2F_4 \\& + 131322744F_0F_2F_6 + 35260974F_0F_4^2 + 448634592F_0F_4F_6 + 2598327900F_0F_6^2 - 8919112F_2^3 - 53767128F_2^2F_4 \\& - 360684408F_2^2F_6 - 138954264F_2F_4^2 - 1384134864F_2F_4F_6 - 7518992376F_2F_6^2 - 197705122F_4^3 \\& - 2997459780F_4^2F_6 - 27075356112F_4F_6^2 - 119842786840F_6^3
\\   
\sum_{i=1}^7 E_i(^2J)^4&=1361367F_0^4 - 20818728F_0^3F_2 - 66160584F_0^3F_4 - 572033448F_0^3F_6 + 179578728F_0^2F_2^2 + 772653168F_0^2F_2F_4 \\& + 5515555248F_0^2F_2F_6 + 1480960908F_0^2F_4^2 + 18842652864F_0^2F_4F_6 + 109129771800F_0^2F_6^2 - 749205408F_0F_2^3 \\& - 4516438752F_0F_2^2F_4 - 30297490272F_0F_2^2F_6 - 11672158176F_0F_2F_4^2 - 116267328576F_0F_2F_4F_6 \\& - 631595359584F_0F_2F_6^2 - 16607230248F_0F_4^3 - 251786621520F_0F_4^2F_6 - 2274329913408F_0F_4F_6^2 \\& - 10066794094560F_0F_6^3 + 1217309808F_2^4 + 9436152896F_2^3F_4 + 62074802496F_2^3F_6 + 33512583200F_2^2F_4^2 \\& + 324552969216F_2^2F_4F_6 + 1645643121696F_2^2F_6^2 + 66390594624F_2F_4^3 + 756470018752F_2F_4^2F_6 \\& + 6061881580608F_2F_4F_6^2 + 27500707795008F_2F_6^3 + 74590801170F_4^4 + 1277437542864F_4^3F_6 \\& + 14139256927376F_4^2F_6^2 + 101980373491136F_4F_6^3 + 360852553075728F_6^4
\end{align*}
\normalsize

For $L_z=6, S_z=\pm 5/2$ there is one state, giving a sextet eigenvalue $E(^6I)=21F_0 - 175F_2 - 504F_4 - 4291F_6$

For $L_z=6, S_z=\pm 3/2$ there are 14 states, giving 5 quartets. The power sums are:
\small
\begin{align*}
\sum_{i=1}^5 E_i(^4I)&=105F_0 - 439F_2 - 1707F_4 - 14959F_6\\    
\sum_{i=1}^5 E_i(^4I)^2&=2205F_0^2 - 18438F_0F_2 - 71694F_0F_4 - 628278F_0F_6 + 52181F_2^2 + 310530F_2F_4 + 2417338F_2F_6 + 649725F_4^2 \\& + 10068114F_4F_6 + 48284453F_6^2\\    
\sum_{i=1}^5 E_i(^4I)^3&=46305F_0^3 - 580797F_0^2F_2 - 2258361F_0^2F_4 - 19790757F_0^2F_6 + 3287403F_0F_2^2 + 19563390F_0F_2F_4 \\& + 152292294F_0F_2F_6 + 40932675F_0F_4^2 + 634291182F_0F_4F_6 + 3041920539F_0F_6^2 - 6742279F_2^3 - 55327185F_2^2F_4 \\& - 425256573F_2^2F_6 - 178598205F_2F_4^2 - 2520158130F_2F_4F_6 - 10758693477F_2F_6^2 - 269979939F_4^3 \\& - 5665693509F_4^2F_6 - 47982906657F_4F_6^2 - 164633340655F_6^3
\\    
\sum_{i=1}^5 E_i(^4I)^4&=972405F_0^4 - 16262316F_0^3F_2 - 63234108F_0^3F_4 - 554141196F_0^3F_6 + 138070926F_0^2F_2^2 + 821662380F_0^2F_2F_4 \\& + 6396276348F_0^2F_2F_6 + 1719172350F_0^2F_4^2 + 26640229644F_0^2F_4F_6 + 127760662638F_0^2F_6^2 - 566351436F_0F_2^3 \\& - 4647483540F_0F_2^2F_4 - 35721552132F_0F_2^2F_6 - 15002249220F_0F_2F_4^2 - 211693282920F_0F_2F_4F_6 \\& - 903730252068F_0F_2F_6^2 - 22678314876F_0F_4^3 - 475918254756F_0F_4^2F_6 - 4030564159188F_0F_4F_6^2 \\& - 13829200615020F_0F_6^3 + 897562661F_2^4 + 9449537220F_2^3F_4 + 73852695476F_2^3F_6 + 41456310990F_2^2F_4^2 \\& + 591094930860F_2^2F_4F_6 + 2456013645246F_2^2F_6^2 + 97741090980F_2F_4^3 + 1907181349500F_2F_4^2F_6 \\& + 14631840630540F_2F_4F_6^2 + 45278203820276F_2F_6^3 + 119678718069F_4^4 + 3103016696388F_4^3F_6 \\& + 35327187963438F_4^2F_6^2 + 214876759892004F_4F_6^3 + 582641538912965F_6^4   
\end{align*}
\normalsize

For $L_z=6, S_z=\pm 1/2$ there are 43 states, giving 9 doublets. The power sums are:
\small
\begin{align*}
\sum_{i=1}^5 E_i(^2I)&=189F_0 - 695F_2 - 2262F_4 - 22211F_6\\    
\sum_{i=1}^5 E_i(^2I)^2&=3969F_0^2 - 29190F_0F_2 - 95004F_0F_4 - 932862F_0F_6 + 101369F_2^2 + 336492F_2F_4 + 2542330F_2F_6 + 722292F_4^2 \\& + 9906876F_4F_6 + 68121809F_6^2\\    
\sum_{i=1}^5 E_i(^2I)^3&=83349F_0^3 - 919485F_0^2F_2 - 2992626F_0^2F_4 - 29385153F_0^2F_6 + 6386247F_0F_2^2 + 21198996F_0F_2F_4 \\& + 160166790F_0F_2F_6 + 45504396F_0F_4^2 + 624133188F_0F_4F_6 + 4291673967F_0F_6^2 - 12633095F_2^3 - 72024066F_2^2F_4 \\& - 612713913F_2^2F_6 - 157672908F_2F_4^2 - 1415831508F_2F_4F_6 - 9597909741F_2F_6^2 - 258948360F_4^3 - 4455144540F_4^2F_6 \\& - 42010465266F_4F_6^2 - 224260928507F_6^3\\    
\sum_{i=1}^5 E_i(^2I)^4&=1750329F_0^4 - 25745580F_0^3F_2 - 83793528F_0^3F_4 - 822784284F_0^3F_6 + 268222374F_0^2F_2^2 + 890357832F_0^2F_2F_4 \\& + 6727005180F_0^2F_2F_6 + 1911184632F_0^2F_4^2 + 26213593896F_0^2F_4F_6 + 180250306614F_0^2F_6^2 - 1061179980F_0F_2^3 \\& - 6050021544F_0F_2^2F_4 - 51467968692F_0F_2^2F_6 - 13244524272F_0F_2F_4^2 - 118929846672F_0F_2F_4F_6 \\& - 806224418244F_0F_2F_6^2 - 21751662240F_0F_4^3 - 374232141360F_0F_4^2F_6 - 3528879082344F_0F_4F_6^2 \\& - 18837917994588F_0F_6^3 + 1832962409F_2^4 + 11626681944F_2^3F_4 + 91883928500F_2^3F_6 + 42377658936F_2^2F_4^2 \\& + 464480740584F_2^2F_4F_6 + 3450037752966F_2^2F_6^2 + 74763940320F_2F_4^3 + 792681057840F_2F_4^2F_6 \\& + 5631841066824F_2F_4F_6^2 + 34646174921876F_2F_6^3 + 97885223568F_4^4 + 2074057479648F_4^3F_6 \\& + 23564909300856F_4^2F_6^2 + 173095821716663F_4F_6^3 + 772801455035225F_6^4
\end{align*}
\normalsize

For $L_z=5, S_z=\pm 5/2$ there are two states, giving another sextet $E(^6H)=21F_0 - 125F_2 - 444F_4 - 4277F_6$.

For $L_z=5, S_z=\pm 3/2$ there are 20 states, giving 5 quartets. The power sums are:
\small
\begin{align*}
\sum_{i=1}^5 E_i(^4H)&=105F_0 - 597F_2 - 1695F_4 - 15141F_6\\    
\sum_{i=1}^5 E_i(^4H)^2&=2205F_0^2 - 25074F_0F_2 - 71190F_0F_4 - 635922F_0F_6 + 80477F_2^2 + 396030F_2F_4 + 3460618F_2F_6 + 631599F_4^2 \\&+ 9561342F_4F_6 + 50217797F_6^2    
\end{align*}
\normalsize

For $L_z=5, S_z=\pm 1/2$ there are 58 states, giving 9 doublets.
The power sums are:
\small
\begin{align*}
\sum_{i=1}^9 E_i(^2H)&=189F_0 - 389F_2 - 2046F_4 - 23177F_6\\    
\sum_{i=1}^9 E_i(^2H)^2&= 3969F_0^2 - 16338F_0F_2 - 85932F_0F_4 - 973434F_0F_6 + 68609F_2^2 + 175044F_2F_4 + 1023610F_2F_6 + 682098F_4^2 + \\& 9138108F_4F_6 + 73063361F_6^2   
\end{align*}
\normalsize

For $L_z=4, S_z=\pm 5/2$ there are three states, giving another sextet $E(^6G)=21F_0 - 120F_2 - 592F_4 - 4368F_6$.

For $L_z=4, S_z=\pm 3/2$ there are 28 states, giving 7 quartets.
The power sums are:
\small
\begin{align*}
\sum_{i=1}^7 E_i(^4G)&=147F_0 - 592F_2 - 2081F_4 - 23968F_6\\    
\sum_{i=1}^7 E_i(^4G)^2&=3087F_0^2 - 24864F_0F_2 - 87402F_0F_4 - 1006656F_0F_6 + 79552F_2^2 + 337040F_2F_4 + 3682448F_2F_6 + 699049F_4^2 \\& + 13758416F_4F_6 + 89486152F_6^2    
\end{align*}
\normalsize

For $L_z=4, S_z=\pm 1/2$ there are 76 states, giving 10 doublets. The power sums are:
\small
\begin{align*}
\sum_{i=1}^{10} E_i(^2G)&=210F_0 - 628F_2 - 2372F_4 - 25060F_6\\    
\sum_{i=1}^{10} E_i(^2G)^2&=4410F_0^2 - 26376F_0F_2 - 99624F_0F_4 - 1052520F_0F_6 + 106760F_2^2 + 278456F_2F_4 + 1871968F_2F_6 + 875566F_4^2 \\& + 9478448F_4F_6 + 83410544F_6^2    
\end{align*}
\normalsize

For $L_z=3, S_z=\pm 5/2$ there are four states, giving another sextet $E(^6F)=21F_0 - 140F_2 - 462F_4 - 4004F_6$.

For $L_z=3, S_z=\pm 3/2$ there are 34 states, giving 5 quartets. The power sums are:
\small
\begin{align*}
\sum_{i=1}^5 E_i(^4F)&=105F_0 - 444F_2 - 1587F_4 - 14700F_6\\    
\sum_{i=1}^5 E_i(^4F)^2&=2205F_0^2 - 18648F_0F_2 - 66654F_0F_4 - 617400F_0F_6 + 55856F_2^2 + 301080F_2F_4 + 2168656F_2F_6 + 598695F_4^2 \\& + 8489208F_4F_6 + 50556632F_6^2    
\end{align*}
\normalsize

For $L_z=3, S_z=\pm 1/2$ there are 92 states, giving 10 doublets. The power sums are:
\small
\begin{align*}
\sum_{i=1}^{10} E_i(^2F)&=210F_0 - 540F_2 - 2160F_4 - 23604F_6\\    
\sum_{i=1}^{10} E_i(^2F)^2&=4410F_0^2 - 22680F_0F_2 - 90720F_0F_4 - 991368F_0F_6 + 111304F_2^2 + 324216F_2F_4 + 891296F_2F_6 + 763638F_4^2 \\& + 7816032F_4F_6 + 87556336F_6^2    
\end{align*}
\normalsize

For $L_z=2, S_z=\pm 5/2$ there are five states, giving one sextet $E(^6D)=21F_0 - 169F_2 - 396F_4 - 5005F_6$.

For $L_z=2, S_z=\pm 3/2$ there are 41 states, giving 6 quartets. The power sums are:
\small
\begin{align*}
\sum_{i=1}^6 E_i(^4D)&=126F_0 - 478F_2 - 2010F_4 - 19642F_6\\    
\sum_{i=1}^6 E_i(^4D)^2&=2646F_0^2 - 20076F_0F_2 - 84420F_0F_4 - 824964F_0F_6 + 75782F_2^2 + 332724F_2F_4 + 2150260F_2F_6 + 816144F_4^2 \\& + 11910108F_4F_6 + 77883638F_6^2
\end{align*}
\normalsize

For $L_z=2, S_z=\pm 1/2$ there are 106 states, giving 7 doublets. The power sums are:
\small
\begin{align*}
\sum_{i=1}^7 E_i(^2D)&=147F_0 - 127F_2 - 1320F_4 - 21175F_6\\    
\sum_{i=1}^7 E_i(^2D)^2&=3087F_0^2 - 5334F_0F_2 - 55440F_0F_4 - 889350F_0F_6 + 47143F_2^2 + 14040F_2F_4 + 321902F_2F_6 + 387222F_4^2 \\& + 7595280F_4F_6 + 71189503F_6^2
\end{align*}
\normalsize

For $L_z=1, S_z=\pm 5/2$ there are 6 states, giving one sextet $E(^6P)=21F_0 - 195F_2 - 528F_4 - 3003F_6$.

For $L_z=1, S_z=\pm 3/2$ there are 44 states, giving 2 quartets:
\small
\begin{align*}
E(^4P)=21F_0 - 27F_2 - 269F_4 - 4095F_6\pm\frac{1}{2}\sqrt{36096F_2^2 - 33344F_2F_4 - 594048F_2F_6 + 72304F_4^2 - 500864F_4F_6 + 4769856F_6^2}    
\end{align*}
\normalsize

For $L_z=1, S_z=\pm 1/2$ there are 114 states, giving 5 doublets. The power sums are:
\small
\begin{align*}
\sum_{i=1}^5 E_i(^2P)&=105F_0 - 243F_2 - 874F_4 - 14259F_6\\    
\sum_{i=1}^5 E_i(^2P)^2&=2205F_0^2 - 10206F_0F_2 - 36708F_0F_4 - 598878F_0F_6 + 28717F_2^2 + 94748F_2F_4 + 1209530F_2F_6 + 283492F_4^2 \\& + 3694460F_4F_6 + 47047693F_6^2
\end{align*}
\normalsize

For $L_z=0, S_z=\pm 7/2$ there is one state, having the octet eigenvalue $E(^8S)=21F_0 - 210F_2 - 693F_4 - 6006F_6$.

For $L_z=0, S_z=\pm 3/2$ there are 47 states, giving 2 quartets:
\small
\begin{align*}
E(^4S)=21F_0 - 90F_2 - 213F_4 - 3276F_6\pm\frac{1}{2}\sqrt{19200F_2^2 + 39360F_2F_4 - 436800F_2F_6 + 26640F_4^2 -952224F_4F_6 + 12322128F_6^2}    
\end{align*}
\normalsize

For $L_z=0, S_z=\pm 1/2$ there are 119 states, giving 2 doublets:
\small
\begin{align*}
E(^2S)=21F_0 - 42F_2 - 336F_4 - 3066F_6\pm\frac{1}{2}\sqrt{16896F_2^2 - 49632F_2F_4 - 73920F_2F_6 + 60336F_4^2 - 178080F_4F_6 - 845917680F_6^2}    
\end{align*}
\normalsize
\end{widetext}
This completes the $f^7$ configuration. The Hund's rule ground state is $^8S$.

This completes all the $f^n$ multiplets. These results reproduce the multiplet-averaged Coulomb interactions \cite{Slater}, the Hund's rule ground states, and the effective $U$ parameter \cite{vanderMarel1988}.

\begin{acknowledgments}
The author would like to thank Frank de Groot and Liu Hao Tjeng for discussions and interest in this work.
\emph{Funding:}
This work was supported by the U.S. Department of Energy (DOE), Office of Basic Energy Sciences,
Division of Materials Sciences and Engineering. 
\end{acknowledgments}
\bibliography{main}

\clearpage
\appendix*

\section{Python code to generate matrix elements}

The python code listed below takes as input the orbital index $l$, the number of particles $n$, and the total azimuthal angular momentum $L_z$ and spin $S_z$, and gives out the basis states along with their matrix elements. If the dimension of the matrix is less than 6, it also prints out the eigenvalues.

\begin{widetext}
\begin{lstlisting}[language=Python]
import sys
from sympy import Rational,sqrt,pi,symbols,zeros,simplify
from sympy.physics.wigner import gaunt
from sympy.combinatorics import Permutation
from itertools import combinations
from itertools import permutations
from collections import Counter

f0, f2, f4, f6 =symbols('f0 f2 f4 f6')
F = symbols('f0 f2 f4 f6')

l=2

# Define the 2*l possible single-electron states for an l-orbital
# Each state is a tuple (m_l, m_s)
single_electron_states = [(m_l, Rational(1,2)) for m_l in range(-l, l+1)] + \
                         [(m_l, Rational(-1,2)) for m_l in range(-l, l+1)]

# Function to find all valid f^n configurations with given total m_l and m_s
def enumerate_fn_configurations(n, target_m_l, target_m_s):
    valid_configs = []

    for config in combinations(single_electron_states, n):
        total_m_l = sum(state[0] for state in config)
        total_m_s = sum(state[1] for state in config)

        if total_m_l == target_m_l and total_m_s == target_m_s:
            valid_configs.append(config)

    return valid_configs

# Example usage
n = 5  # number of electrons in the l-shell
target_m_l = 3
target_m_s = Rational(1,2)

configs = enumerate_fn_configurations(n, target_m_l, target_m_s)

# Print results
print(f"Total configurations for f^{n} with M_L = {target_m_l}, M_S = {target_m_s}: {len(configs)}\n")
for cfg in configs:
    print(cfg)

# Count configurations
count = 0
for config in combinations(single_electron_states, n):
    ml = sum(s[0] for s in config)
    ms = sum(s[1] for s in config)
    if ml == target_m_l and ms == target_m_s:
        count += 1
print(count)

# Print results
print(f"Total configurations with M_L = {target_m_l}, M_S = {target_m_s}: {len(configs)}\n")

H = zeros(len(configs))

def coulomb_contribution(m1, m2, m3, m4, l):
    results = []

    if l ==3:
        # Sign conventions
        pre01 = -1 if m1 in [-3, -1, 1, 3] else 1
        pre02 = -1 if m3 in [-3, -1, 1, 3] else 1
        
        for k, fk in zip([0, 2, 4, 6], [f0, f2, f4, f6]):
            if k == 0:
                pre = 1
            elif k == 2:
                pre = 15
            elif k == 4:
                pre = 33
            elif k == 6:
                pre = Rational(429, 5)

            T1 = pre01 * pre * sqrt(4 * pi) * gaunt(l, k, l, -m1, m1 - m4, m4) / sqrt(2 * k + 1)
            T2 = pre02 * pre * sqrt(4 * pi) * gaunt(l, k, l, -m3, m3 - m2, m2) / sqrt(2 * k + 1)

            results.append(fk*T1 * T2)

    elif l ==2:
        # Sign conventions
        pre01 = -1 if m1 in [-1, 1,] else 1
        pre02 = -1 if m3 in [-1, 1] else 1
        
        for k, fk in zip([0, 2, 4], [f0, f2, f4]):
            if k == 0:
                pre = 1
            elif k == 2:
                pre = 7
            elif k == 4:
                pre = 21

            T1 = pre01 * pre * sqrt(4 * pi) * gaunt(l, k, l, -m1, m1 - m4, m4) / sqrt(2 * k + 1)
            T2 = pre02 * pre * sqrt(4 * pi) * gaunt(l, k, l, -m3, m3 - m2, m2) / sqrt(2 * k + 1)

            results.append(fk*T1 * T2)

    elif l == 1:
        # Sign conventions
        pre01 = -1 if m1 in [-1, 1] else 1
        pre02 = -1 if m3 in [-1, 1] else 1
        
        for k, fk in zip([0, 2], [f0, f2]):
            if k == 0:
                pre = 1
            elif k == 2:
                pre = 5

            T1 = pre01 * pre * sqrt(4 * pi) * gaunt(l, k, l, -m1, m1 - m4, m4) / sqrt(2 * k + 1)
            T2 = pre02 * pre * sqrt(4 * pi) * gaunt(l, k, l, -m3, m3 - m2, m2) / sqrt(2 * k + 1)

            results.append(fk*T1 * T2)
        
    return sum(results)

def get_required_coulomb_terms(state1, state2):
    """
    Given two Slater determinants (lists of n (m_l, m_s) tuples),
    return the list of (m1, m2, m3, m4) entries needed for
    antisymmetrized Coulomb terms:
    
    Each returned pair corresponds to:
        coulomb_contribution(m1, m2, m3, m4) - coulomb_contribution(m1, m2, m4, m3)
    
    If the matrix element is 0, returns an empty list.
    """

    terms = []

    # Case 1: Diagonal element
    if state1 == state2:
        terms = []
        for i in range(len(state1)):
            for j in range(i + 1, len(state1)):
                (ml1, ms1) = state1[i]
                (ml2, ms2) = state1[j]
                terms.append((+1, "direct", (ml1, ml2, ml1, ml2)))
                if ms1 == ms2:
                    terms.append((+1, "exchange", (ml1, ml2, ml2, ml1)))
        return terms

# Off-diagonal: compare as multisets to find differing orbitals
    c1 = Counter(state1)
    c2 = Counter(state2)

    removed = list((c1 - c2).elements())
    added   = list((c2 - c1).elements())

    if len(removed) != 2 or len(added) != 2:
        return []  # Not a 2-body transition

    # Try all matching permutations of removed→added
    for (a, b) in permutations(removed, 2):
        for (c, d) in permutations(added, 2):
            if a[1] == c[1] and b[1] == d[1]:
                trial_state = list(state1)
                try:
                    idx_a = trial_state.index(a)
                    idx_b = trial_state.index(b)
                except ValueError:
                    continue

                trial_state[idx_a] = c
                trial_state[idx_b] = d

                # Now check if this gives state2 (possibly up to ordering)
                if Counter(trial_state) != Counter(state2):
                    continue
                # Permutation needed to turn trial_state into state2
                # Convert to index mapping
                try:
                    perm = [trial_state.index(x) for x in state2]
                except ValueError:
                    continue  # mismatch

                sign = Permutation(perm).signature()
                ml1, ml2 = a[0], b[0]
                ml3, ml4 = c[0], d[0]

                terms.append((sign, "direct", (ml1, ml2, ml3, ml4)))
                if a[1] == b[1]:  # ← add this check!
                    terms.append((sign, "exchange", (ml1, ml2, ml4, ml3)))
                return terms
    
    return []

for i, cfg1 in enumerate(configs):
    for j, cfg2 in enumerate(configs):
        print(cfg1,"::::",cfg2)
        state1 = cfg1
        state2 = cfg2
        total_contrib=0
        terms = get_required_coulomb_terms(state1, state2)

        for sign, kind, (m1, m2, m3, m4) in terms:
            if kind == "direct":
                print(f"coulomb_contribution({m1}, {m2}, {m3}, {m4})", end="")
                print(sign,m2,m1,m3,m4)
                total_contrib=total_contrib+sign*coulomb_contribution(m2, m1, m3, m4,l)
            elif kind == "exchange":
                print(f" - coulomb_contribution({m1}, {m2}, {m4}, {m3})", end="")
                print(sign,m2,m1,m3,m4)
                total_contrib=total_contrib-sign*coulomb_contribution(m2, m1, m3, m4,l)

        H[i, j] = total_contrib
        print(i, j, H[i,j])
        print()
print(H.trace())

if len(configs) <6:
    print(H.eigenvals())
else:
    print('Matrix too large to diagonalize symbolically')

sys.exit()

\end{lstlisting}
\end{widetext}

\end{document}